\documentclass[twocolumn]{aastex6}
\usepackage{graphicx}
\usepackage{amssymb,amsfonts,amsmath,amstext,amsgen,amsopn,amsxtra,indentfirst,times,subfigure,natbib}
\begin{document}
\title{THREE-DIMENSIONAL CIRCULATION DRIVING CHEMICAL DISEQUILIBRIUM IN WASP-43b}
\author{Jo\~{a}o M. Mendon\c{c}a\altaffilmark{1,2}}
\author{Shang-min Tsai\altaffilmark{2}}
\author{Matej Malik\altaffilmark{2}}
\author{Simon L. Grimm\altaffilmark{2}}
\author{Kevin Heng\altaffilmark{2}}
\altaffiltext{1}{Astrophysics and Atmospheric Physics, National Space Institute, Technical University of Denmark, Elektrovej, 2800 Kgs. Lyngby, Denmark}
\altaffiltext{2}{University of Bern, Center for Space and Habitability, Gesellschaftsstrasse 6, CH-3012, Bern, Switzerland.  Emails: joao.mendonca@space.dtu.dk, shang-min.tsai@space.unibe.ch, matej.malik@csh.unibe.ch, simon.grimm@csh.unibe.ch, kevin.heng@csh.unibe.ch}

\begin{abstract}

Spectral features in the observed spectra of exoplanets depend on the composition of their atmospheres. A good knowledge of the main atmospheric processes that drive the chemical distribution is therefore essential to interpret exoplanetary spectra. An atmosphere reaches chemical equilibrium if the rates of the forward and backward chemical reactions converge to the same value. However, there are atmospheric processes, such as atmospheric transport, that destabilize this equilibrium. In this work we study the changes in composition driven by a 3D wind field in WASP-43b using our Global Circulation Model, \texttt{THOR}. Our model uses validated temperature- and pressure-dependent chemical timescales that allow us to explore the disequilibrium chemistry of CO, CO$_2$, H$_2$O and CH$_4$. In WASP-43b the formation of the equatorial jet has an important impact in the chemical distribution of the different species across the atmosphere. At low latitudes the chemistry is longitudinally quenched, except for CO$_2$ at solar abundances. The polar vortexes have a distinct chemical distribution since these are regions with lower temperature and atmospheric mixing. Vertical and latitudinal mixing have a secondary impact in the chemical transport. We determine graphically the effect of disequilibrium on observed emission spectra. Our results do not show any significant differences in the emission spectra between the equilibrium and disequilibrium solutions for C/O = 0.5. However, if C/O is increased to 2.0, differences in the spectra due to the disequilibrium chemistry of CH$_4$ become non-negligible. In some spectral ranges the emission spectra can have more than 15$\%$ departures from the equilibrium solution. 

\end{abstract}

\section{Introduction}
\subsection{Background}

WASP-43b is a hot Jupiter planet with twice the mass and roughly the same size as Jupiter (\citealt{2011Hellier}). With a semi-major axis of approximately 0.0153 AU, WASP-43b orbits around WASP-43 in just 19.2 hours (\citealt{2012Gillon}). WASP-43 is a K7 star with a mass of 0.73 M$_{\odot}$. The atmosphere of WASP-43b has an equilibrium temperature close to 1440 K (e.g., \citealt{2014Blecic}) and its properties have been studied in previous works. Some of the highlights are the non-detection  of thermal inversion on the dayside of the planet (e.g., \citealt{2012Gillon}; \citealt{2013Wang}; \citealt{2014Blecic}; \citealt{2014Stevenson}) and physical constraints on water abundances from \cite{2014Kreidberg}. In this work, we assume WASP-43b to be tidally locked due to its proximity to the parent star, which induces large tidal stresses on the planet (e.g., \citealt{1996Guillot} and \citealt{2014Showman}). The main planet bulk parameters are shown in Table \ref{tab:model}. \cite{2015Kataria} was the first study to explore WASP-43b with a Global Circulation Model (GCM) and to find a reasonable fit of the model dayside emission spectra to the \textit{Hubble Space Telescope }(HST) data from Stevenson et al. 2014.  However, in the nightside, the results from \cite{2015Kataria} under-predict the emission fluxes (the nightside is too bright compared to the measured HST data). In \cite{2017Mendoncaa}, we showed that a permanent gray cloud structure in the nightside of WASP-43b is consistent with HST \citep{2014Stevenson} and {\it Spitzer} data (\citealt{2017Mendoncaa}).  \cite{2017Mendoncaa} also refers to the possibility that the phase-curve at 4.5 $\mu$m is better fitted by a GCM-generated phase curve that is enhanced in carbon dioxide relative to its chemical equilibrium. Chemical disequilibrium processes occur in the atmosphere mainly due to atmospheric transport and photochemistry processes. However, at high temperatures in the dayside of hot Jupiter planets, the effects of photochemistry become relatively minor compared to thermochemical processes (e.g., \citealt{2011Moses}, \citealt{2012Kopparapu}, \citealt{2014Moses}). Atmospheric dynamics becomes then the dominant process leading to chemical disequilibrium in hot Jupiter atmospheres. This process can be understood in terms of timescales: disequilibrium happens when the fast transport process (by mean motion and eddies) has a shorter timescale ($\tau_{dyn}$) than the chemical timescale ($\tau_{che}$), and the chemical processes are not fast enough to push the chemical constituents back to chemical equilibrium. The chemical timescale refers to the mean characteristic time for production or destruction of a chemical species by local chemical processes. Different scenarios are possible from the competition between dynamical and chemical processes:
\begin{itemize}
\item $\tau_{dyn}$ $\ll$ $\tau_{che}$. The distribution of the chemical species is dominated by the dynamical transport compared to the localized sources and sinks. In this situation, the dispersive effects of dynamical transport causes the chemical species to be well mixed in the atmosphere 
\item $\tau_{dyn}$ $\gg$ $\tau_{che}$. Large variability is expected in the distribution of the chemical species, which is kept close to chemical equilibrium (depends solely on local temperature, pressure and elemental abundance).
\item $\tau_{dyn}$ and $\tau_{che}$ are the same order of magnitude. This scenario is more difficult to analyse than the two cases above because both chemistry and dynamics have an important role determining the distribution of the chemical species. 
\end{itemize}

The thermochemical equilibrium is maintained in the hot deep regions where $\tau_{che}$ is typically shorter than $\tau_{dyn}$, but at lower pressures the two timescales start to become comparable, and the departures from equilibrium in the atmospheres become larger. When $\tau_{che}$ becomes larger than $\tau_{dyn}$ we say that the chemical component became ``quenched''. A description of the quenching effect was first described in \cite{1977Prinn} to explain the distribution of CO in Jupiter's troposphere, which is associated with a rapid upward mixing from the deeper atmosphere. 


\subsection{Motivation}
\label{subsec:Motiv}

A good understanding of the chemical distribution across the atmosphere is essential to improve our 3D simulations and interpretation of exoplanet atmospheric spectra. Advances on this topic will lead us into a deeper knowledge on how different atmospheric processes control the climate in the planets. Theoretical and observational studies of hot Jovian atmospheres (e.g., \citealt{2009Showman}, \citealt{2009knutson}) show that these atmospheres are very dynamically active with equatorial jets of a few km/s speed, which indicates that a simple thermochemical equilibrium calculation based on local pressure and temperature is not enough to determine its chemical composition across the atmosphere.

Studies with different levels of complexity have explored the problem of chemical disequilibrium driven by atmospheric transport. One-dimensional models, for example, include complex kinetic codes with hundreds to thousands of reactions (e.g., \citealt{2012Kopparapu}; \citealt{2011Moses}; \citealt{2015Venot}; \citealt{2014Hu}; \citealt{2016Rimmer}; \citealt{2017Tsai}), but very simplified representations of atmospheric transport (e.g., eddy diffusion parameterisation). To interpret the spectra from observational results it is necessary that we start looking into the impact of three-dimensional atmospheric transport processes, however, the chemical kinetic schemes used in the simple one-dimensional models are too computational expensive to be coupled directly with three-dimensional models (e.g., GCMs). Self-consistent three-dimensional simulations require the computation of the chemical kinetic code for each grid box (usually thousands of boxes covering the entire model domain), over a long time integration until the steady state solution is reached. This latter property is very important in numerical simulations to avoid the final conclusions being influenced by the initial conditions. Despite the complexity, several works have broken through the three-dimensional problem using approaches of intermediate complexity. This hierarchy in model complexity is very important to guide us through the learning process of complex atmospheric mechanisms. Examples are the work of \cite{2012Agundez} and \cite{2014Agundez}. In these two works the complex chemical kinetics model was coupled with a simplified model to represent the dynamics based on a constant solid-body rotation. The prescribed circulation mimicked the broad equatorial jet predicted by theoretical GCMs (e.g., \citealt{2009Showman}; \citealt{2015Heng}). The simplification in the model made it possible to explore conceptually how the atmospheric transport (along the longitude) impacts the chemical distribution across the atmosphere.

Another approach was the pioneering work of \cite{2005Cooper, 2006Cooper}, who used a GCM but with a simplified chemical kinetics model. In this work, the complex chemical network is replaced by a simple parameterization called the chemical relaxation method, which consists of a single Newtonian equation where the chemical species are forced towards an equilibrium value. The strength of the source/sink term in the equation depends on a chemical timescale. \cite{2005Cooper, 2006Cooper}, studied the molecular distribution of CO across the HD 209458b planet atmosphere. Their study showed that the vertical winds have an important role in the distribution of CO (vertical quenching). Their chemical relaxation scheme is based on the assumption that a single rate-limiting reaction can describe the interconversion between CO and CH$_4$ over a broad range of pressure and temperature. In \cite{2018Drummond} the simple relaxation method developed in \cite{2006Cooper} was coupled with a nongray radiation scheme. The results show that in HD 209458b the horizontal transport also has an important impact in the final chemical distribution of methane.

\cite{2018Tsai} extended and validated the model of \cite{2006Cooper}, and developed a pathway analysis tool that identifies the rate-limiting reaction as a function of temperature, pressure and chemical abundances. \cite{2006Cooper} used essentially a shorter chemical timescale of CO, which was determined by a single rate-limiting reaction. \cite{2006Cooper} and \cite{2018Drummond} only calculate CO using the relaxation method and relate CH$_4$ through mass balance, assuming that all the carbons are locked in either CH$_4$ and CO. We emphasize that this mass balance relation is only valid when the system is in or close to chemical equilibrium and hence not applicable to the relaxation method (see \citealt{2018Tsai}). Our method makes it possible to study chemical disequilibrium of several chemical species using a GCM at very low computational cost. In this work, we apply the formulation developed in \cite{2018Tsai} to study the chemical distribution of CO, CH$_4$, CO$_2$ and H$_2$O on WASP-43b, with our GCM, \texttt{THOR} (\citealt{2016Mendoncab}). This implementation is part of a model hierarchy (such as the others mentioned above) devoted to studying disequilibrium chemistry in the atmosphere of hot Jupiters. The model presented still does not include a self-consistent representation of the chemistry in the radiative transfer code. The representation of the radiation in the GCM simulations are also simplified (gray radiative transfer code) as we explain below. However, our model has an intermediate level of complexity necessary to help us gain physical intuition on the chemical disequilibrium problem. Changes in the atmospheric composition due to atmospheric transport have an impact on the observed atmospheric spectra. Coupling the new relaxation method with \texttt{THOR} allow us to have better knowledge of the distribution of the different chemical species, which is a powerful tool in the interpretation of atmospheric spectra. We also explore and analyse the distribution of different chemical species, however, further detailed analyses of the main dynamical mechanisms transporting these components by, for example, eddies, wave breaking, diffusion, and the impact of different spatial resolutions is beyond the scope of this paper.

\subsection{Structure of the present study}
In the next section \ref{sec:model}, we explain the theoretical tools, including the GCM and chemical relaxation method used to explore the 3D chemical distribution in WASP-43b. In section \ref{sec:basesimu}, we analyse the results for multiple scenarios integrated with the GCM: with and without thermal inversion in the dayside. In section \ref{sec:des_che}, we study the chemical distribution across the atmosphere for the different species and compare the physical timescales involved to determine the main process controlling the chemical distribution. In section \ref{sec:phcrv_spec}, we compare the emission spectra from equilibrium and disequilibrium chemistry. Final concluding remarks and future prospects are discussed in section \ref{sec:conclu}.

\section{Global Circulation Model}
\label{sec:model}
In this work a comprehensive GCM is used to explore the disequilibrium chemistry in the atmosphere of WASP-43b. This model includes a deep non-hydrostatic code, \texttt{THOR}, a simplified radiative transfer scheme and chemical network, and a convection scheme that mixes entropy in statically unstable atmospheric columns. Note that in this work the radiative transfer does not take into account the changes in chemistry during the model integration.

In Table \ref{tab:model}, we show the main model parameters that represent the parent star (WASP-43) and the planet (WASP-43b). Below, we describe dynamical and physical packages which are included in the GCM.

\begin{table}
\begin{center}
\caption{Model parameters used in the baseline simulation of WASP-43b.}
\begin{tabular}{ | l | l | l |}
\hline
 Parameters &  & Units  \\ \hline \hline
 Star Temperature & 4520 & K \\ \hline
 Planet distance & 0.015 & AU \\ \hline
 Mean Radius & 72427 & km \\ \hline
 Gravity & 47.0 & m/s$^2$ \\ \hline
 Gas constant & 3714 & J/K/kg \\ \hline
 Specific heat & 13000 & J/K/kg \\ \hline
 Bond albedo & 0.18 & - \\ \hline
 Highest pressure & $\sim$100 & bar \\ \hline
 Interior flux & $\sim$ 27 & kW/m$^2$ \\ \hline
 Rotation rate & $9.09\times 10^{-5}$ & s$^{-1}$ \\ \hline
 Orbit inclination & 0 & deg \\ \hline
 Orbit eccentricity & 0 & deg \\ \hline
 \end{tabular}
\end{center}
\label{tab:model}
\end{table}

\subsection{Dynamical Core}
\label{subsec:dcore}
The dynamical core solves the equations that represent the dynamical behaviour of the resolved global atmospheric flow. For this work we use our planetary GCM \texttt{THOR}\footnote{\texttt{THOR} is an open-source software designed to run on Graphics Processing Units (GPUs): https://github.com/exoclime/THOR or https://bitbucket.org/jmmendonca/thor.} (\citealt{2016Mendoncab}). \texttt{THOR} was developed from scratch to explore a large diversity of planetary conditions, and it solves the full atmospheric fluid equations in a rotating sphere (fully compressible non-hydrostatic system). Numerical solvers based on finite-differences or finite-volume as the one used in \texttt{THOR} have to be horizontally discretized on a spherical grid. A simple latitude-longitude grid is associated with the convergence of the meridians at the poles which largely constrains the time-step at high latitudes to maintain the model stability (known as the ``pole problem''). In order to relax the Courant-Friedrichs-Lewy condition at high latitudes the equations in \texttt{THOR} are solved on an icosahedral grid, which is a quasi-uniform grid. The icosahedral grid is also modified to increase the numerical accuracy. In this step, the grid distortions are smoothed using a method called ``spring dynamics'' (\citealt{2001Tomita}). 

To preserve numerical stability we include fourth-order hyperdiffusion and divergence damping as explained in \citealt{2016Mendoncab}. This parameterization removes the numerical noise and also represents the physical phenomena of eddy viscosity and turbulence in the sub-grid scale. The diffusion timescale applied was 940 s. The horizontal and vertical resolutions are roughly 4 degrees for the horizontal and 40 equally spaced layers for the vertical, which covers the atmosphere from roughly 100 bar up to a pressure level of roughly 0.01 hPa. 

\subsection{Physics Core} 
\label{subsec:phys}

Coupled with the dynamical core we include three parameterizations which represent the radiation, chemical kinetics and convection processes in the atmosphere. We have also implemented a  ``sponge'' layer scheme to improve the stability of the numerical simulations. The time-step in the physical core was set to 100 seconds when running without the chemical kinetic code, and 30 seconds with chemistry. The required short time-step when using the chemical relaxation timescale is associated with the short chemical timescales obtained in the very deep atmosphere, which are of the order of tens of seconds.

\subsubsection{Radiative transfer scheme}
\label{subsec:rtm}
The radiation scheme is based on a two-band model: one band solves the equations that represent the stellar radiation and the other the thermal emission from the different atmospheric layers and planet interior. For the thermal radiation we calculated an opacity ($k_{planet}$) of 0.05 cm$^2$g$^{-1}$, which sets the peak emission of the photosphere at roughly 100 mbar consistent with \cite{2014Stevenson}. In this work we define the greenhouse parameter $G$ as,
\begin{equation}
G = \frac{k_{star}}{k_{planet}}.
\end{equation}
Varying the value of $k_{star}$, which is the opacity for the equation that solves the stellar radiation across the atmosphere, we can control the dayside temperature gradient in the upper atmosphere. The observed dayside thermal structure in WASP-43b is well represented by G equals to 0.5, as shown in \cite{2017Mendoncaa}.  The 3D temperature structure produced with G equals to 0.5 is also quantitatively similar to the results produced in \cite{2015Kataria}, that simulated WASP-43b with a more complex radiative transfer model code. In this work we also explore $G$ equal to 2, which produces a thermal inversion  (the temperature increases with altitude) in the dayside due to the larger deposition of stellar energy at lower pressures. We include a bond albedo of 0.18 that was estimated in \cite{2014Stevenson}. The globally averaged flux coming from the planet's interior ($\sim 50$ kW/m$^2$) was calculated to represent an equilibrium temperature consistent with the observed thermal phase-curve of WASP-43b (\citealt{2014Stevenson} and \citealt{2017Mendoncaa}). We include the same gray cloud opacity structure in the nightside as in the reference simulation of \cite{2017Mendoncaa}, which sets the photosphere of the permanet nightside to roughly 10 mbar. We showed in \cite{2017Mendoncaa} that a cloud deck in the nightside of WASP-43b produces a thermal emission spectra consistent with HST - Wide Field Camera 3 (WFC3) (\citealt{2014Stevenson}) and Spitzer (\citealt{2017Mendoncaa}) data. Note that the gray clouds forming due to the cooler temperatures of the nightside, represent cloud particles that are large compared to the wavelengths of thermal emission and that the timescale for them to condense out of the atmospheric gas is short compared to any radiative or dynamical timescales (\citealt{2017Mendoncaa}).

The zenith angle is the angle between the zenith point and the center of the star's disc, and the amount of incoming stellar flux at the top of the models domain is weighted by the cosine of the zenith angle. The positive values of the cosine are related to the dayside of the planet, negative to the nightside and zero at the terminator. As in \cite{2017Mendoncaa}, we correct the solar path-length to take into account the effect of the atmospheric spherical curvature. See \cite{2017Mendoncaa} for details on the equations solved.

In section \ref{sec:phcrv_spec}, we present the multi-wavelength phase-curves that are obtained from post-processing the results from \texttt{THOR} with a more sophisticated radiative transfer model. The spectra
are generated combining the radiative emission solution from \cite{2015Mendonca} and \cite{2017Malik}. For these results we include the main absorbers in the infrared from these line lists --- EXOMOL: H$_2$O (\citealt{2006Barber}, CH$_4$ (\citealt{2014Yurchenko}), NH$_3$ (\citealt{2011Yurchenko}), HCN (\citealt{2006Harris}), H$_2$S (\citealt{2016Azzam}) -- HITEMP (\citealt{2010Rothman}): CO$_2$, CO -- HITRAN (\citealt{2013Rothman}): C$_2$H$_2$. We also add the resonance lines for Na and K as described in \cite{2011Draine} and add CIA H$_2$-H$_2$ H$_2$-He absorption (\citealt{2012Richard}). We also include in the post-processing analysis the cloud structure used in GCM simulations and defined in \cite{2017Mendoncaa}. As in \cite{2017Mendoncaa}, spectra from the planet atmosphere is obtained from projecting the outgoing intensity for each geographical location of the observed hemisphere that moves with the orbital phase. The spectral resolution in our plots is  3000 spectral bins covering a spectral range from 0.3 $\mu$m to 10 cm. The stellar flux was interpolated from the PHOENIX model database (\citealt{1995Allard}; \citealt{2013Husser}). The species in chemical equilibrium were computed using FastChem (\citealt{2017Stock}). 

\subsubsection{Chemistry - relaxation time scale scheme}
\label{subsubsec:chem}

The chemical kinetics calculations in the 3D simulations are performed using the chemical timescales developed in \cite{2018Tsai}. The abundances of the chemical species are updated during the simulations with a backward Euler scheme which is an implicit formulation defined by:
\begin{equation}
\label{eq:eq_che}
\chi_{new} = \frac{\chi+\chi_{EQ}\times\frac{dt}{\tau_{che}}}{1+\frac{dt}{\tau_{che}}},
\end{equation}
where $\chi$ represents the current volume mixing ratio of the chemical species, which is forced linearly towards its chemical equilibrium $\chi_{EQ}$ with a timescale of $\tau_{che}$ (\citealt{1998Smith}; \citealt{2006Cooper}). 

In the particular case of CO$_2$, we use a slightly different version of Eq. \ref{eq:eq_che}:
\begin{equation}
\label{eq:eq_che_co2}
[CO_2]_{new} = \frac{[CO_2]+[CO_2]_{pseudo}\times\frac{dt}{\tau_{che}}}{1+\frac{dt}{\tau_{che}}},
\end{equation}
where the the CO$_2$ concentration ($[CO_2]$) is forced towards $[CO_2]_{pseudo}$ defined by:
\begin{equation}
\label{eq:eq_che_pseudo}
[CO_2]_{pseudo} = [CO_2]_{EQ}\times\frac{[CO][H_2O]}{[CO]_{EQ}[H_2O]_{EQ}}.
\end{equation}
These two equations are based on the efficient conversion of CO$ + $H$_2$O $\leftrightarrow$ CO$_2 + $H$_2$. CO$_2$ relaxes towards a pseudo equilibrium because its production scheme is relatively fast, which means it can keep a pseudo equilibrium while some ingredients for its production are already quenched (\citealt{2018Tsai}). 

The distribution across the atmosphere of each chemical tracer is defined by the flux form equation plus a net source term (\texttt{S}) in the right side of the equation:
\begin{equation}
\label{eq:transp_che}
\frac{\partial \rho \chi}{\partial t} + \nabla \cdot (\rho \textbf{v} \chi) = \texttt{S},
\end{equation}
where $\rho$ is the atmospheric density and \textbf{$v$} the 3D wind field. \texttt{S} includes the chemical production and loss, and the turbulent diffusion processes working at sub-grid scales. The turbulent diffusion is parameterized with a horizontal fourth-order hyperdiffusion with a timescale of 940 s. The finite volume method used to solve the equations is the same as used to solve the entropy equation from the fluid equations (see \citealt{2016Mendoncab} for more information). Without the \texttt{S} term, the chemical tracer would be completely well mixed by the atmospheric dynamics after a long integration period. However, the chemical source/sink term forces the system to maintain spatial variability, and as it was pointed out in section \ref{subsec:Motiv}, the chemical distribution will depend on the balance between the chemical timescale ($\tau_{che}$) and the dynamical timescale ($\tau_{dyn}$). 

Our model has the limitation of not including photochemical effects. Read more about photochemistry effects on the atmosphere of hot jupiters in \cite{2014Moses}.

\subsubsection{Convection}
\label{subsubsec:conv}
A simple convective adjustment scheme was included. This simple parameterization represents the vertical mixing of potential temperature in a buoyantly unstable atmospheric column. The instability condition is satisfied when $\delta \theta/\delta p > 0$. During the mixing of the potential temperature in an unstable column the total enthalpy ($Cp\times T$) is conserved. The final mean potential temperature is found assuming a dry atmosphere,
\begin{equation}
\label{equ:mixpt}
\bar{\theta} =\frac{\int^{p_{top}}_{p_{bot}} C_p\theta \Pi d p}{\int^{p_{top}}_{p_{bot}} C_p\Pi dp}
\end{equation}
where $\Pi$ is the exener function ($(p/p_0)^{Rd/Cp}$), and $p_{top}$ and $p_{bot}$ are the pressures at the top and bottom of the unstable column (e.g., \citealt{1965Manabe}; \citealt{2016Mendoncab}). 

\subsubsection{Boundary layer}
\label{subsubsec:BL}
At the top of the model's domain a sponge parameterization is used to damp the \emph{eddy} component of the wind field to zero. These sponges avoid any non-physical wave reflections at the model's boundaries. More details about the scheme used is included in Appendix \ref{apxd:sponge}.

\section{Reference simulations}
\label{sec:basesimu}

\begin{figure}
\label{fig:dyn_spin}
\begin{centering}
\includegraphics[width=1.0\columnwidth]{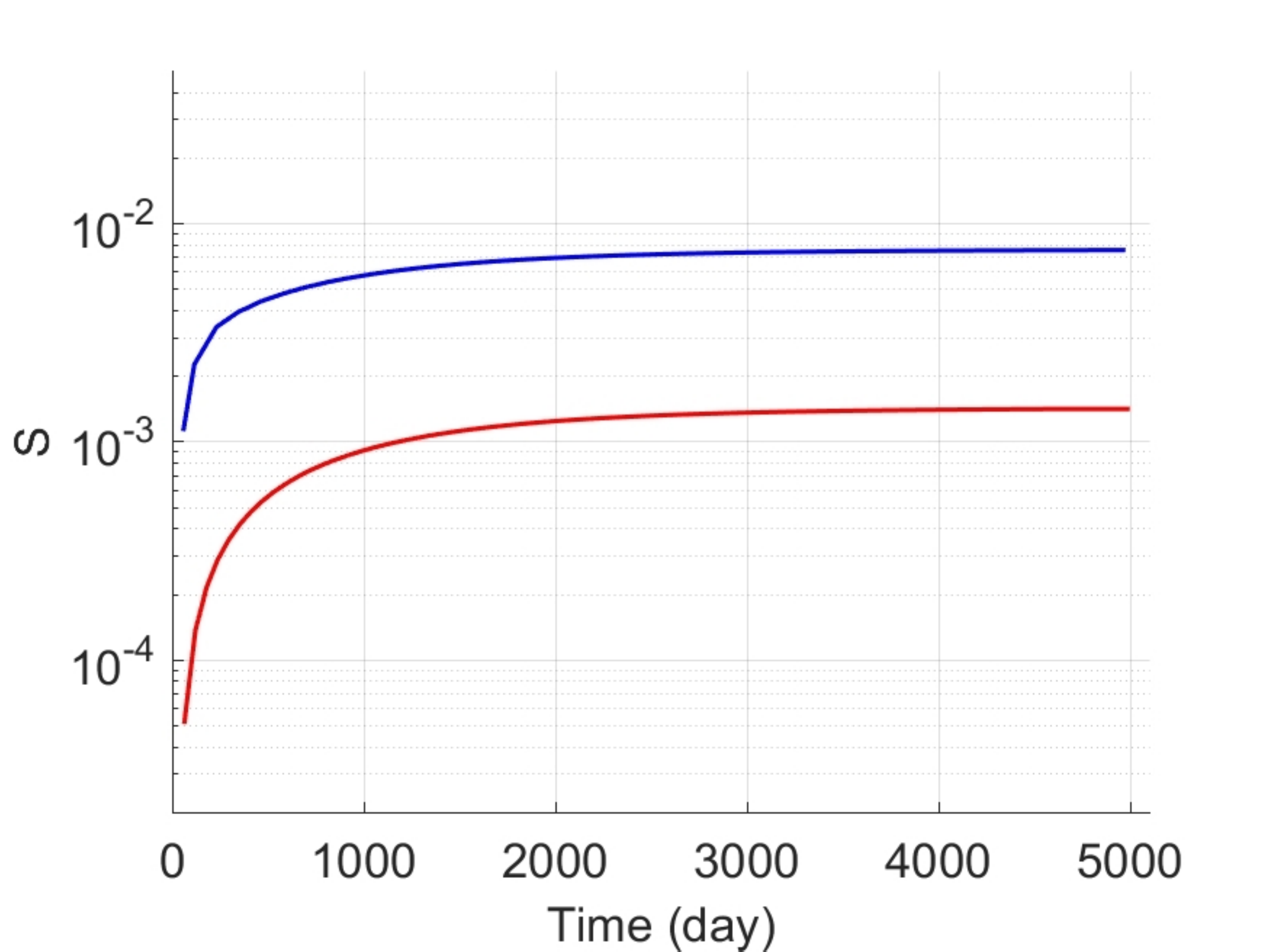}
\caption{Global super-rotation index ($S$) for the reference simulations: G = 0.5 (blue) and G = 2 (red). $S$ measures the excess of total angular momentum in the atmosphere (\citealt{1986Read}) and is used in this work to analyse the spin-up phase of the GCM simulations.}
\label{fig:spinup}
\end{centering}
\end{figure}

\begin{figure*}
\begin{centering}
\subfigure[G = 0.5]{
\includegraphics[width=0.5\columnwidth]{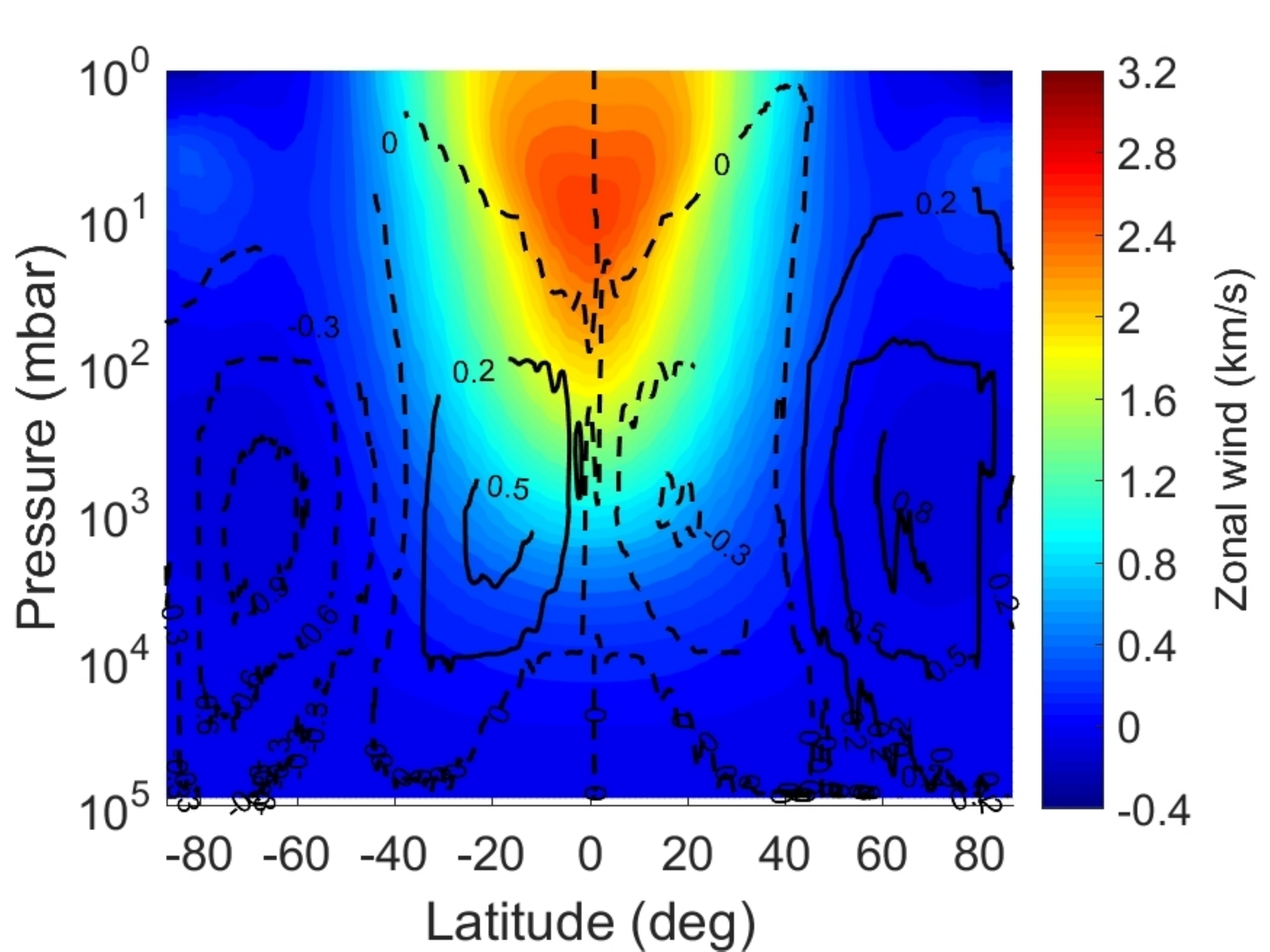}}
\subfigure[G = 0.5]{
\includegraphics[width=0.5\columnwidth]{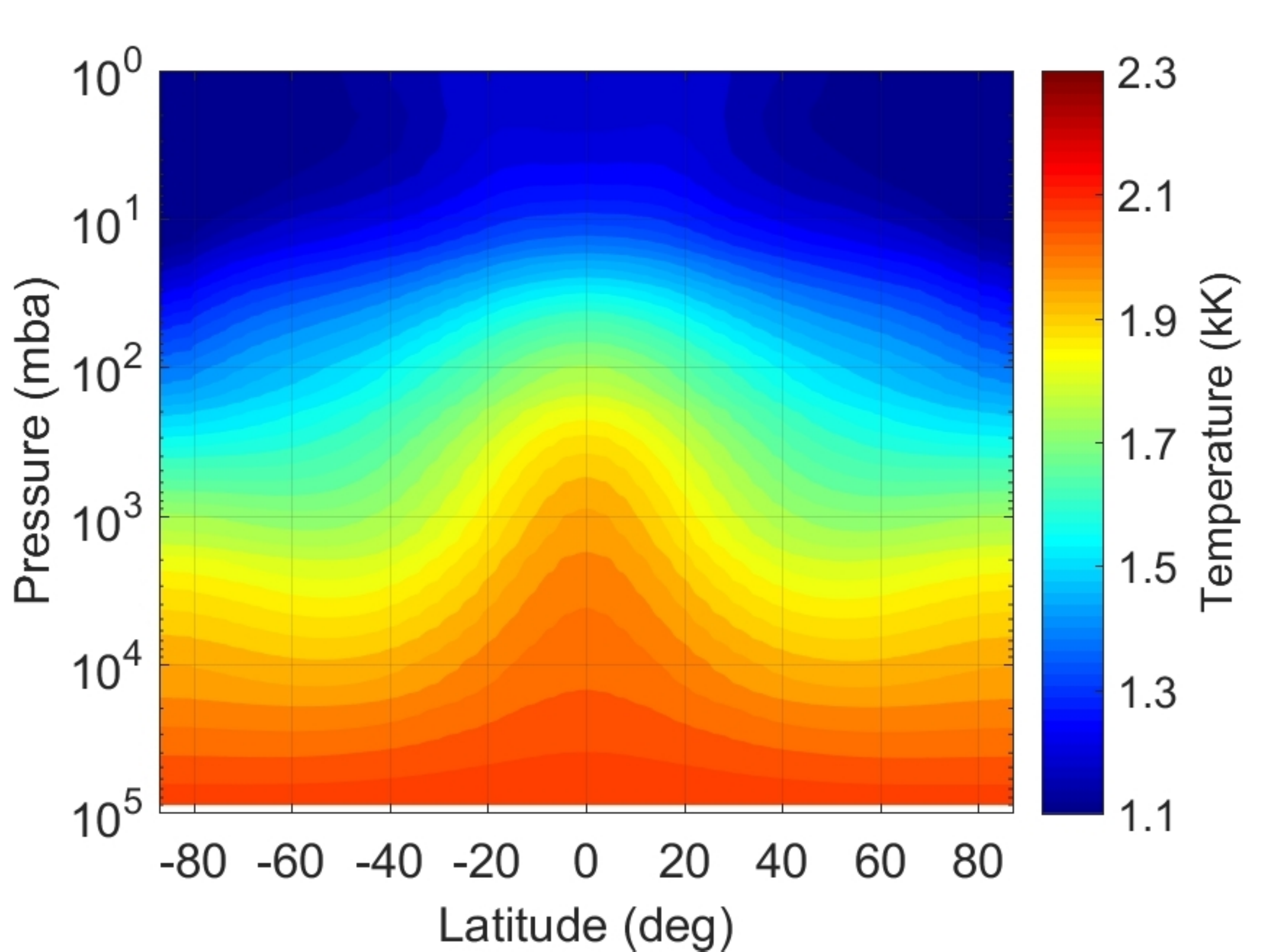}}
\subfigure[G = 0.5]{
\includegraphics[width=0.5\columnwidth]{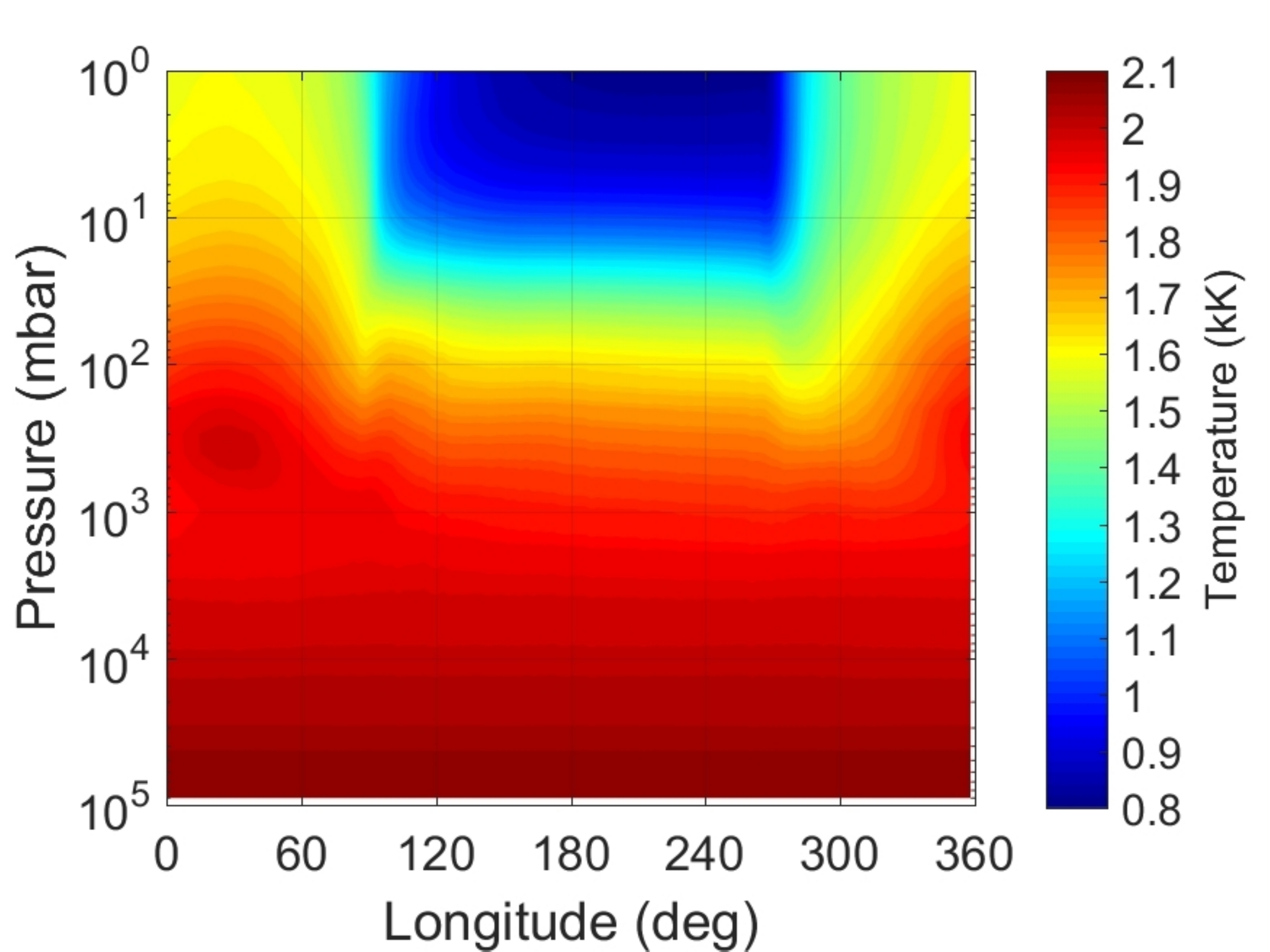}}
\subfigure[G = 0.5]{
\includegraphics[width=0.5\columnwidth]{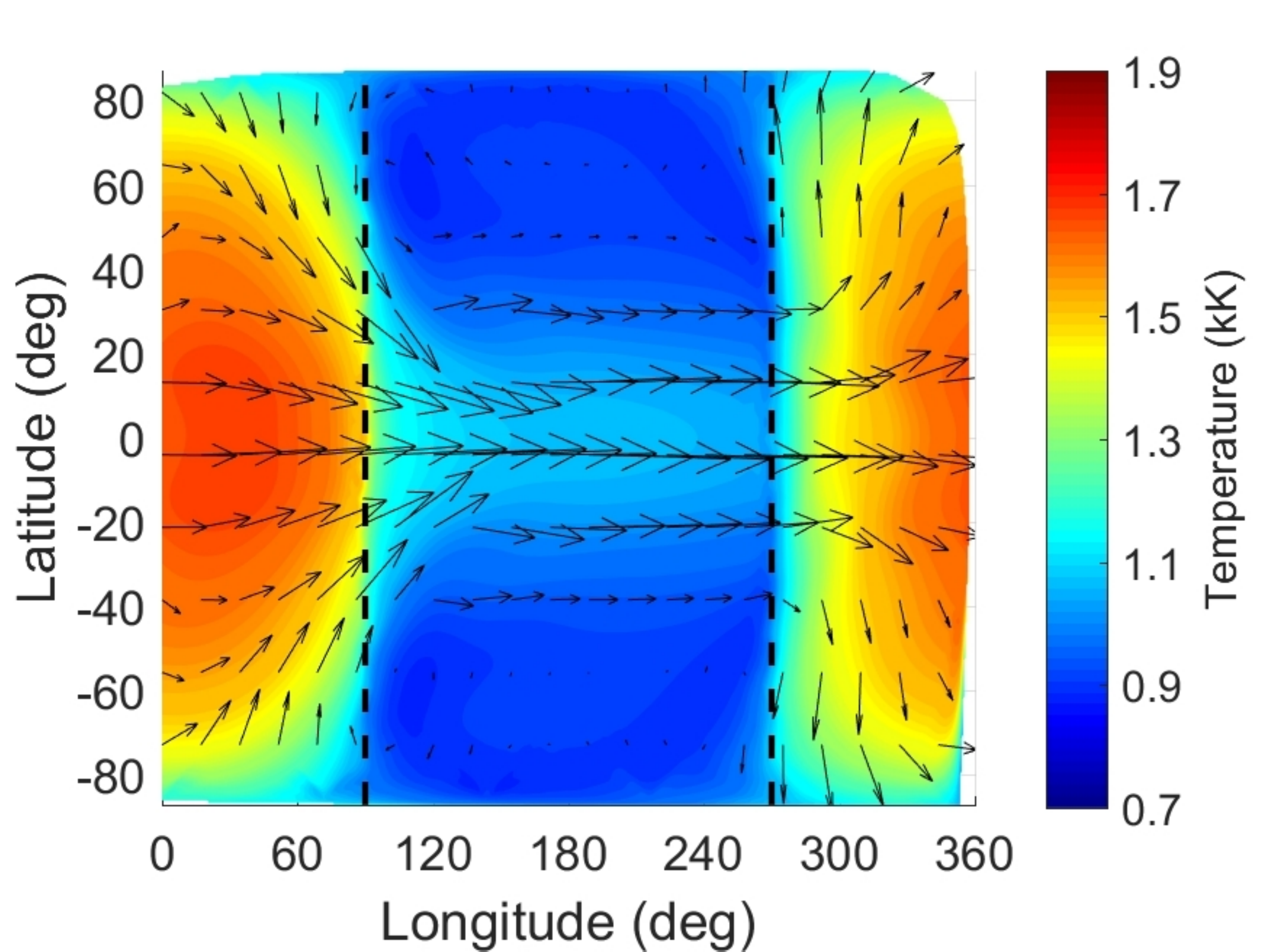}}
\subfigure[G = 2.0]{
\includegraphics[width=0.5\columnwidth]{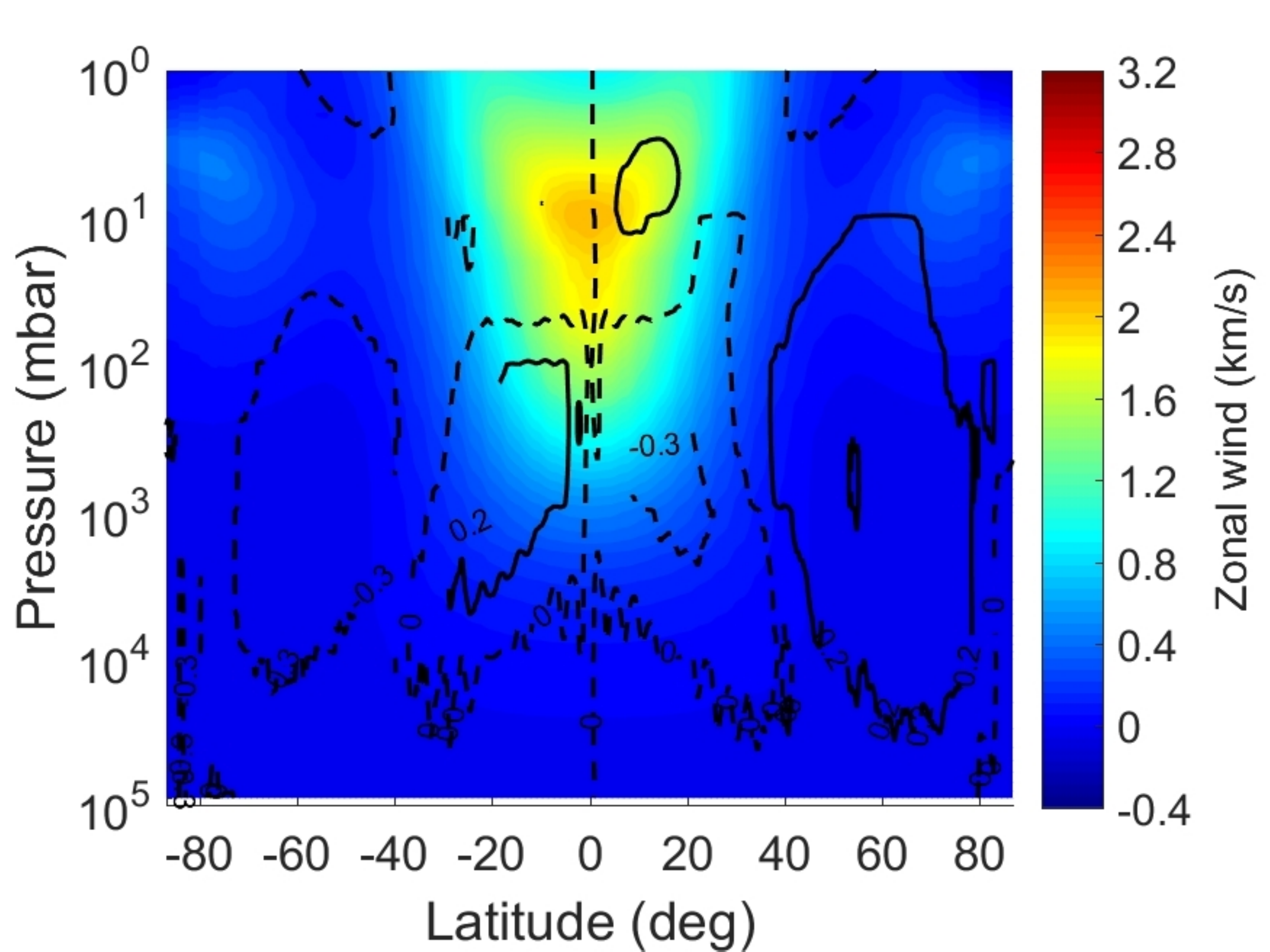}}
\subfigure[G = 2.0]{
\includegraphics[width=0.5\columnwidth]{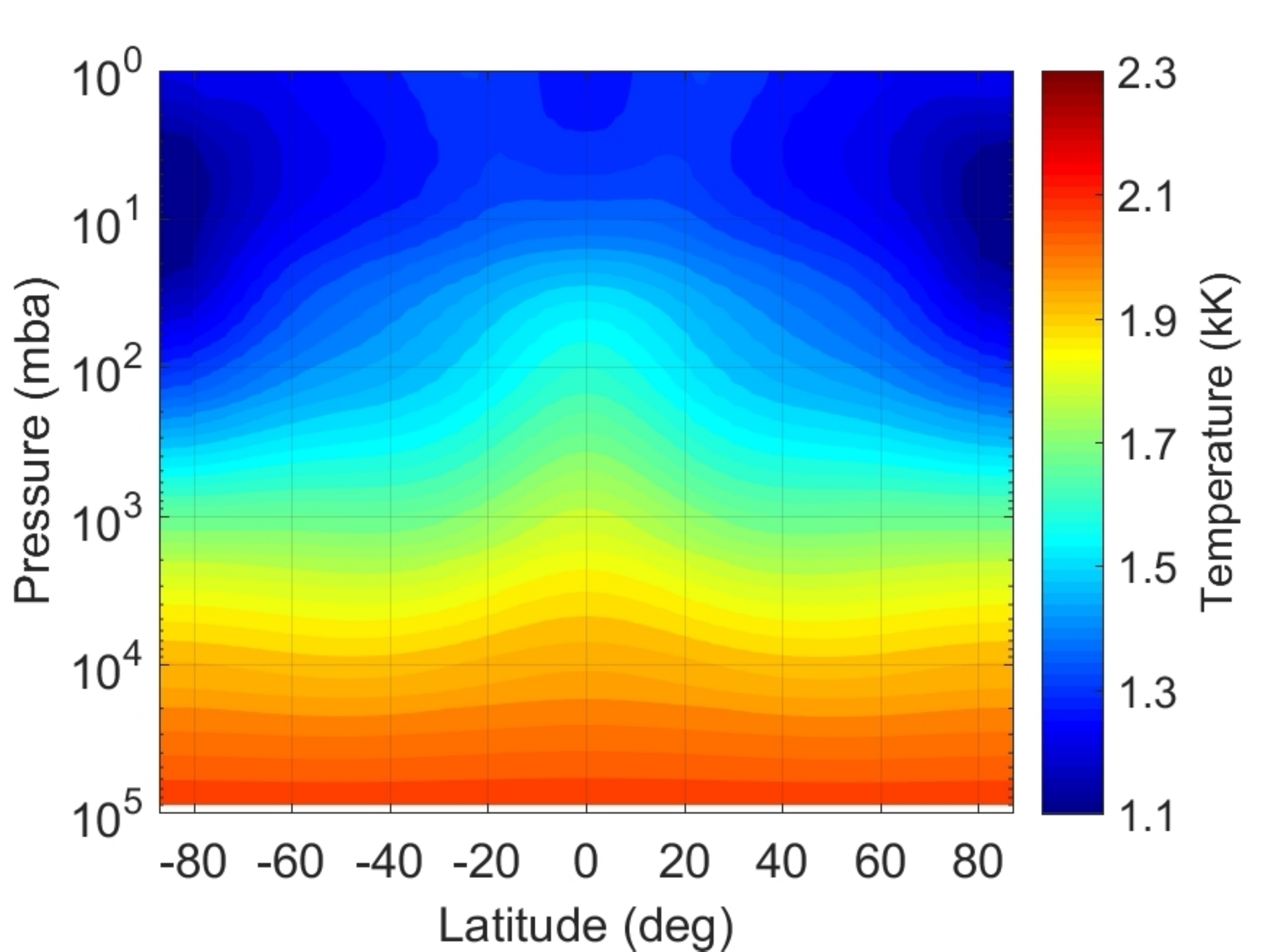}}
\subfigure[G = 2.0]{
\includegraphics[width=0.5\columnwidth]{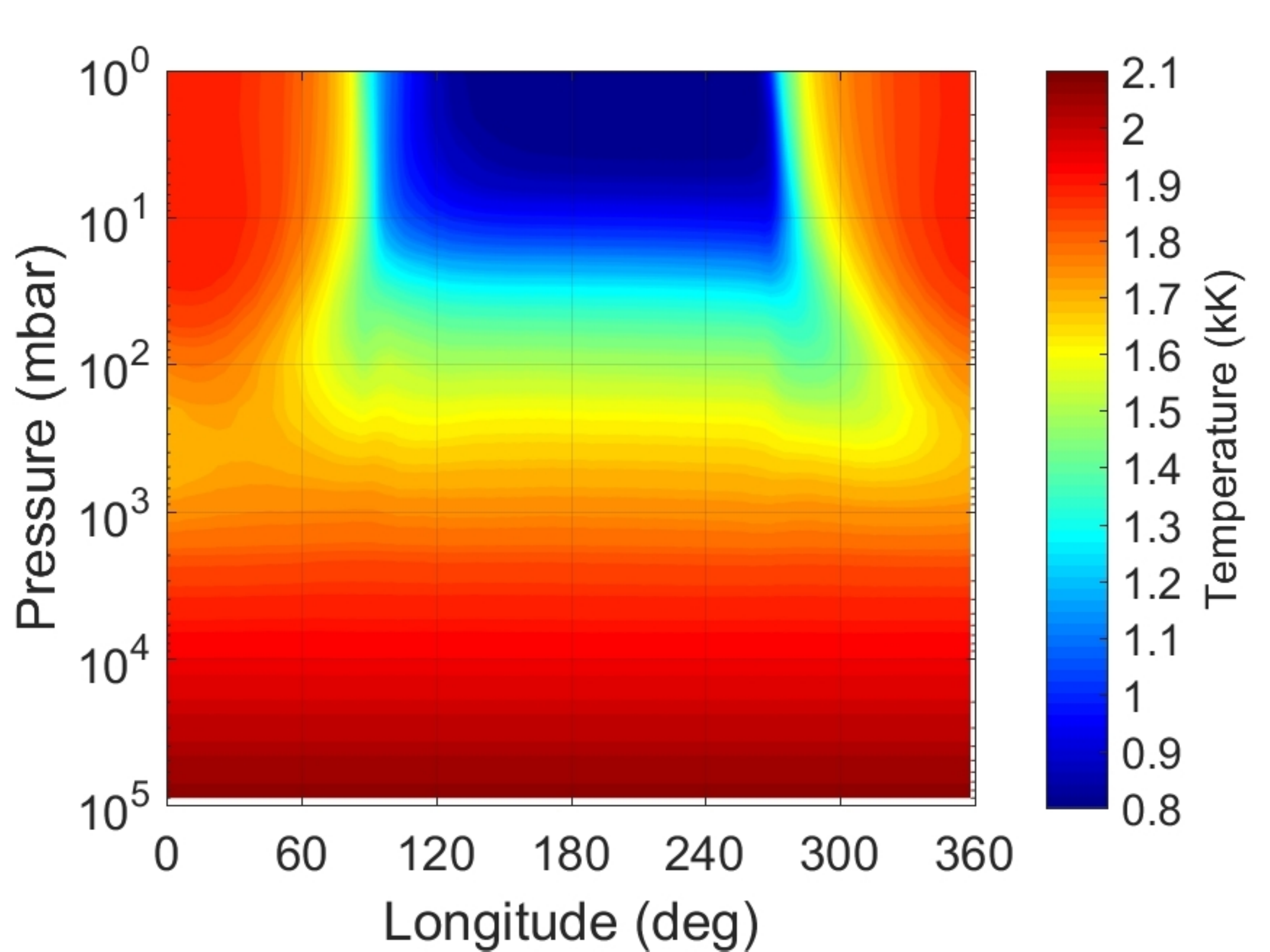}}
\subfigure[G = 2.0]{
\includegraphics[width=0.5\columnwidth]{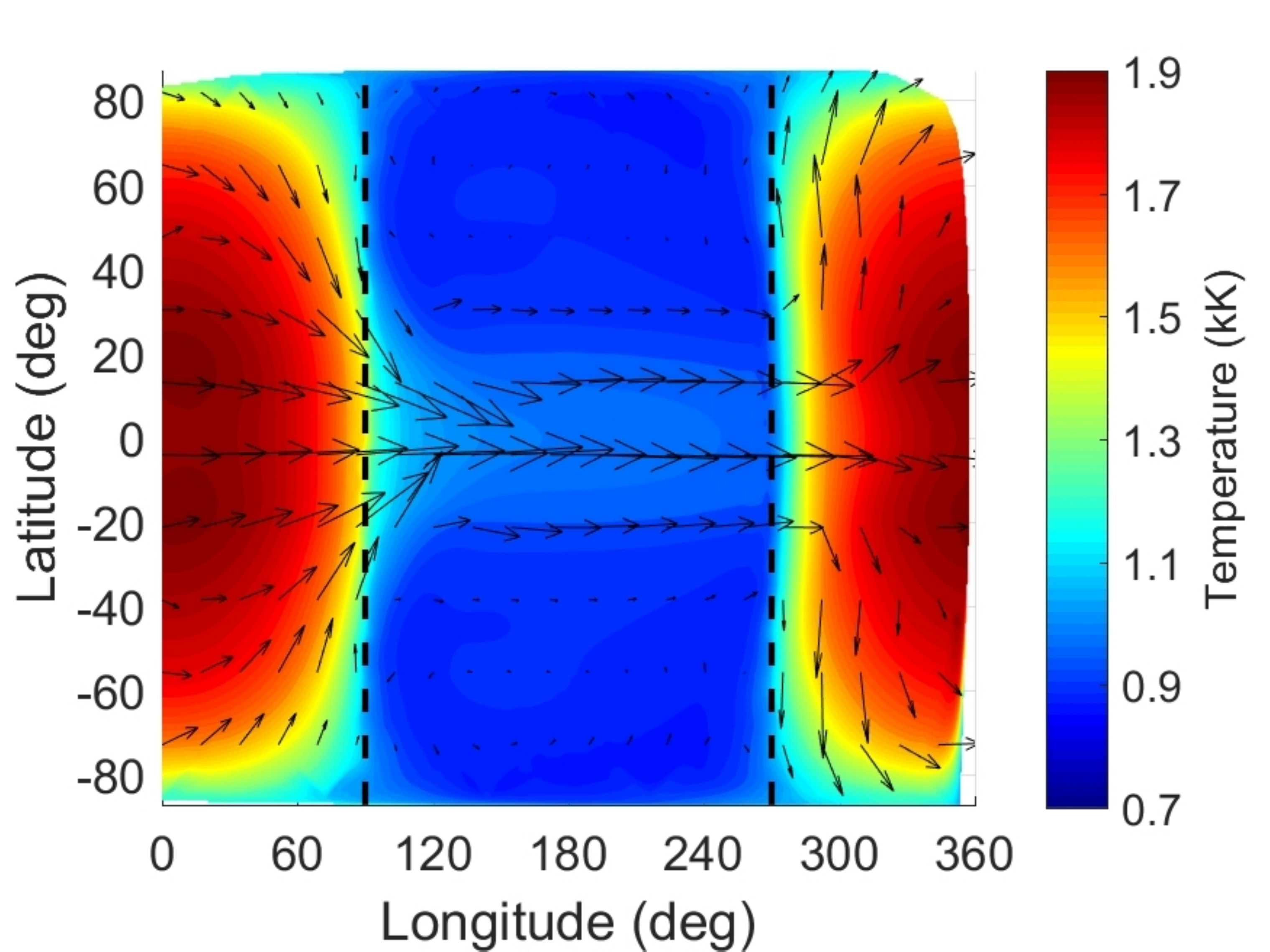}}
\caption{Zonal winds and temperature maps. \textbf{(a)} and \textbf{(e)} are the longitude and time averaged zonal winds for the simulations with $G = 0.5$ and $G = 2$, respectively. The lines are contours of the averaged mass-stream function (in units of 10$^{13}$kg/s). The dashed lines represent the anti-clockwise circulation and the solid lines the clockwise. \textbf{(b)} and \textbf{(f)} are the maps of temperature averaged in time and longitude. \textbf{(c)} and \textbf{(g)} are the maps of temperature averaged in time and latitude. The latitudinal averaging was calculated between 20 and -20 degrees latitude. \textbf{(d)} and \textbf{(h)} are horizontal maps of temperature at 10 mbar. The arrows show the time averaged direction of the wind speed. All the results shown in this figure were averaged over 500 Earth days. The vertical dashed lines in \textbf{(d)} and \textbf{(h)} are the terminators of the planet and the nightside is from longitudes of 60 to 240 degrees.}
\label{fig:ref_results_u_temp}
\end{centering}
\end{figure*}

The two reference simulations of WASP-43b with $G$ equals 0.5 and 2.0, are started from an isothermal atmosphere at 1400 K and from rest. Both models were integrated for 5000 Earth days, and as we show below, it is long enough to complete the spin-up phase of the simulation, and avoid being biased towards the initial conditions. Note that at this point we have not turned on the chemistry parameterizations in the 3D simulations yet, in order to make the model more efficient (see next section). Fig. \ref{fig:spinup}, shows the time evolution of the super-rotation index or also known as Read number (\citealt{1986Read}). This number is defined in the literature to classify how strong is the super-rotation phenomenon in planetary atmospheres, but it can also be used as a good indicator of the spin-up phase of simulations (e.g., \citealt{2013Lebonnois}; \citealt{2016Mendonca}). The super-rotation index is calculated as the total axial angular momentum of the atmosphere ($M_t$) as a function of time over the total axial angular momentum of the same atmosphere at rest state ($M_0$) minus one:
\begin{equation}
S=  \frac{M_t}{M_0}-1;                                        
\end{equation}
\begin{equation}
M= \int\int\int\frac{ma^2}{g} \cos\phi \ d\phi \ d\lambda \ dp,            
\end{equation}
where $\phi$ is the latitude, $\lambda$ is the longitude, $p$ is pressure, $g$ is the gravitational acceleration, $a$ is the radius of the planet and $m$ is the angular momentum per unit mass:
\begin{equation}
\label{eq:m}
m= a \cos\phi⁡ (a\Omega \cos\phi⁡+u).       
\end{equation}
In Eq. \ref{eq:m}, $\Omega$ is the rotation rate of the planet and $u$ is the zonal component of the wind velocity. $M_0$ is calculated in the same way as $M_t$ but setting the zonal wind ($u$) to zero. $S$ is a mass-weighted number, so the deepest regions of the atmosphere have the largest contributions. The increase of $S$, is associated with the dynamical adjustment of the atmosphere towards a steady state combined with processes working in the simulations that do not conserve angular momentum, such as the ``sponge'' layer, divergence damping and inaccuracies from the numerical methods. The sponge layer does not make a significant contribution to $S$ since this latter quantity is mass weighted and the sponge layer just acts in the upper layers. Prograde angular momentum is brought from the deeper layers in the atmosphere to form the strong prograde equatorial jet, which results in negative wind speeds in the lower atmosphere. $S$ is then increased because the divergence damping and numerical inaccuracies act on the negative wind field. The numbers in both simulations start at zero and reach roughly 0.007 for the simulation with $G$ = 0.5 and 0.001 for $G$ = 2.0. The lower values obtained in the simulation with $G$ = 2.0 are associated with the weaker atmospheric wind speeds obtained. The long convergence of $S$ is expected due to the large thermal inertia of the deep atmosphere.

\begin{figure*}
\begin{centering}
\subfigure[G = 0.5 $\&$ C/O = 0.5]{
\includegraphics[width=0.8\columnwidth]{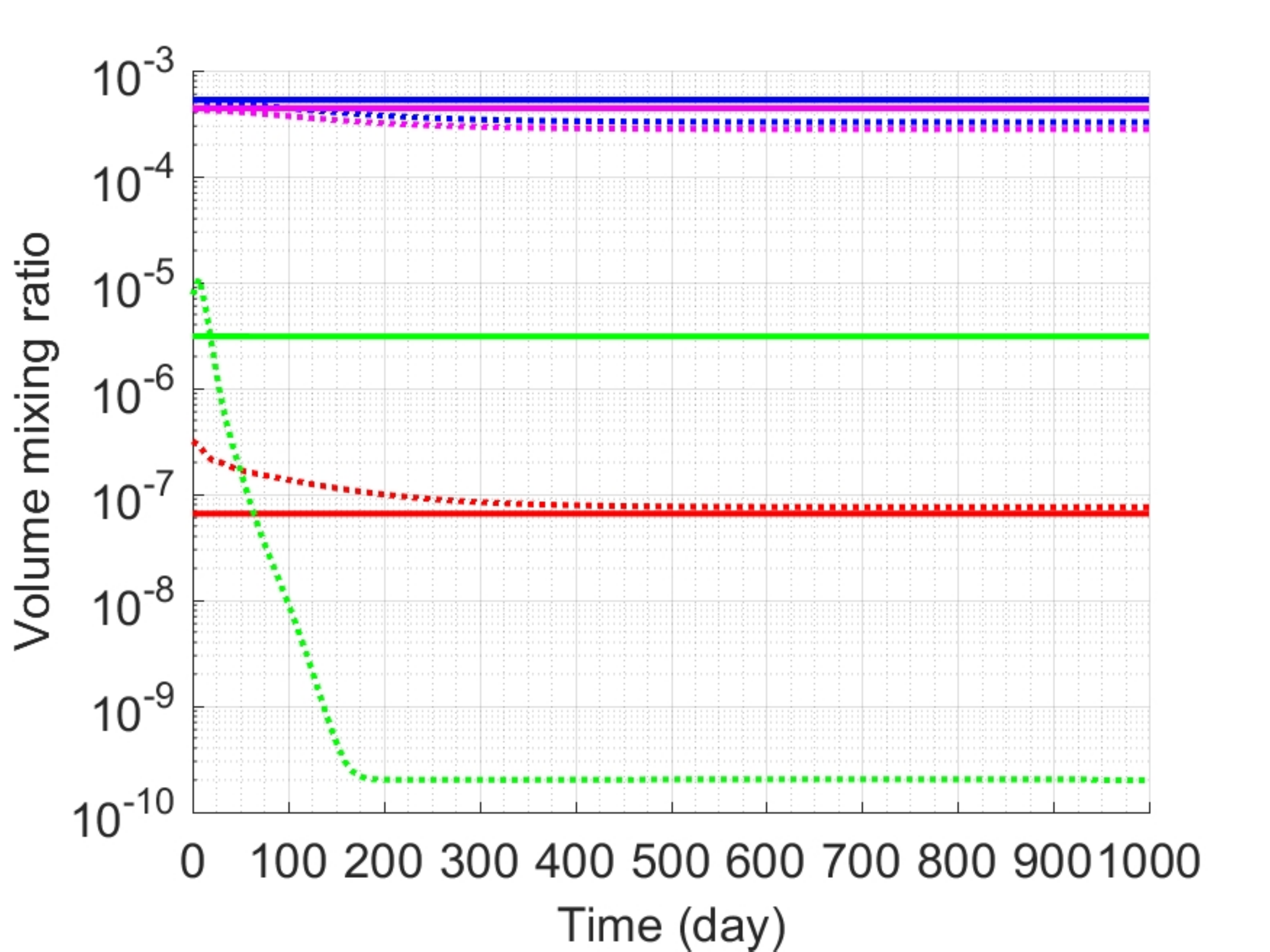}}
\subfigure[G = 2.0 $\&$ C/O = 0.5]{
\includegraphics[width=0.8\columnwidth]{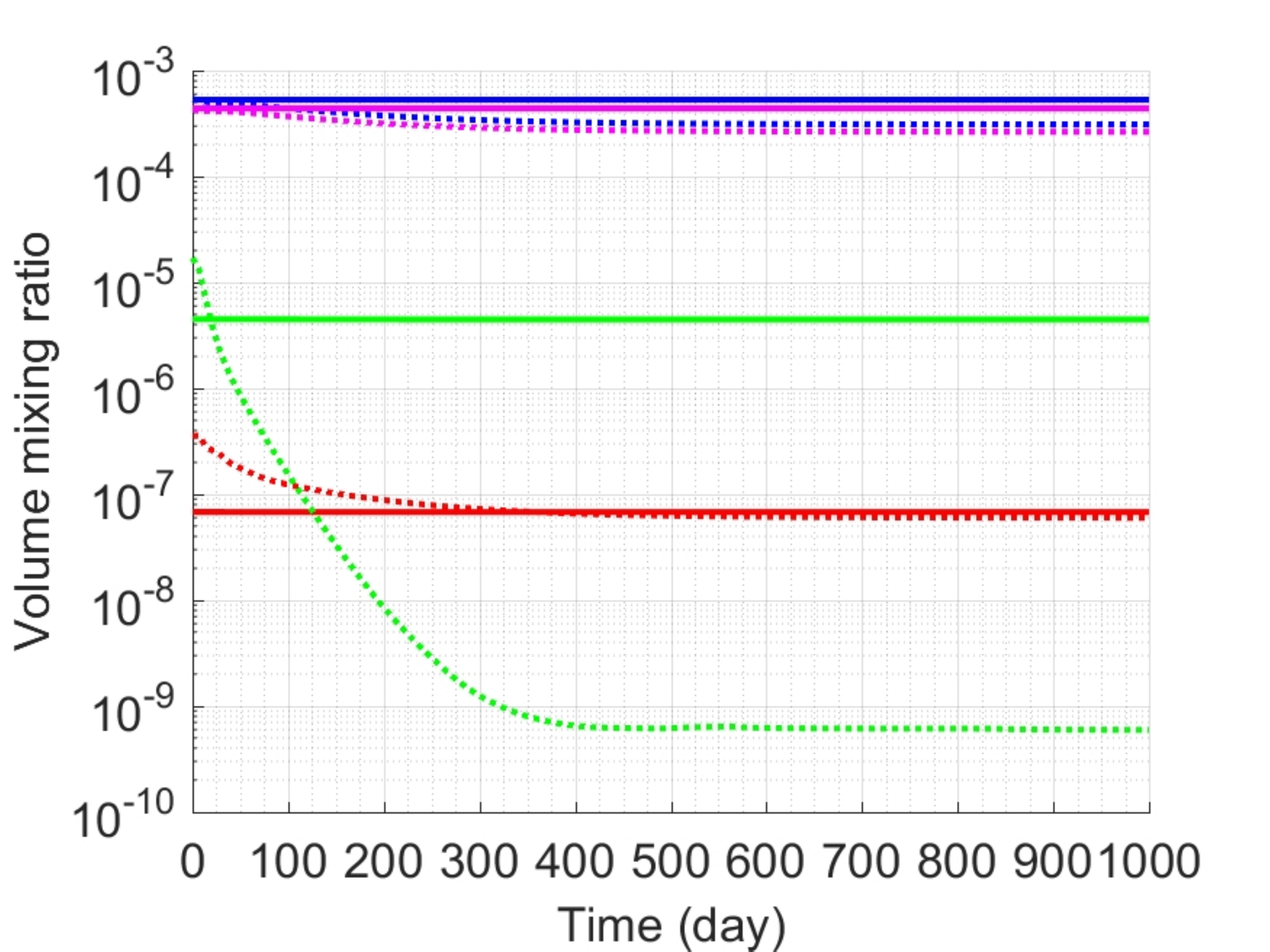}}
\subfigure[G = 0.5 $\&$ C/O = 2]{
\includegraphics[width=0.8\columnwidth]{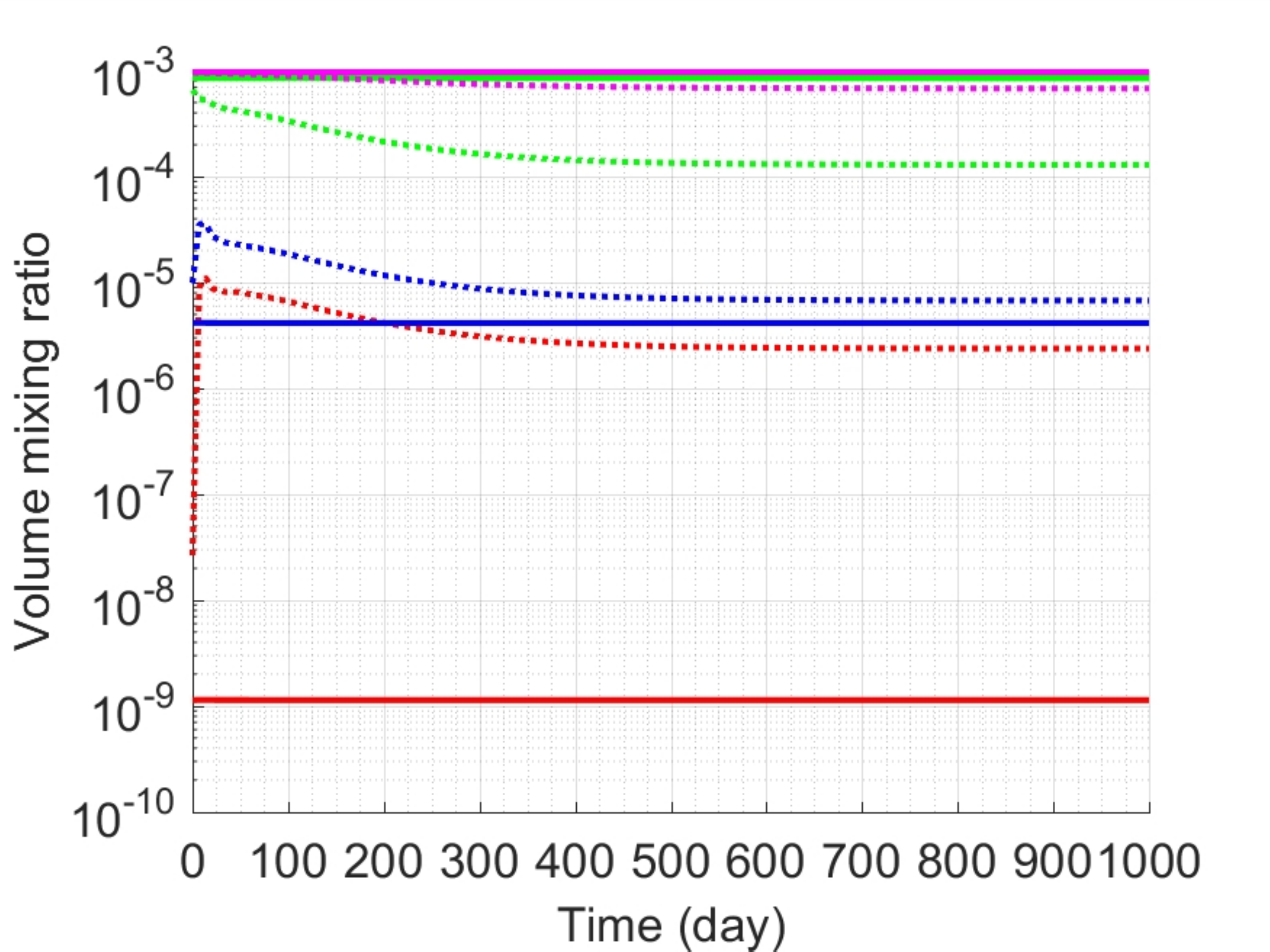}}
\subfigure[G = 2 $\&$ C/O = 2]{
\includegraphics[width=0.8\columnwidth]{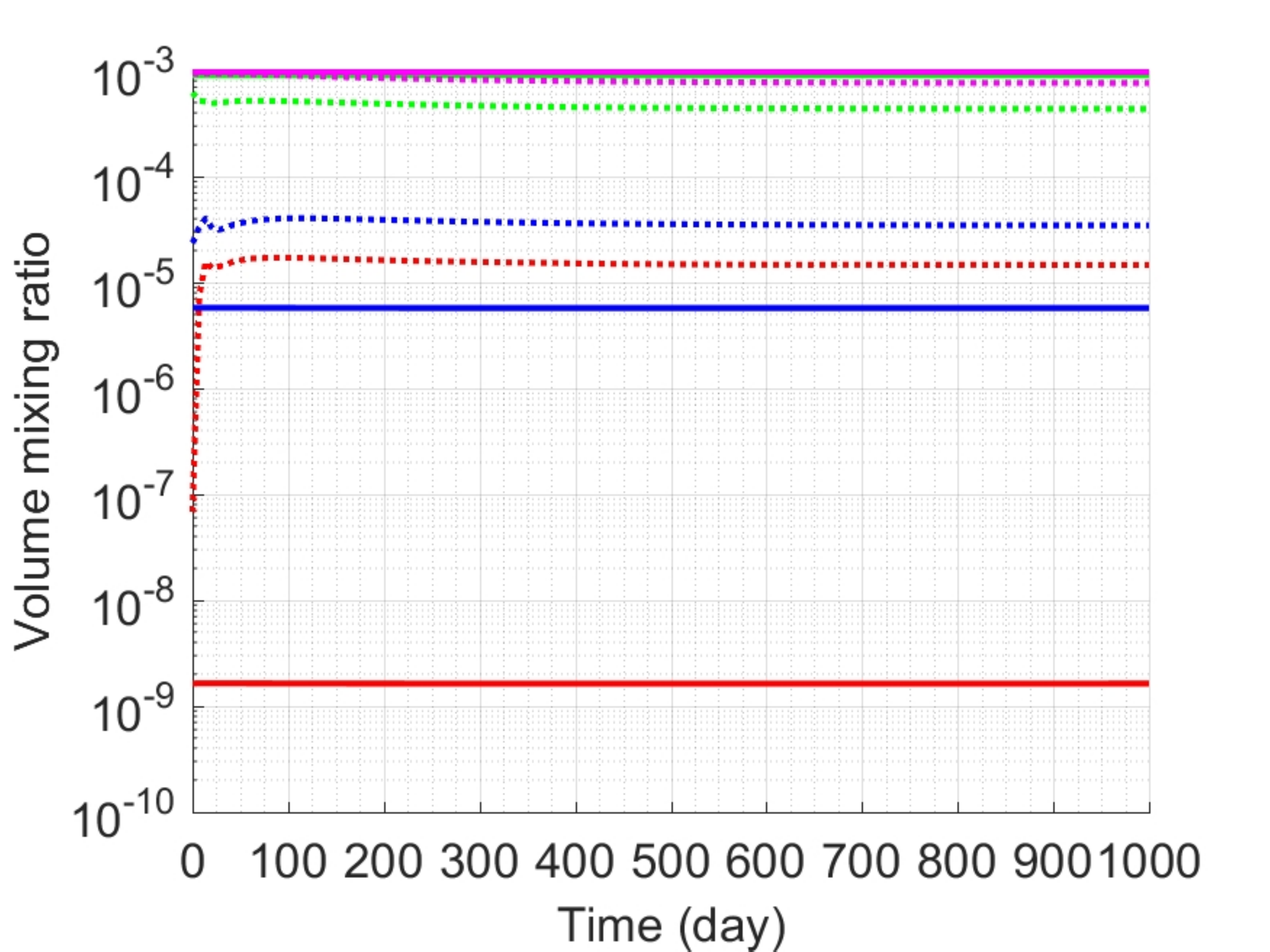}}
\caption{The different lines represent the time evolution of the chemical species averaged in longitude and latitude from the simulations with different $G$ and $C/O$ at two pressure layers : 10 mbar (dotted line) and 10 bar (solid line). The line colors are associated to different species: CO$_2$ (red), H$_2$O (blue), CO (magenta) and CH$_4$ (green).}
\label{fig:che_spinup}
\end{centering}
\end{figure*}

Fig. \ref{fig:ref_results_u_temp}, shows the results on atmospheric circulation and thermal structure of the two experiments. Panels (a) and (e) show the presence of a broad equatorial jet in both simulations. This is a well known feature (e.g., \citealt{2009Showman}; \citealt{2009Menou}; \citealt{2011aHeng}; \citealt{2015Heng}) that has an important role transporting heat from the dayside to the nightside. The jet in the atmosphere with thermal inversion in the dayside ($G = 2$) is weaker. The less efficient mechanism driving the strong wind at the equator in this scenario is related to the shorter radiative timescales in the upper atmosphere when the $k_{star}$ is doubled (e.g., \citealt{2013Perez-Becker}). The case $G = 0.5$ shows less steeper gradients in latitude in the longitudinally averaged temperature field and no thermal inversion in the dayside. Panels (d) and (h), show a weaker day-night contrast in the case $G = 0.5$, which also shows a broader equatorial jet. The long time averaged in these figures (500 Earth days) is to avoid having transient features in the figure. The radiative timescales of the deeper atmosphere (below the pressure level 10 bar) are of the order of hundreds of Earth days (\citealt{2005Iro}).  

\section{Disequilibrium chemistry}
\label{sec:des_che}
We have integrated \texttt{THOR} coupled with the chemical relaxation method for four different scenarios: with and without thermal inversion in the upper atmosphere ($G = 0.5$ and $G = 2.0$) and with different $C$ to $O$ ratios ($C/O = 0.5$ and $C/O = 2.0$). The four simulations started after the reference simulations analysed in the previous section have reached a statistical steady state. The initial chemical abundances in the simulations were set to chemical equilibrium, and integrated for extra 1000 Earth days of simulation as shown in Fig. \ref{fig:che_spinup}. The different colours in Fig. \ref{fig:che_spinup} represent the different chemical species: CO$_2$ (red), H$_2$O (blue), CO (magenta) and CH$_4$ (green). The different line styles refer to the abundances averaged in longitude and latitude at different pressure levels: 10 mbar (dashed line) and 10 bar (solid line). The solid lines in Fig. \ref{fig:che_spinup} almost do not change with time because the simulations are starting from chemical equilibrium and the chemical relaxation timescales are shorter than the dynamical timescales in the deep atmosphere. At 10 mbar, the winds become very strong and the chemical timescales increase, leading to larger departures from the chemical equilibrium state. The dashed lines are good indicators on how long the dynamical-chemical simulations have to be integrated until the distribution of the chemical species across the atmosphere have reached steady state (similar to the spin-up phase described in section \ref{sec:basesimu}). Our plots show that all the chemical species have reached equilibrium at roughly 500 Earth days. All the results obtained before 500 days are discarded because they are biased towards the initial chemical equilibrium condition (see trends in Fig. \ref{fig:che_spinup}). A striking feature from this figure is the sudden drop of CH$_4$ in the two simulations with $C/O$ equal to 0.5, which is associated with the dominant ``quenching effect'' present in WASP-43b as explained in the next section.

In our simulations the tracer's mixing is not considered in the convective adjustment scheme that represents the small-scale convection due to buoyant instability. The thermal structure becomes closer to super-adiabatic in the cloud region of the planet's night side, however, we find that the convective adjustments do not have a relevant role driving the atmospheric dynamics in the case studied. Nevertheless, further studies need to be done with more complex convection schemes that represent the tracers mixing during small-scale convection.

\begin{figure*}
  \begin{tabular}{m{0.1cm} c c c c}
     & G = 0.5 $\&$ C/O = 0.5 & G = 2.0 $\&$ C/O = 0.5 & G = 0.5 $\&$ C/O = 2 & G = 2 $\&$ C/O = 2 \\
    \vspace{-3cm}\rotatebox{90}{Dayside} & {\includegraphics[width=4.2cm]{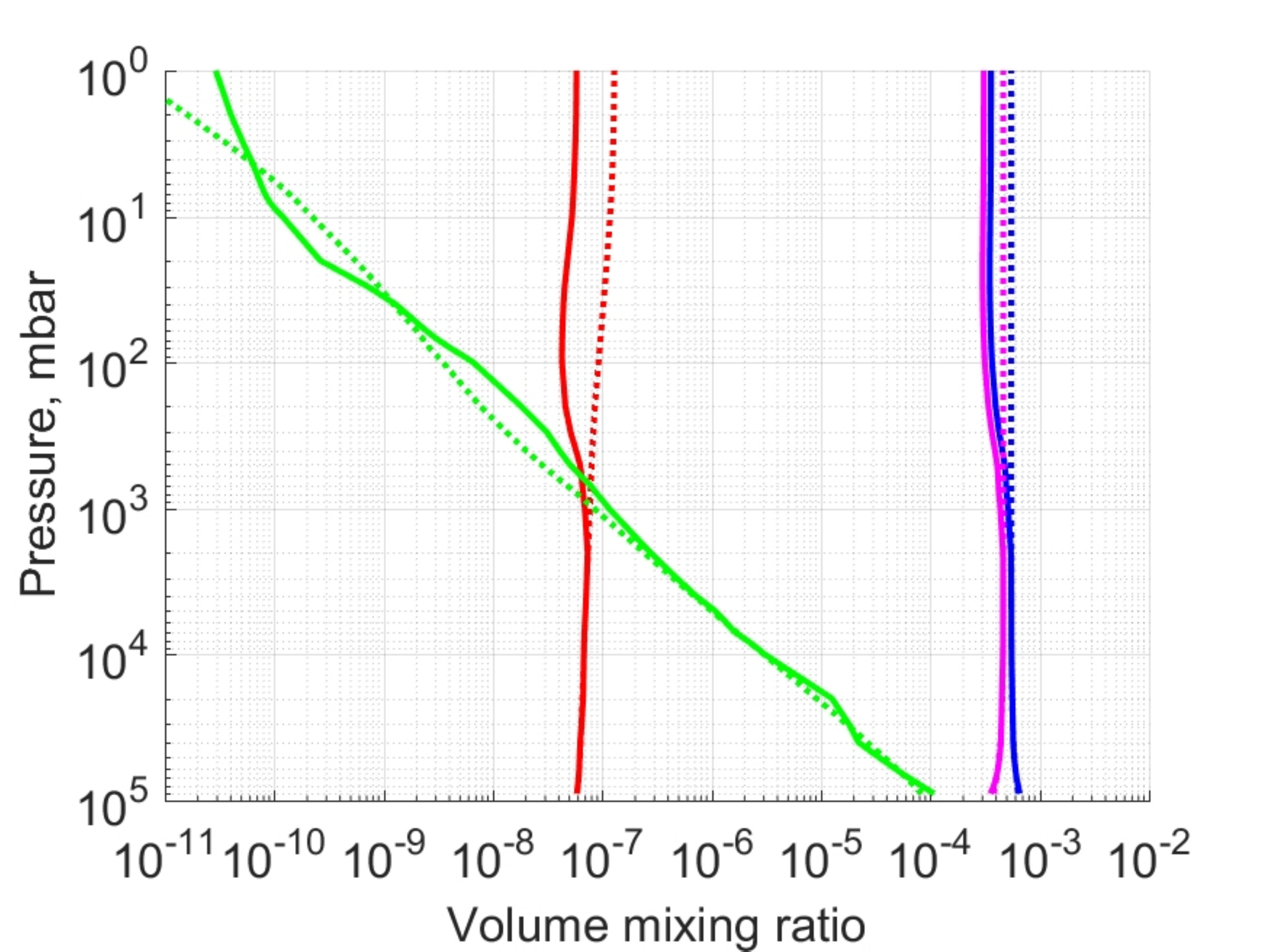}} & {\includegraphics[width=4.2cm]{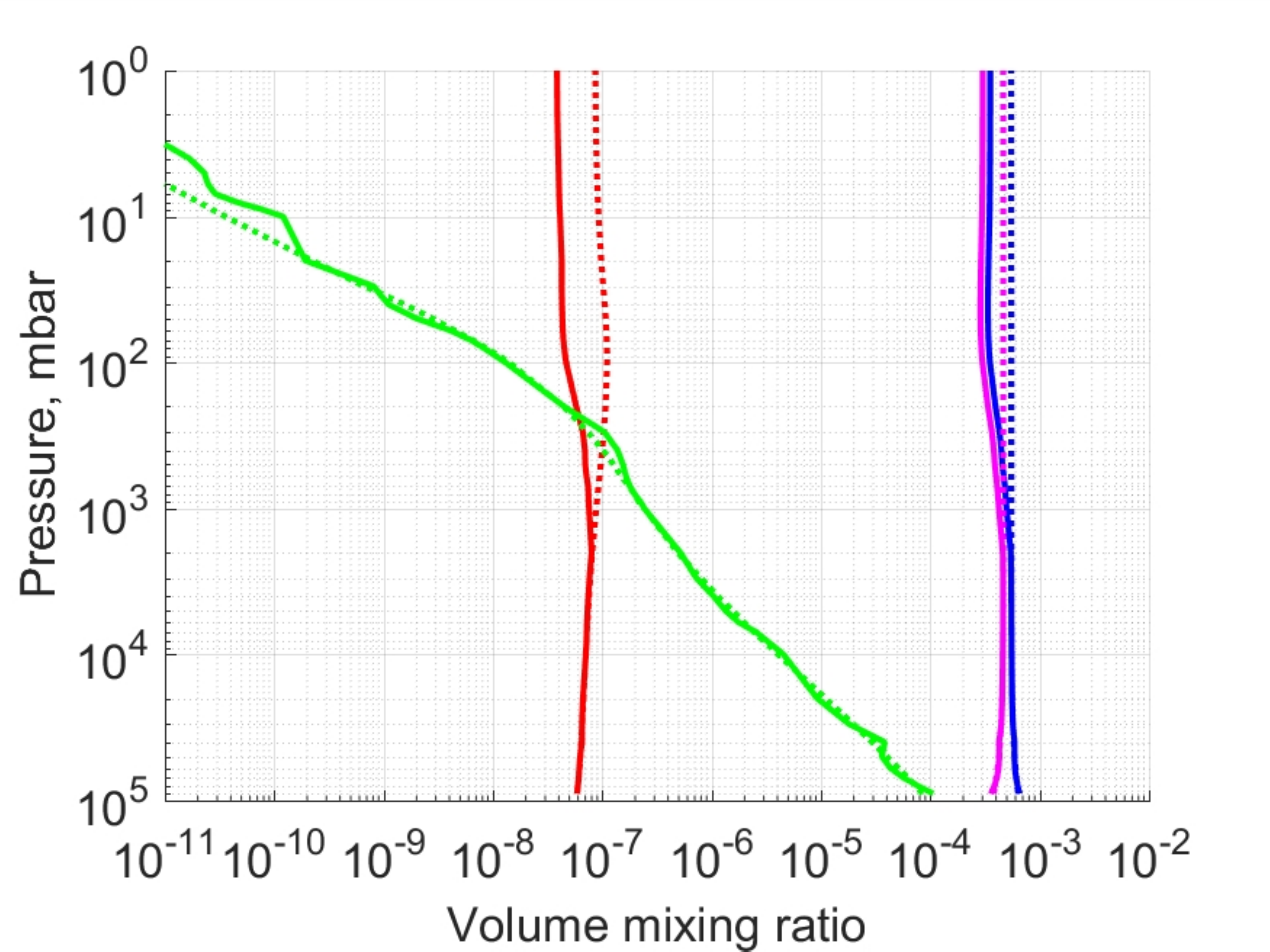}} &
{\includegraphics[width=4.2cm]{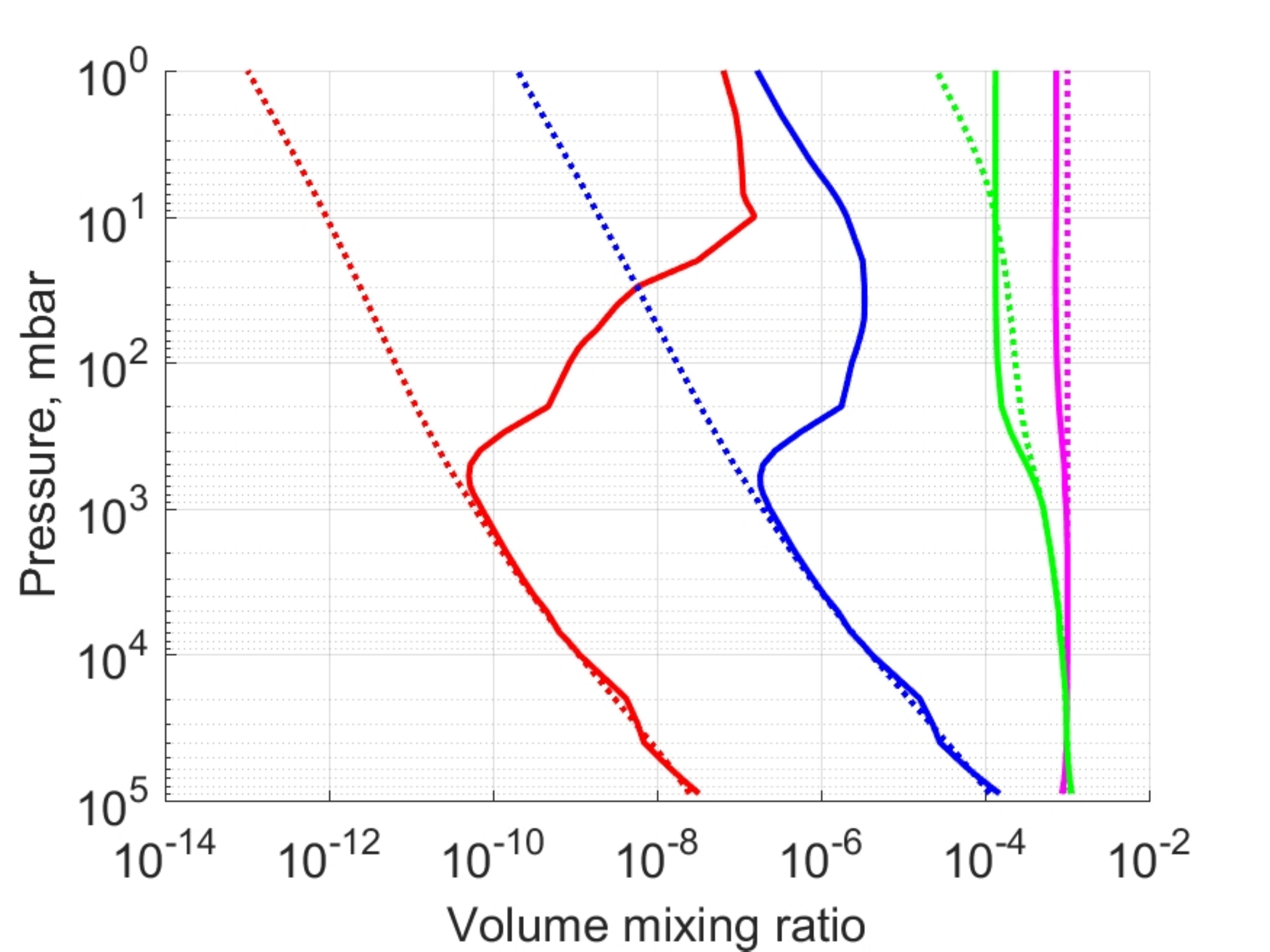}} & 
{\includegraphics[width=4.2cm]{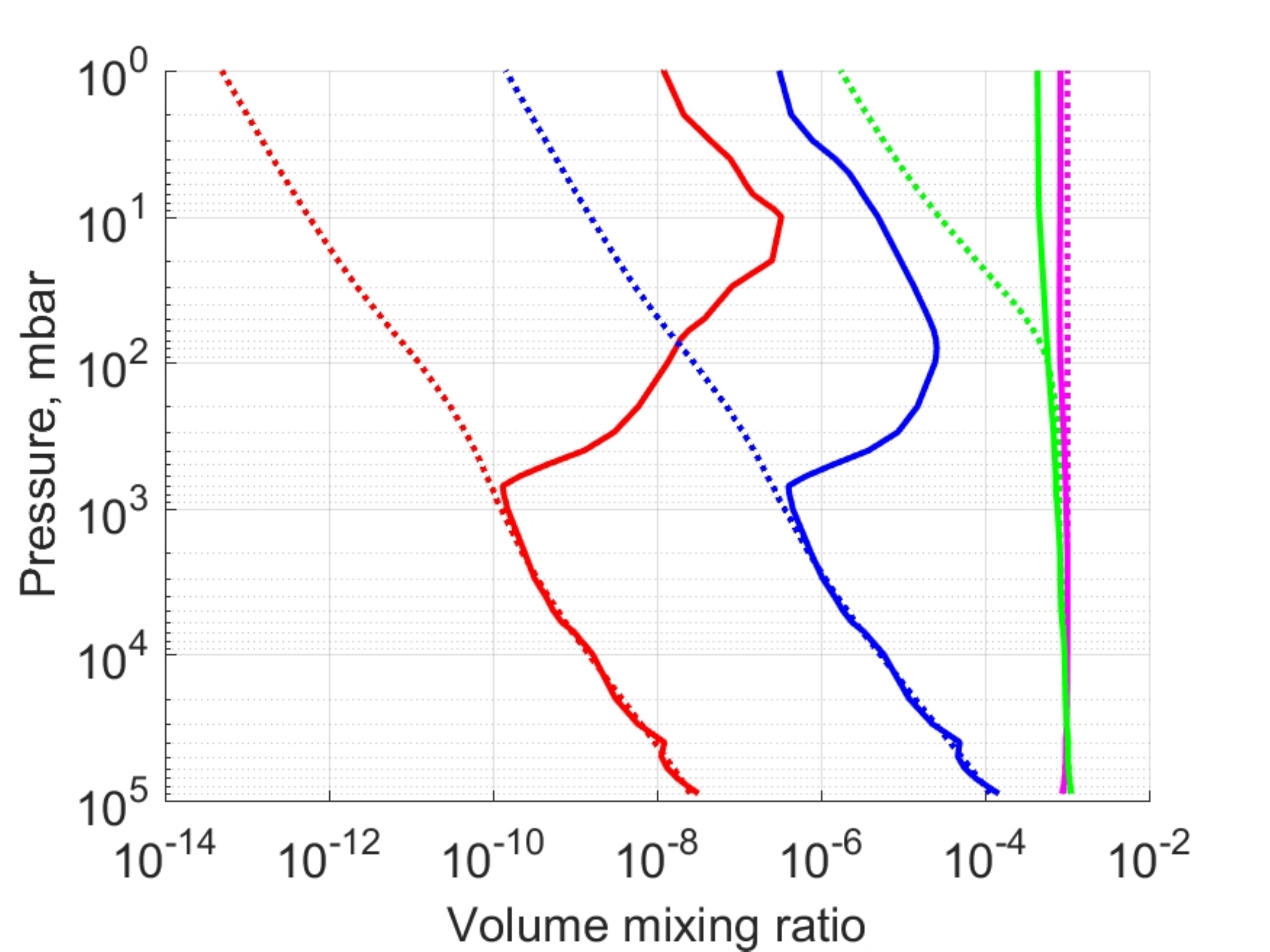}} \\ 
    \vspace{-3cm}\rotatebox{90}{Nightside} & {\includegraphics[width=4.2cm]{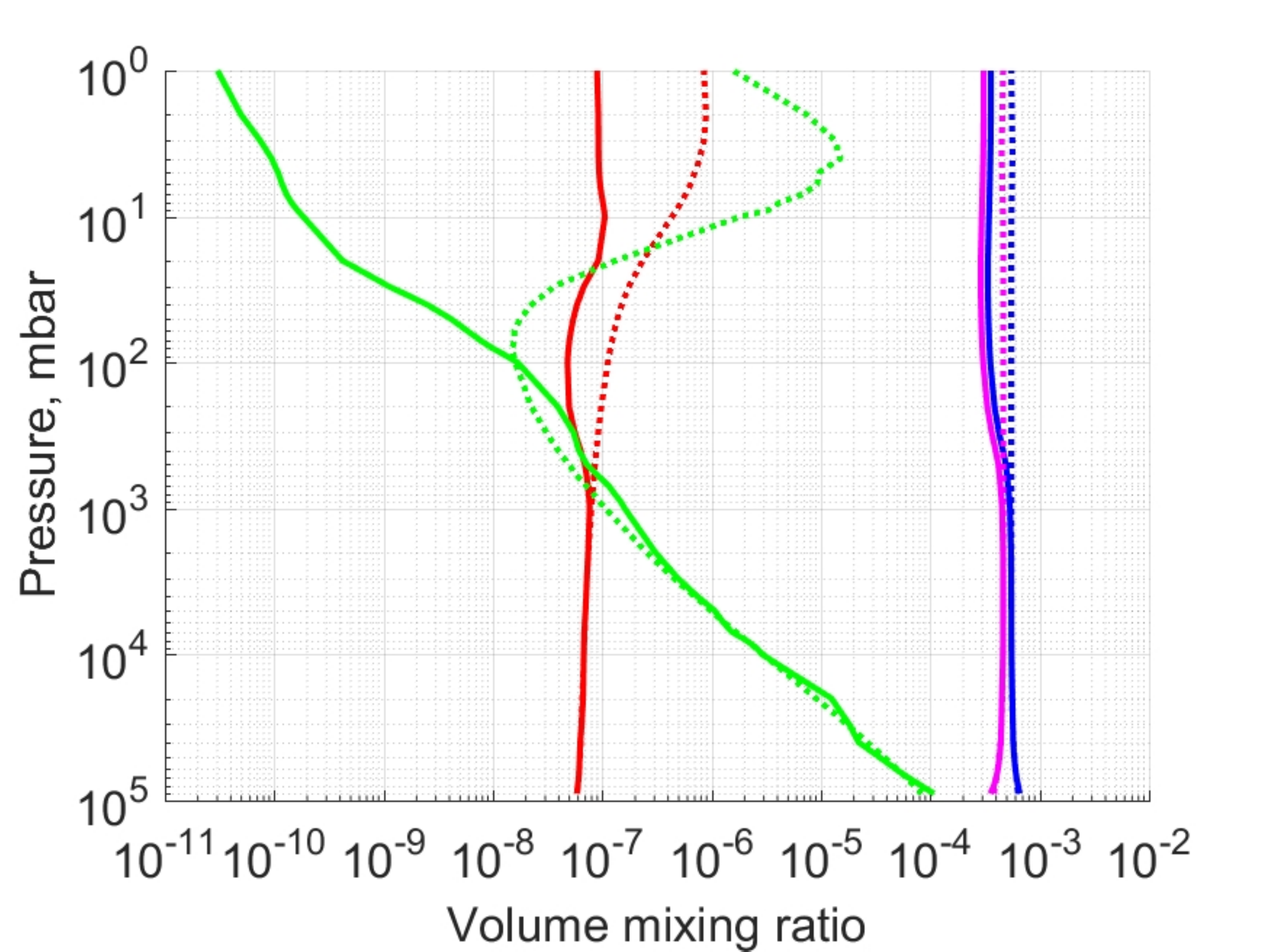}} & {\includegraphics[width=4.2cm]{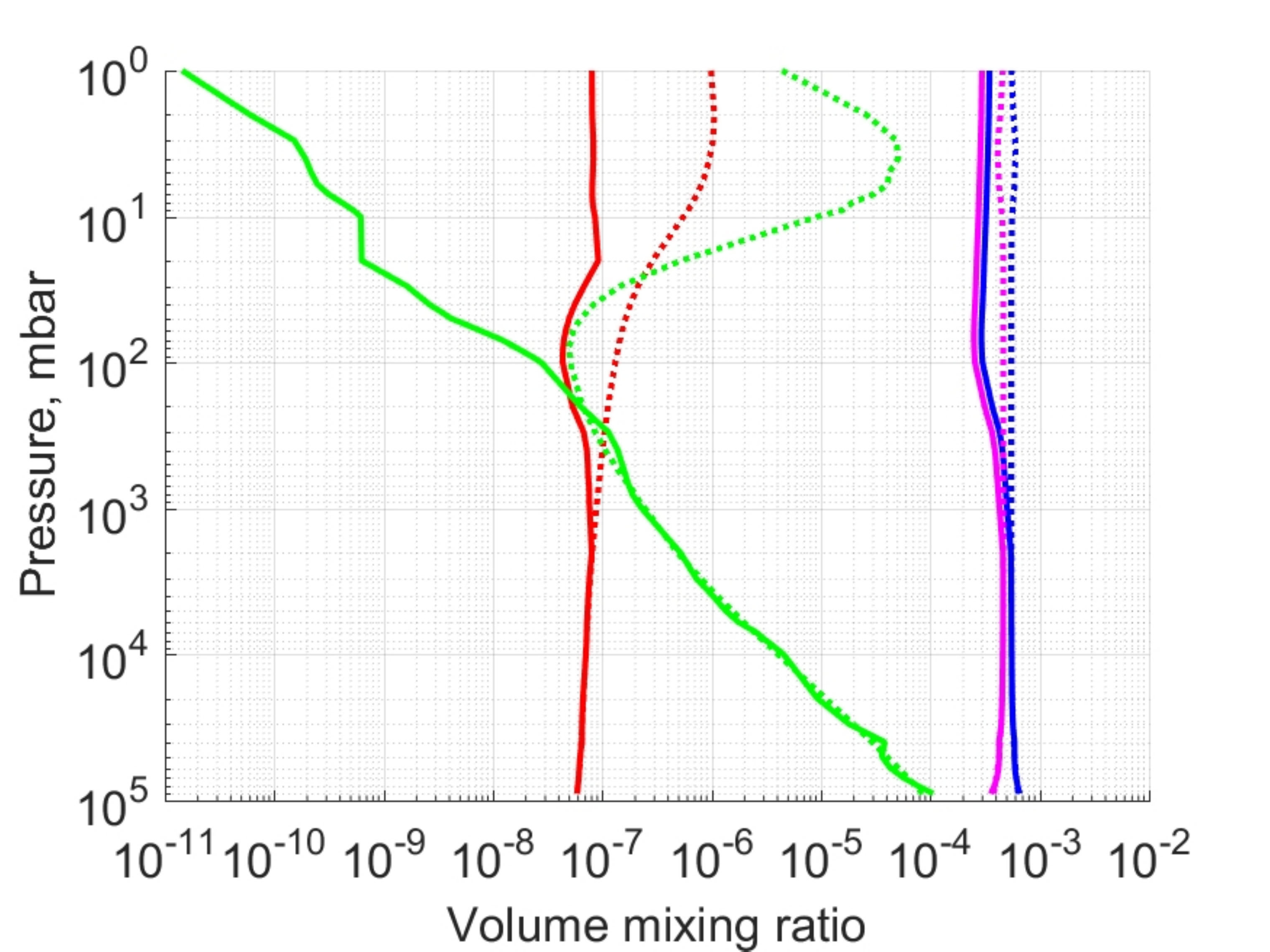}} &
{\includegraphics[width=4.2cm]{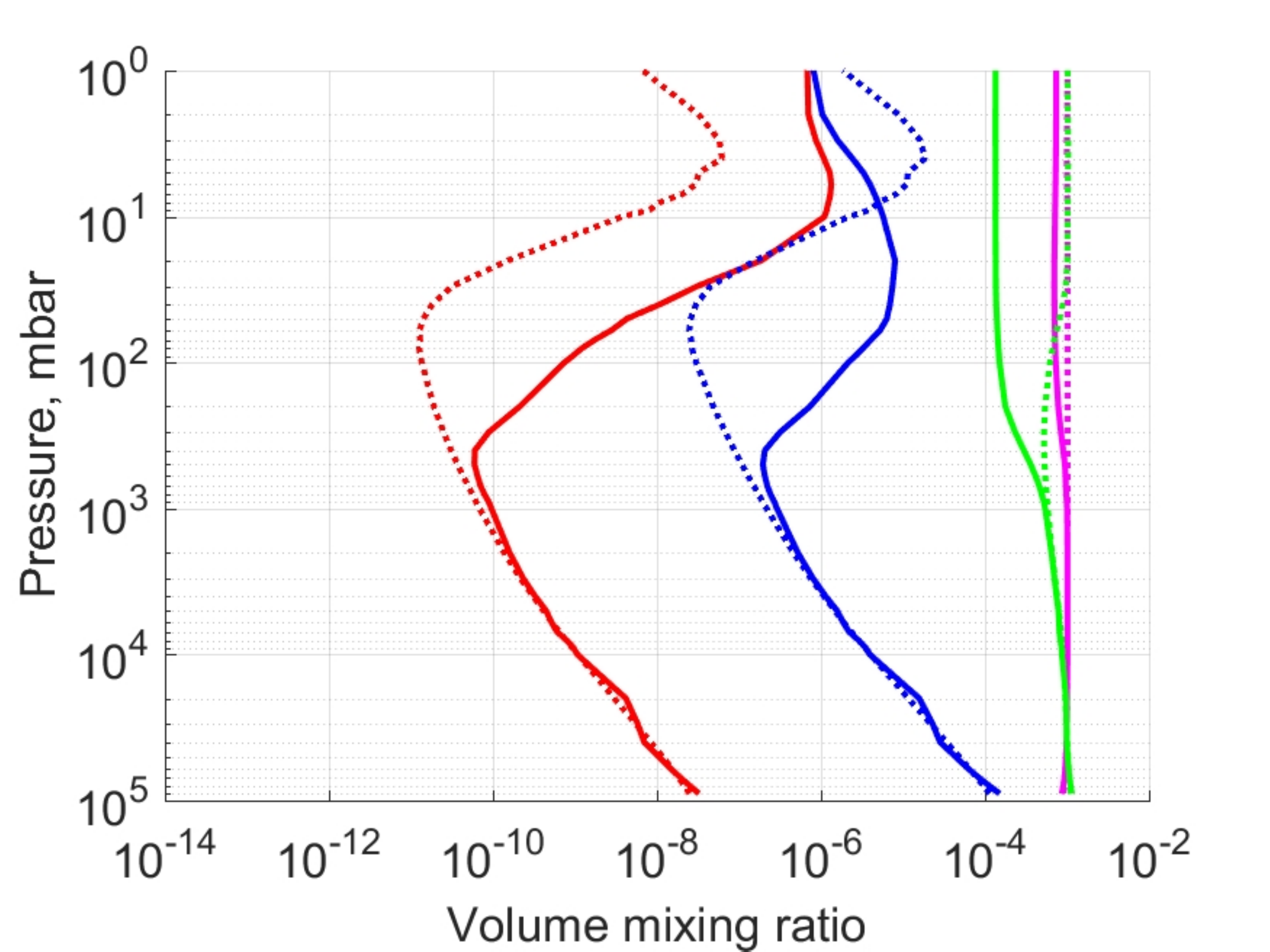}} & 
{\includegraphics[width=4.2cm]{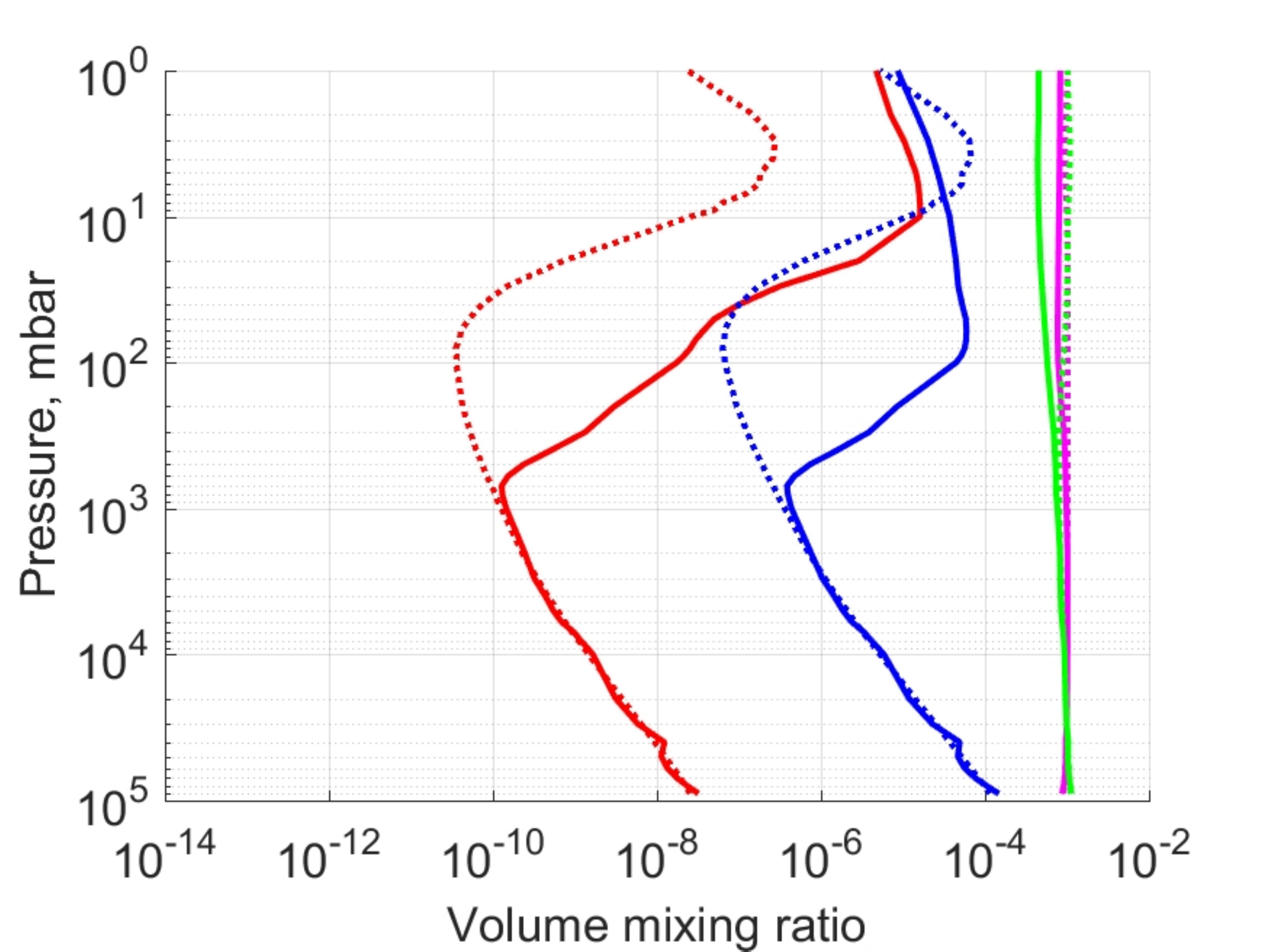}} \\ 
    \vspace{-3cm}\rotatebox{90}{Morning terminator} & {\includegraphics[width=4.2cm]{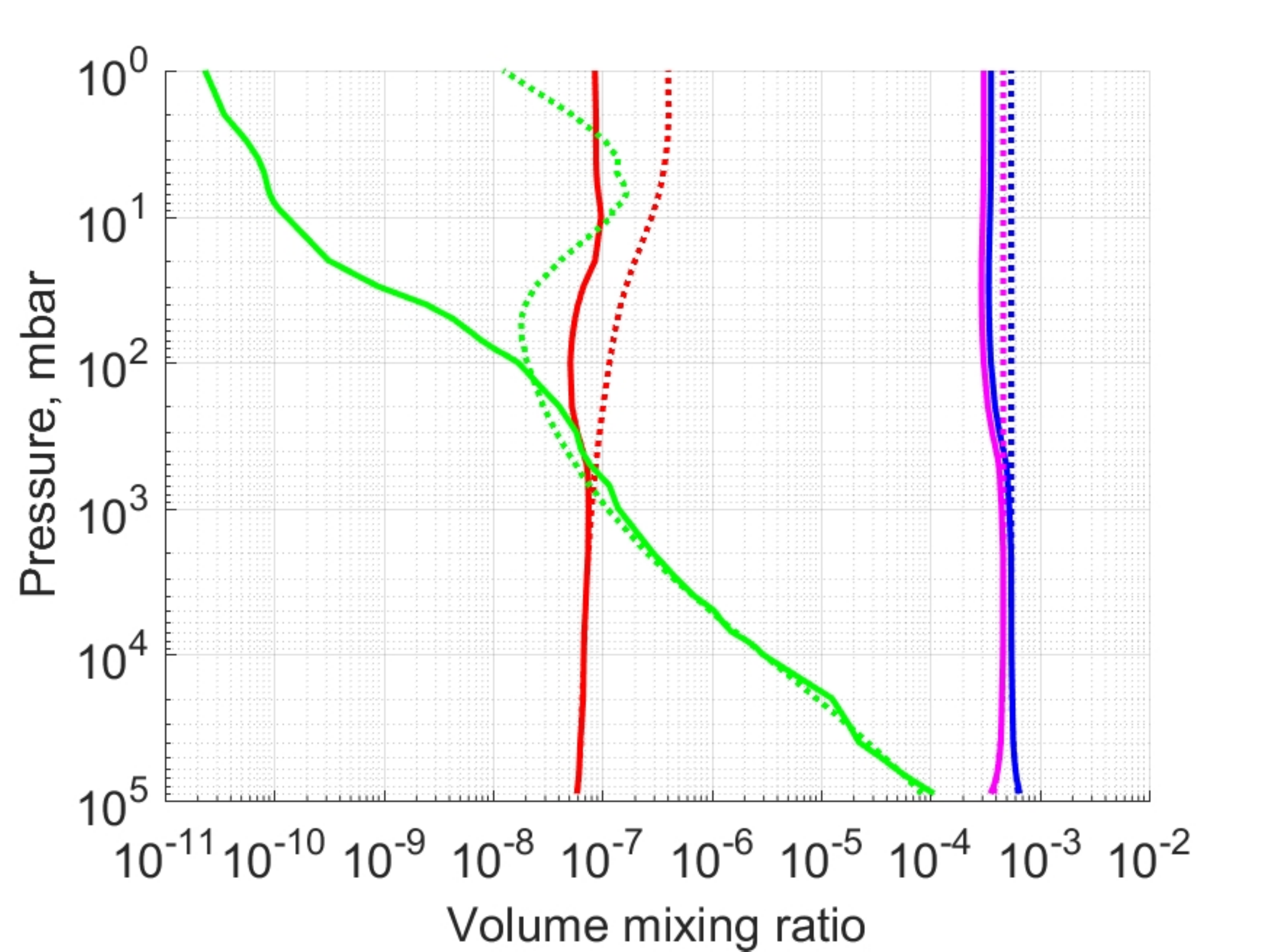}} & {\includegraphics[width=4.2cm]{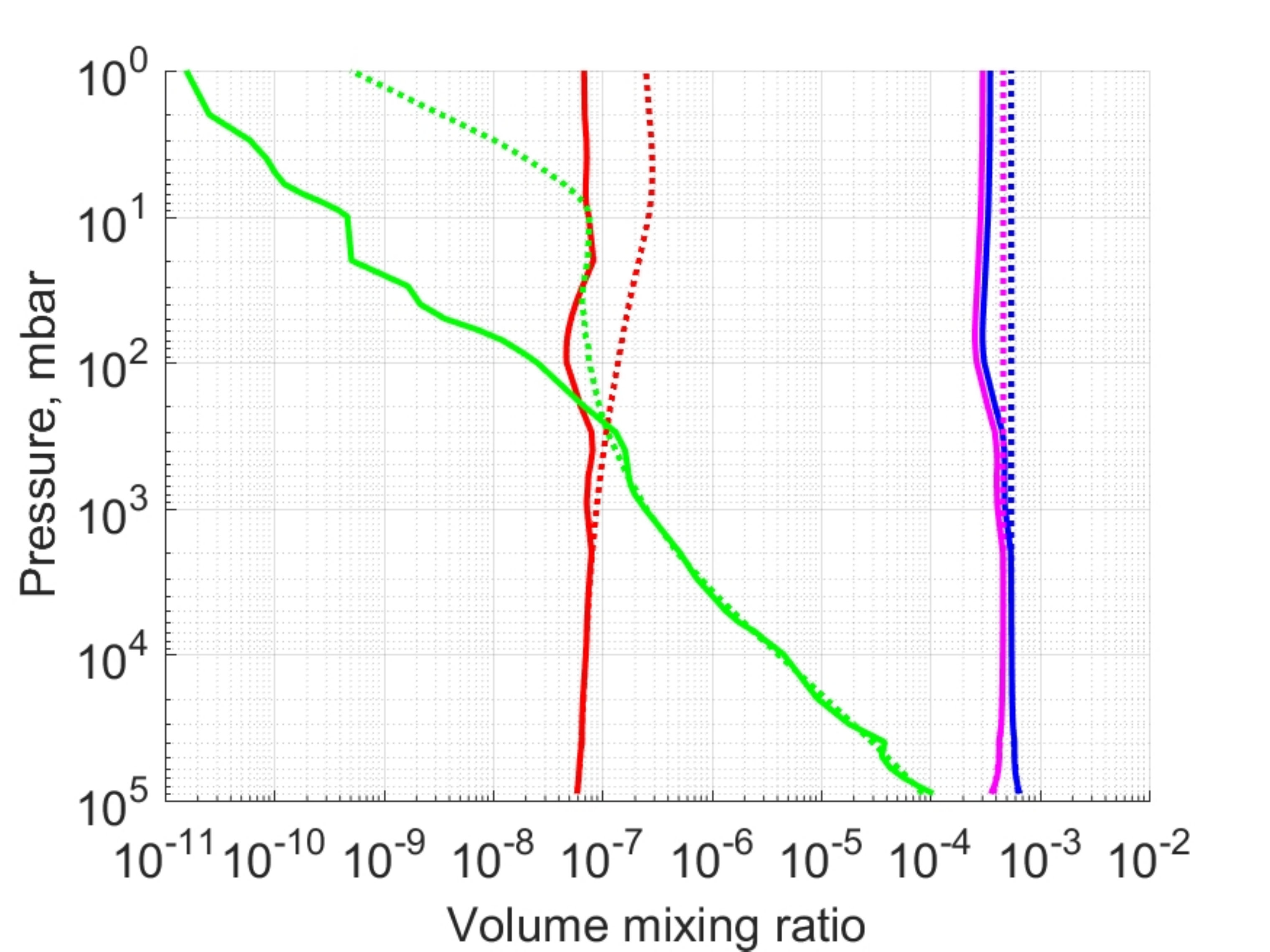}} &
{\includegraphics[width=4.2cm]{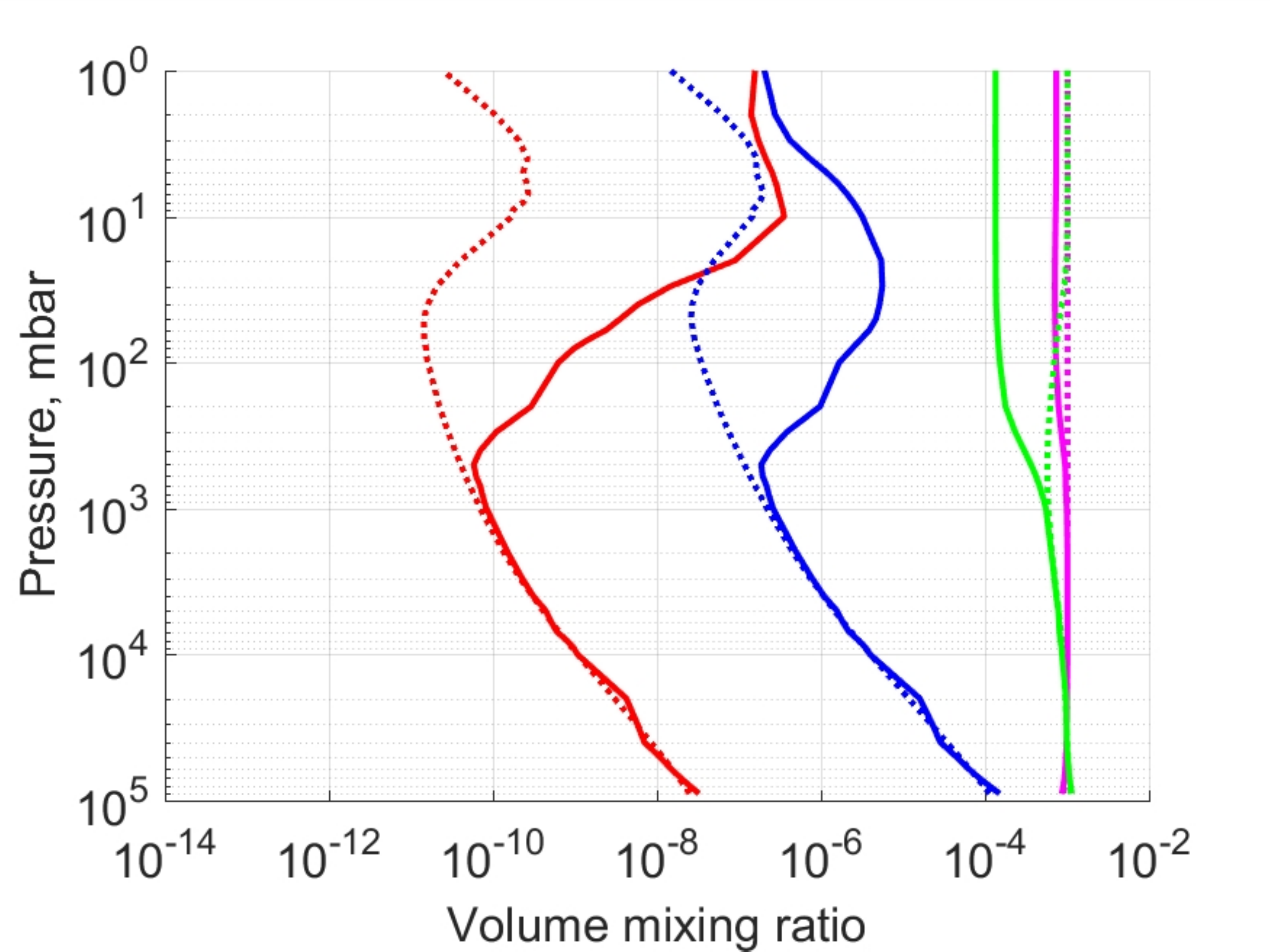}} & 
{\includegraphics[width=4.2cm]{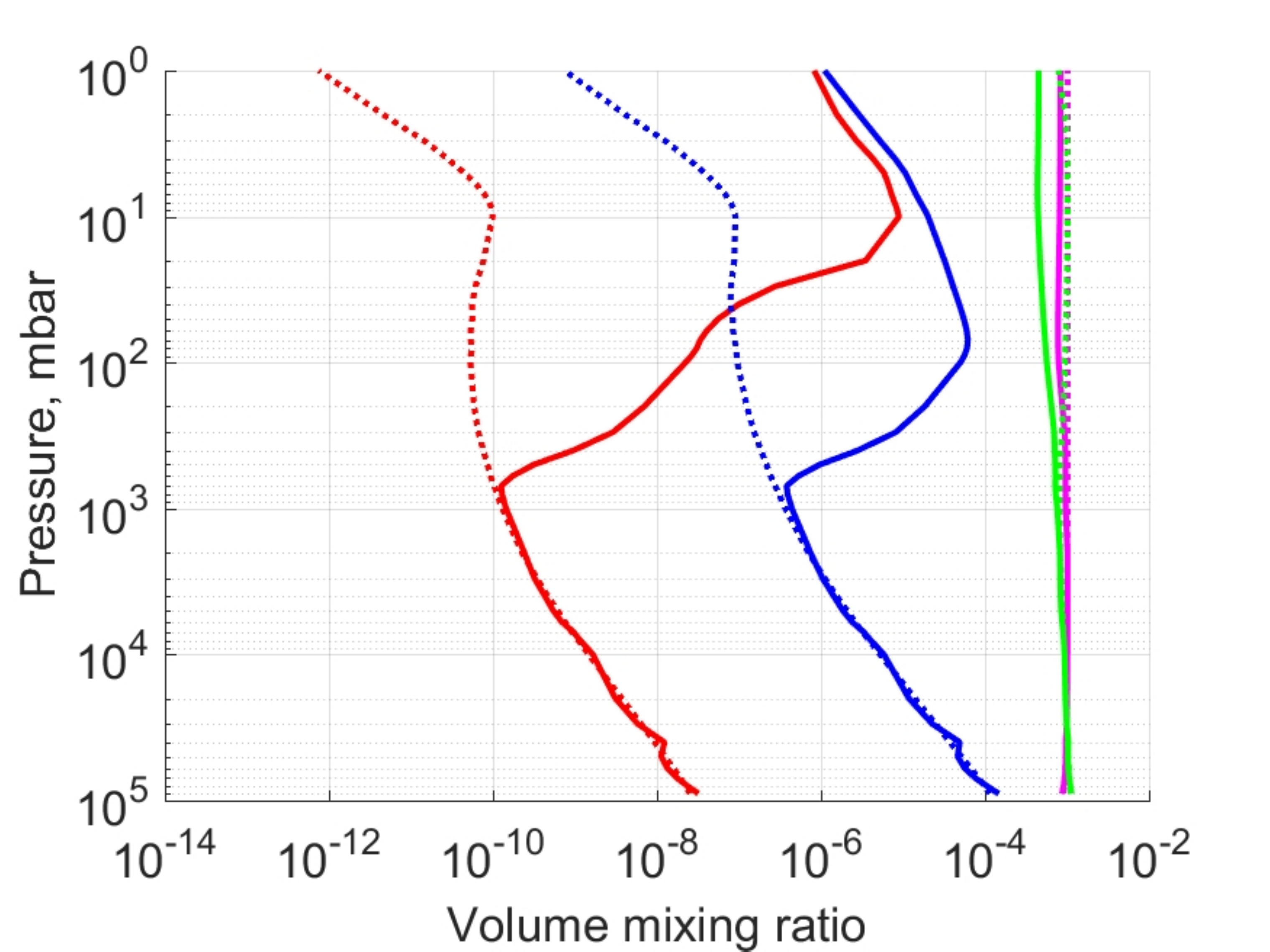}} \\ 
    \vspace{-3cm}\rotatebox{90}{Evening terminator} & {\includegraphics[width=4.2cm]{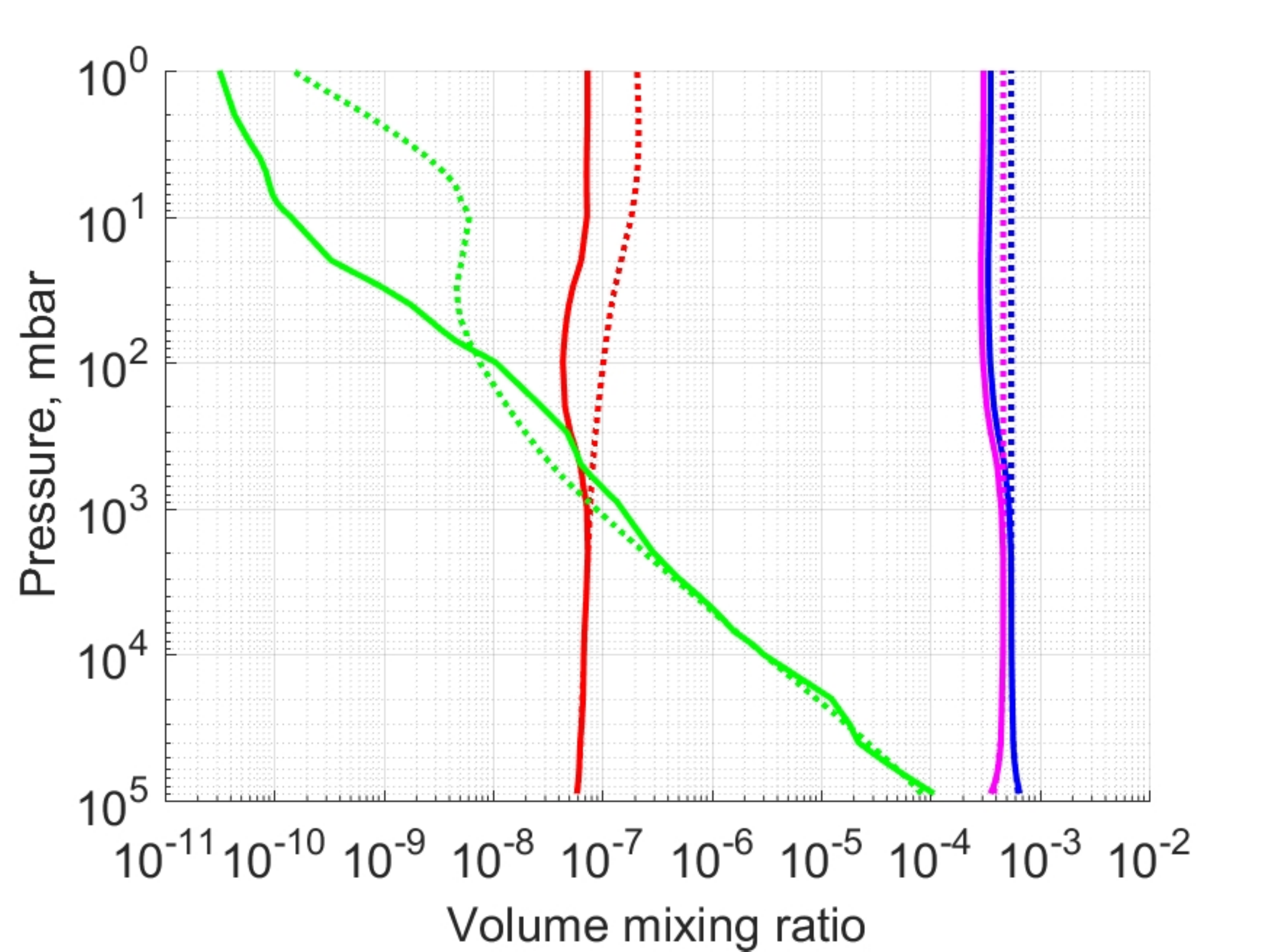}} & {\includegraphics[width=4.2cm]{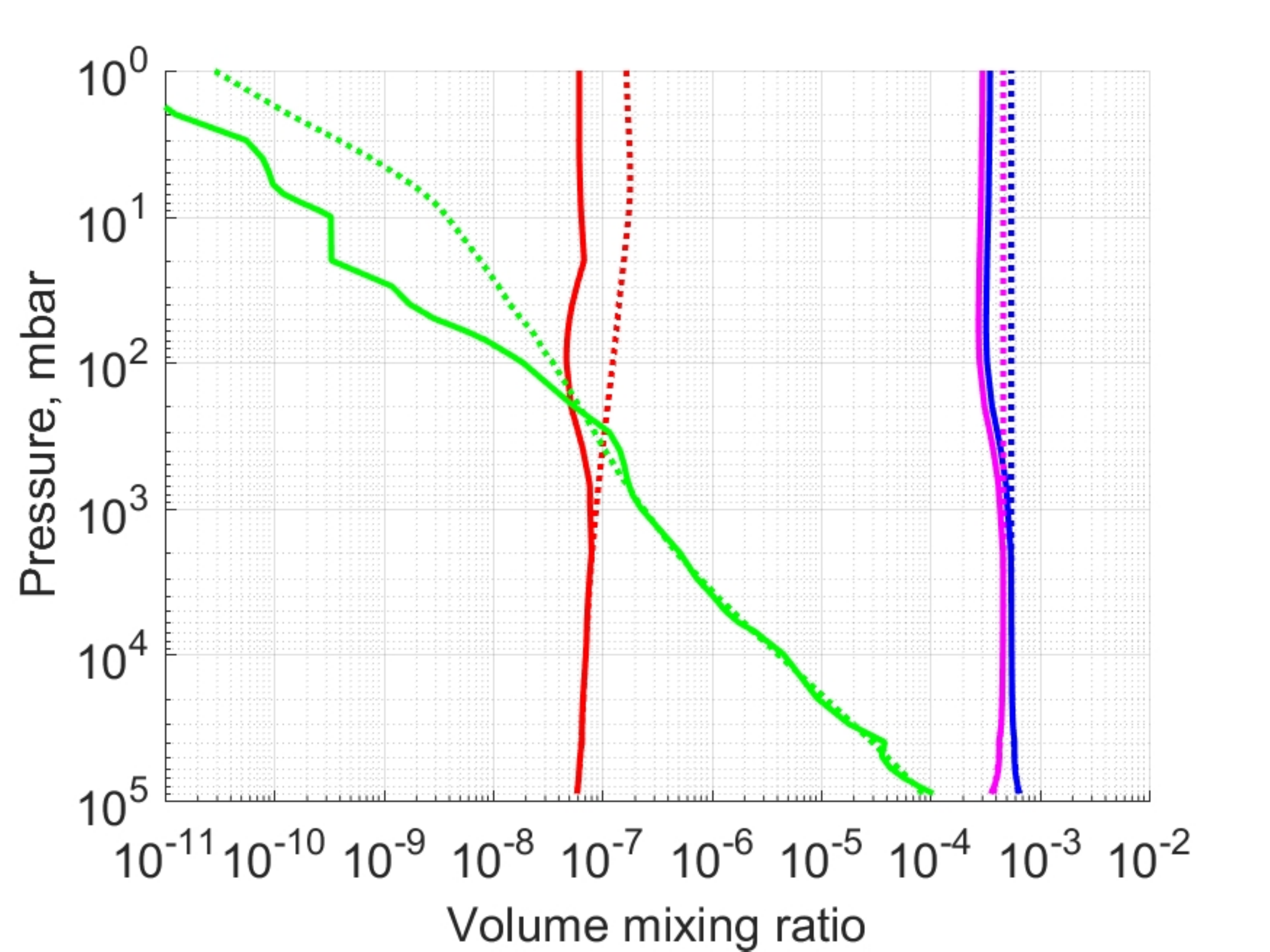}} &
{\includegraphics[width=4.2cm]{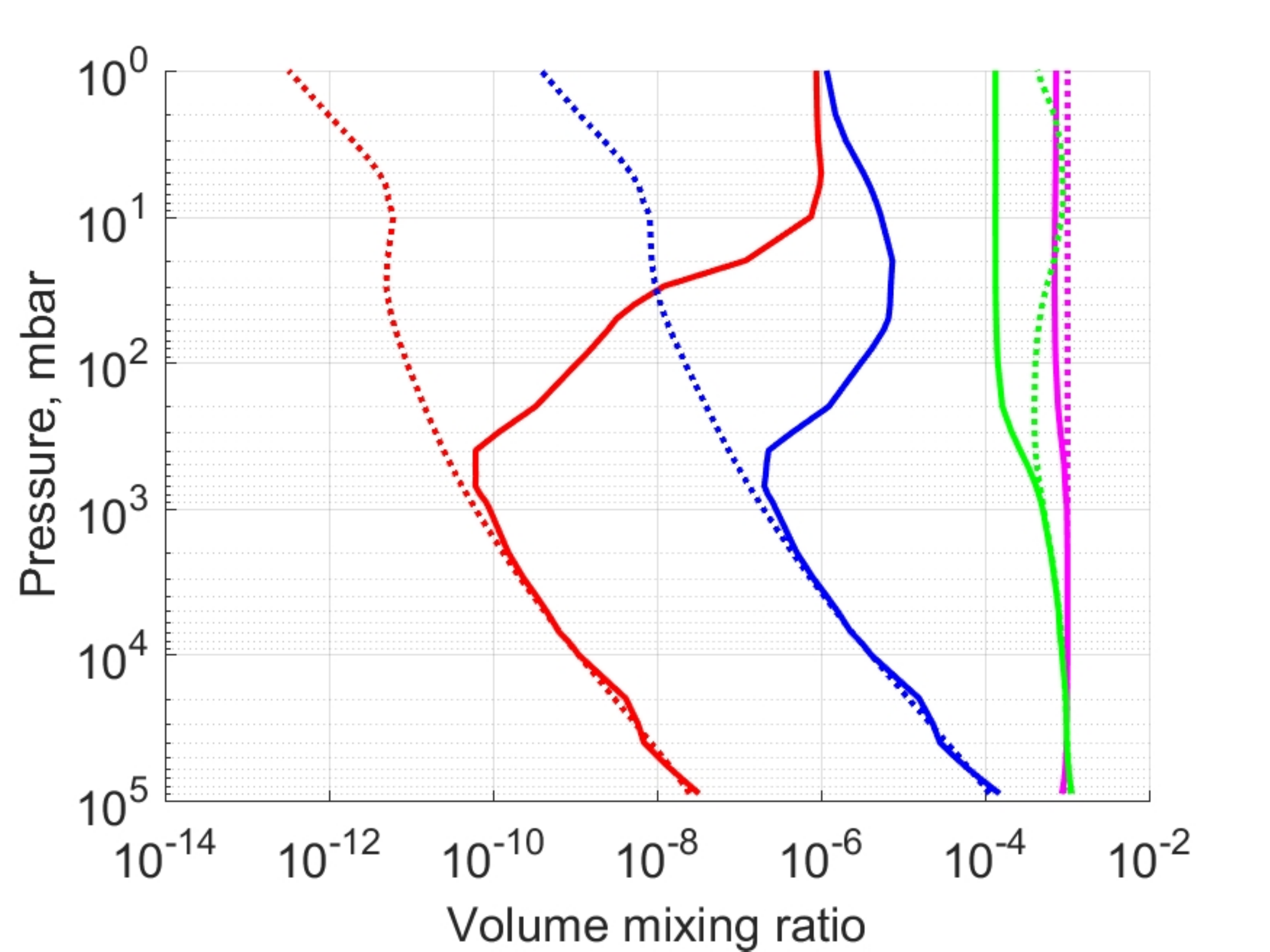}} & 
{\includegraphics[width=4.2cm]{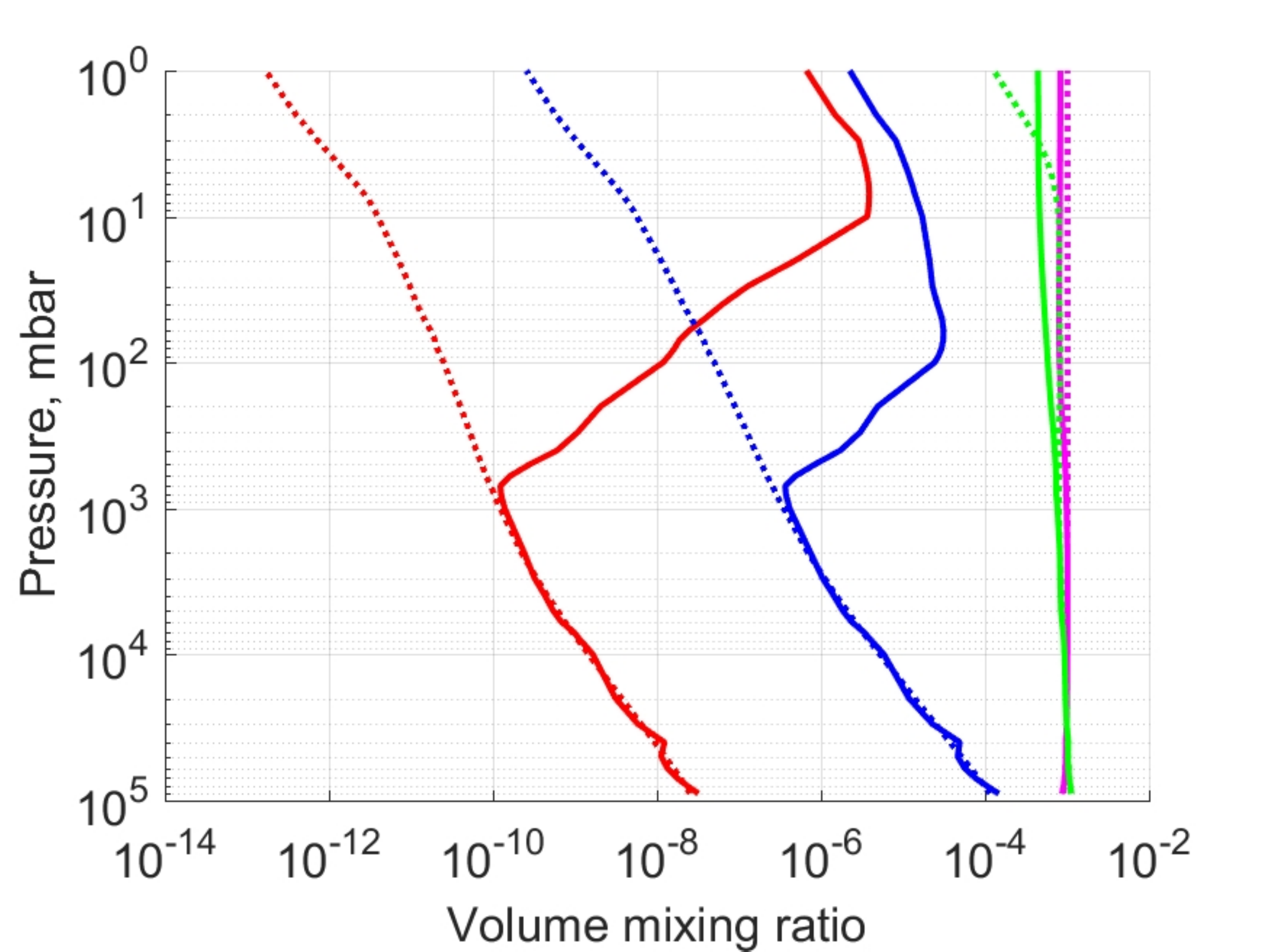}} \\     
  \end{tabular}
  \caption{Vertical distribution of CH$_4$, CO, H$_2$O and CO$_2$ from the simulations with different $G$ and $C/O$. The line colors are associated with different species: CO$_2$ (red), H$_2$O (blue), CO (magenta) and CH$_4$ (green). The different rows correspond to different spatial averaging. In the first row the values were averaged over the dayside, second row over the nightside, third row over the morning terminator region (between 225 and 305 degrees longitude) and last row over the evening (between 45 and 135 degrees longitude) terminator region. The evening terminator is defined as the leading limb in the first transit. The dotted lines are the abundances of the chemical species in chemical equilibrium and the solid lines are the abundances obtained from the 3D dynamical simulations. All the results shown in this figure were averaged over the last 100 days of the simulations.}
\label{fig:che_ref_v} 
\end{figure*}

\begin{figure*}
\begin{centering}
\subfigure[G = 0.5 $\&$ C/O = 0.5]{
\includegraphics[width=0.8\columnwidth]{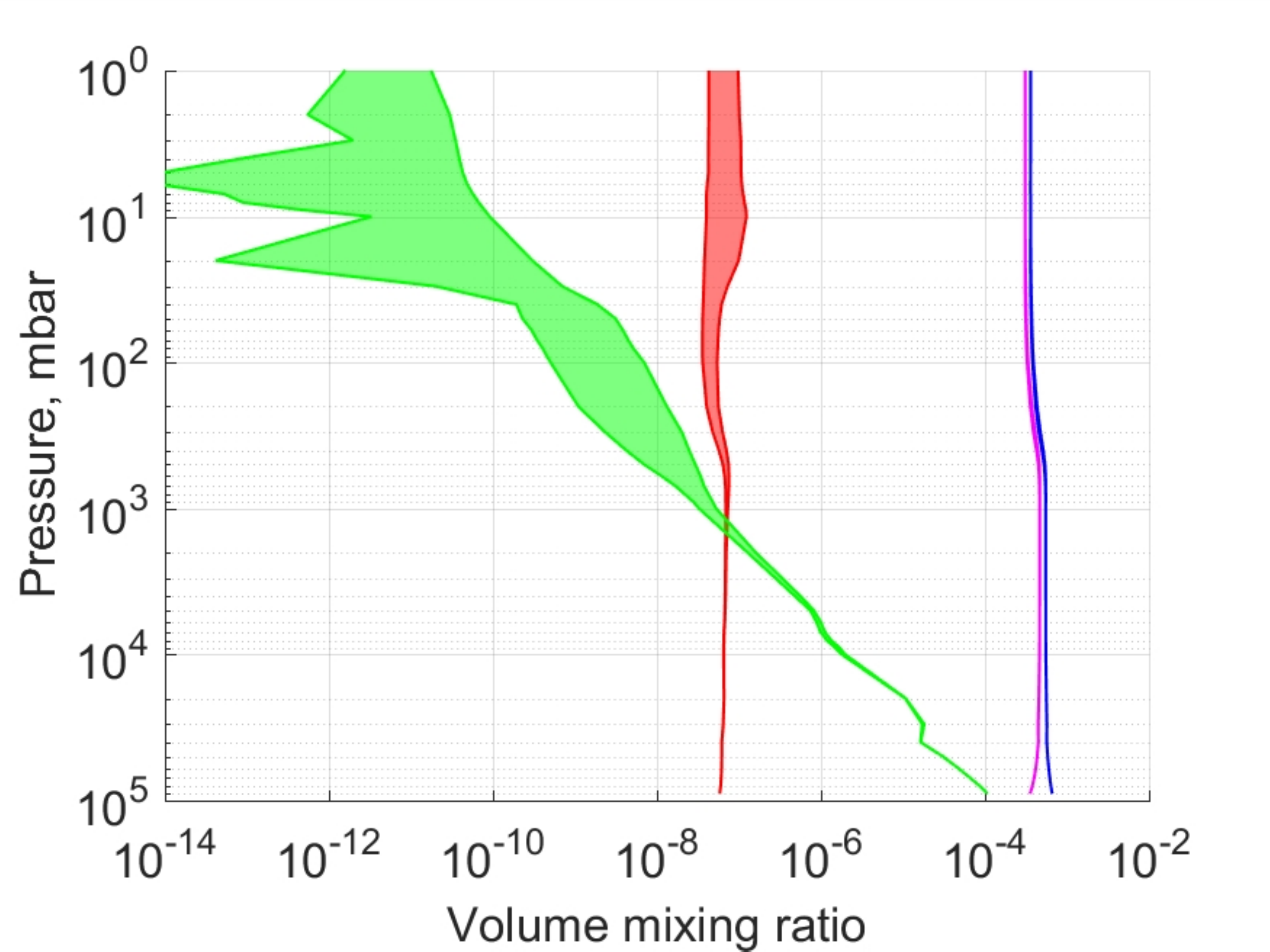}}
\subfigure[G = 2.0 $\&$ C/O = 0.5]{
\includegraphics[width=0.8\columnwidth]{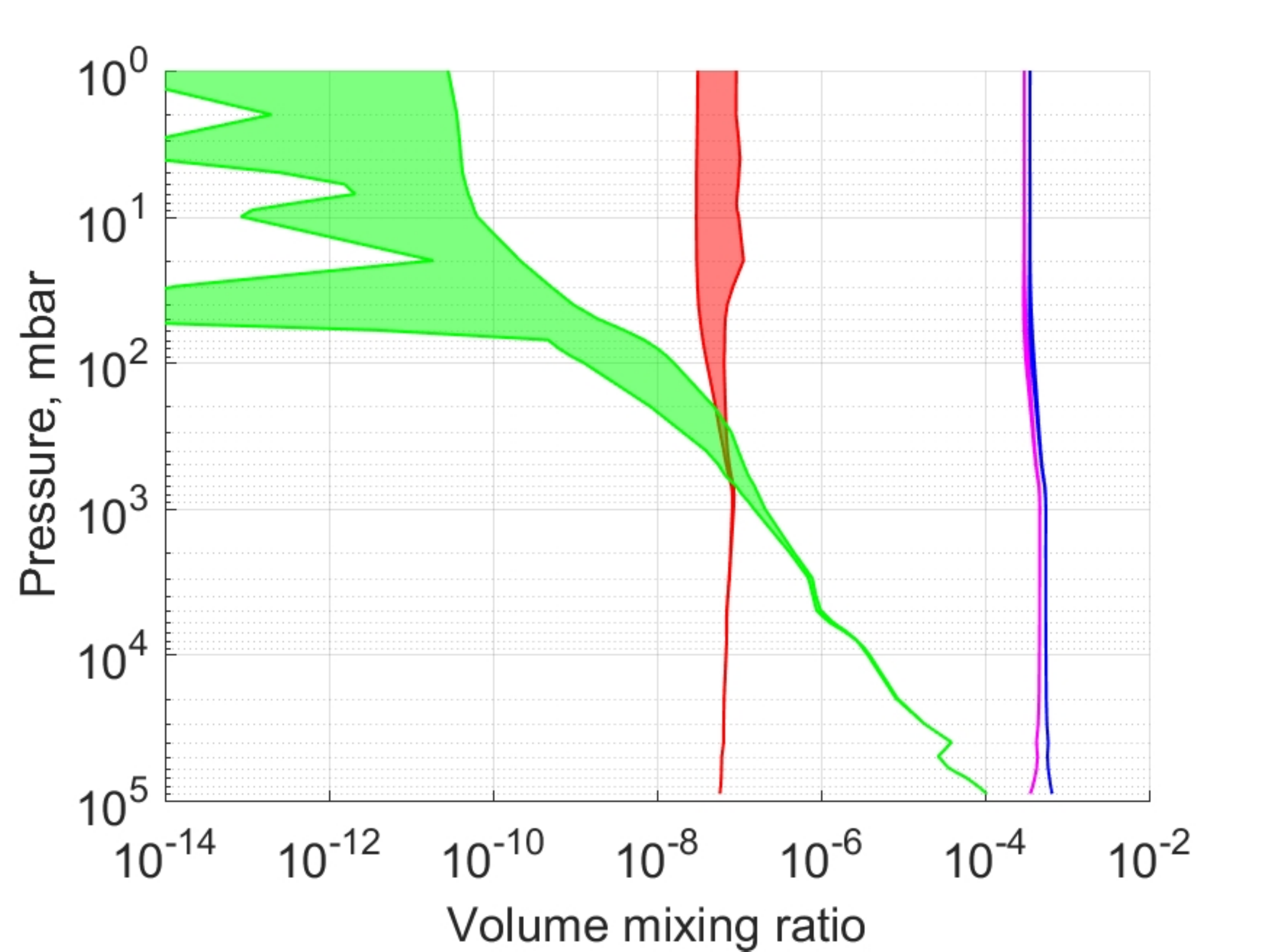}}
\subfigure[G = 0.5 $\&$ C/O = 2]{
\includegraphics[width=0.8\columnwidth]{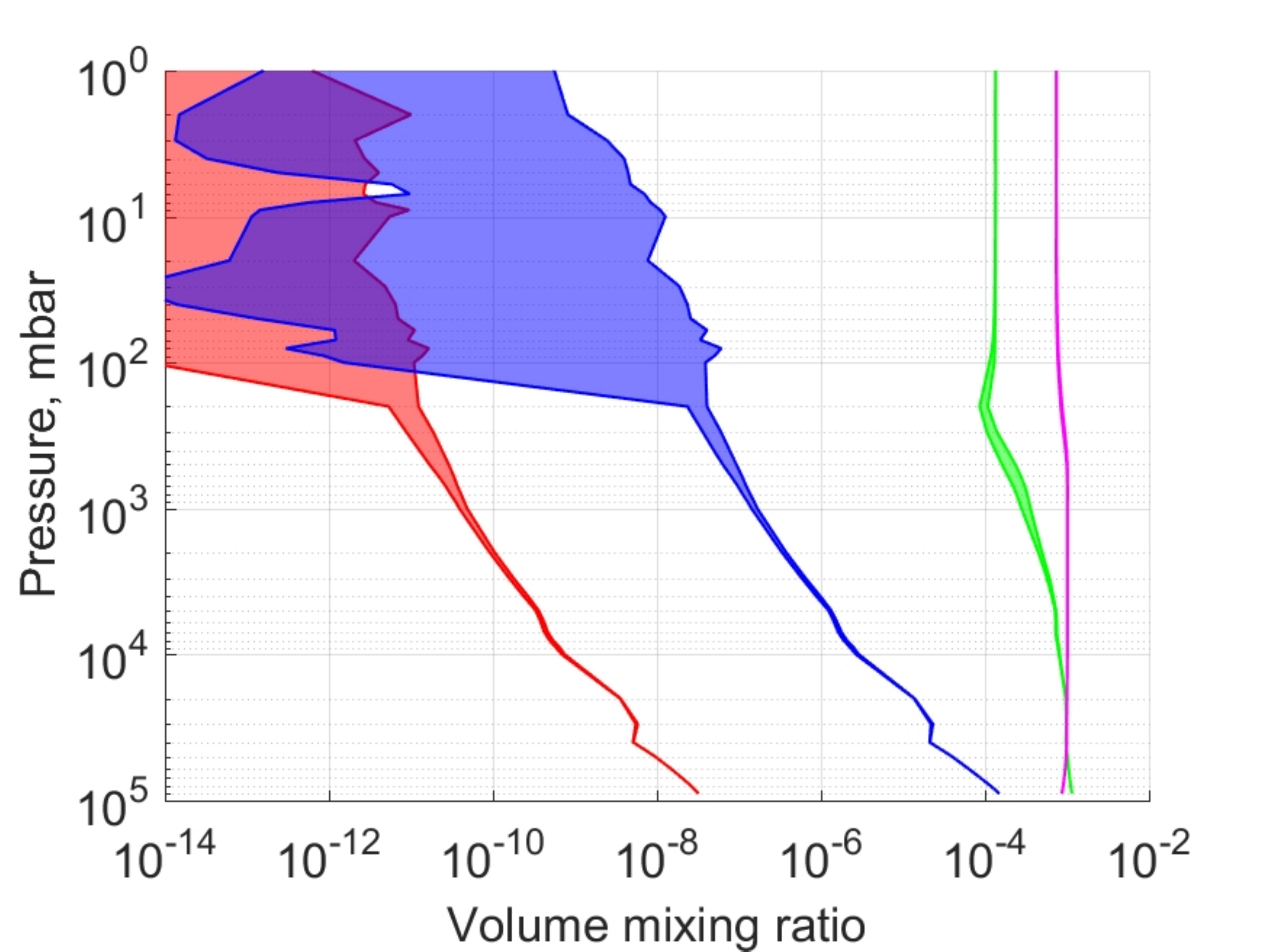}}
\subfigure[G = 2 $\&$ C/O = 2]{
\includegraphics[width=0.8\columnwidth]{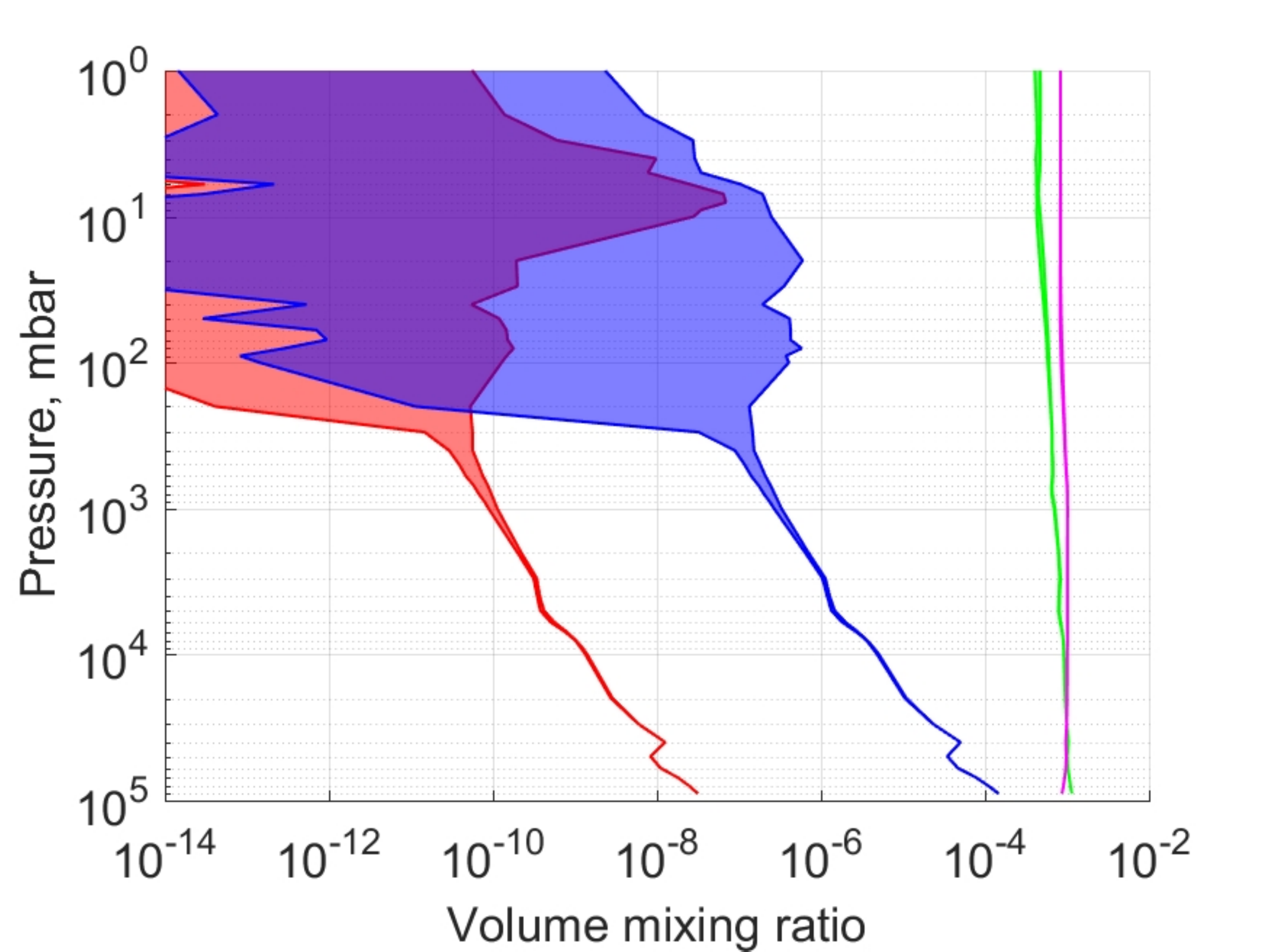}}
\caption{The colors represent the dispersion of the volume mixing ratios close to the equator for the different chemical species: CO$_2$ (red), H$_2$O (blue), CO (magenta) and CH$_4$ (green). All the results shown in this figure were averaged over the last 100 days of the simulations, and in latitude between 20 and -20 degrees.}
\label{fig:che_disp}
\end{centering}
\end{figure*}
\begin{figure*}
  \begin{tabular}{m{0.1cm} c c c c}
     & G = 0.5 $\&$ C/O = 0.5 & G = 2.0 $\&$ C/O = 0.5 & G = 0.5 $\&$ C/O = 2 & G = 2 $\&$ C/O = 2 \\
    \vspace{-3cm}\rotatebox{90}{CH$_4$} & {\includegraphics[width=4.2cm]{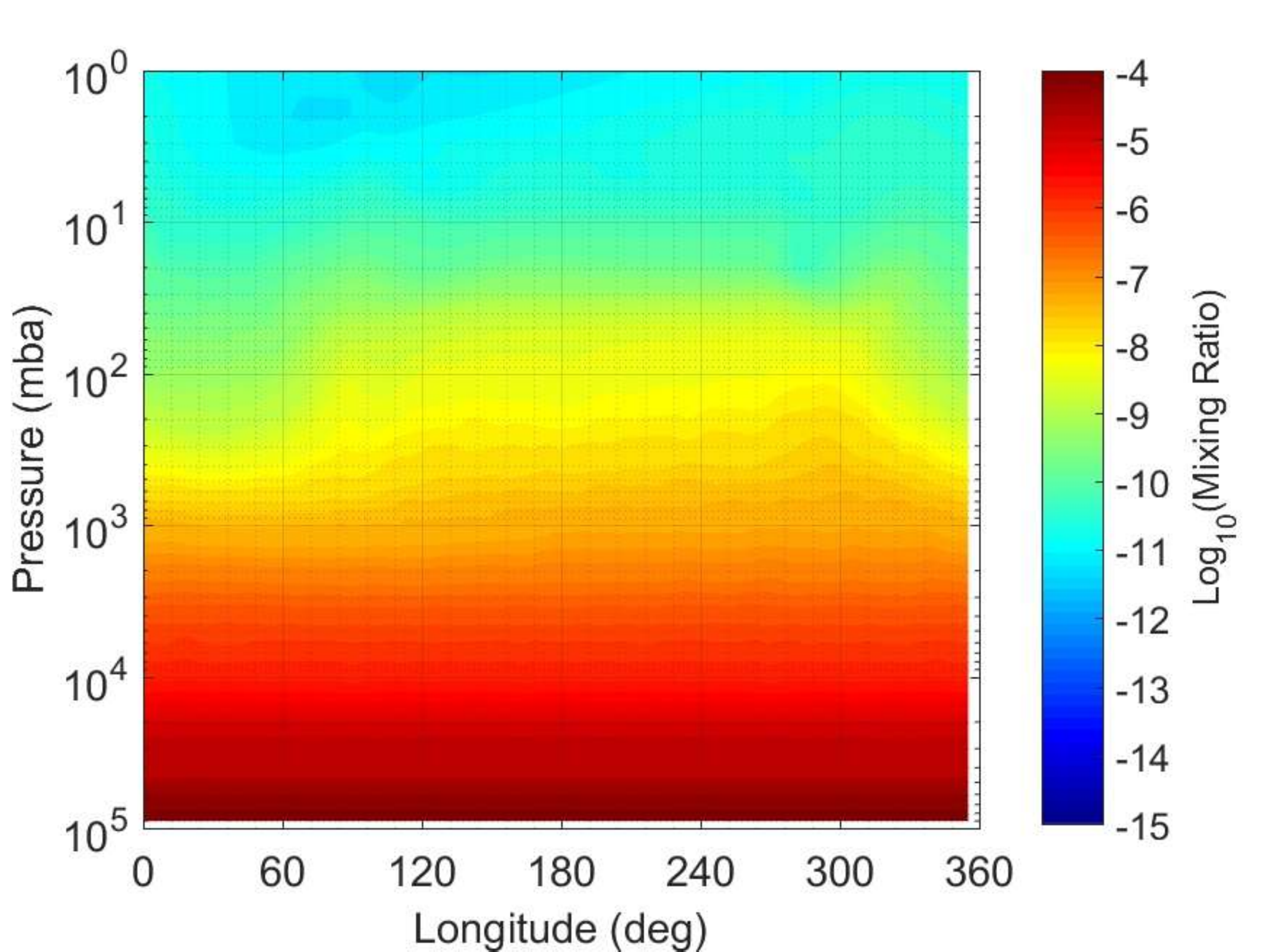}} & {\includegraphics[width=4.2cm]{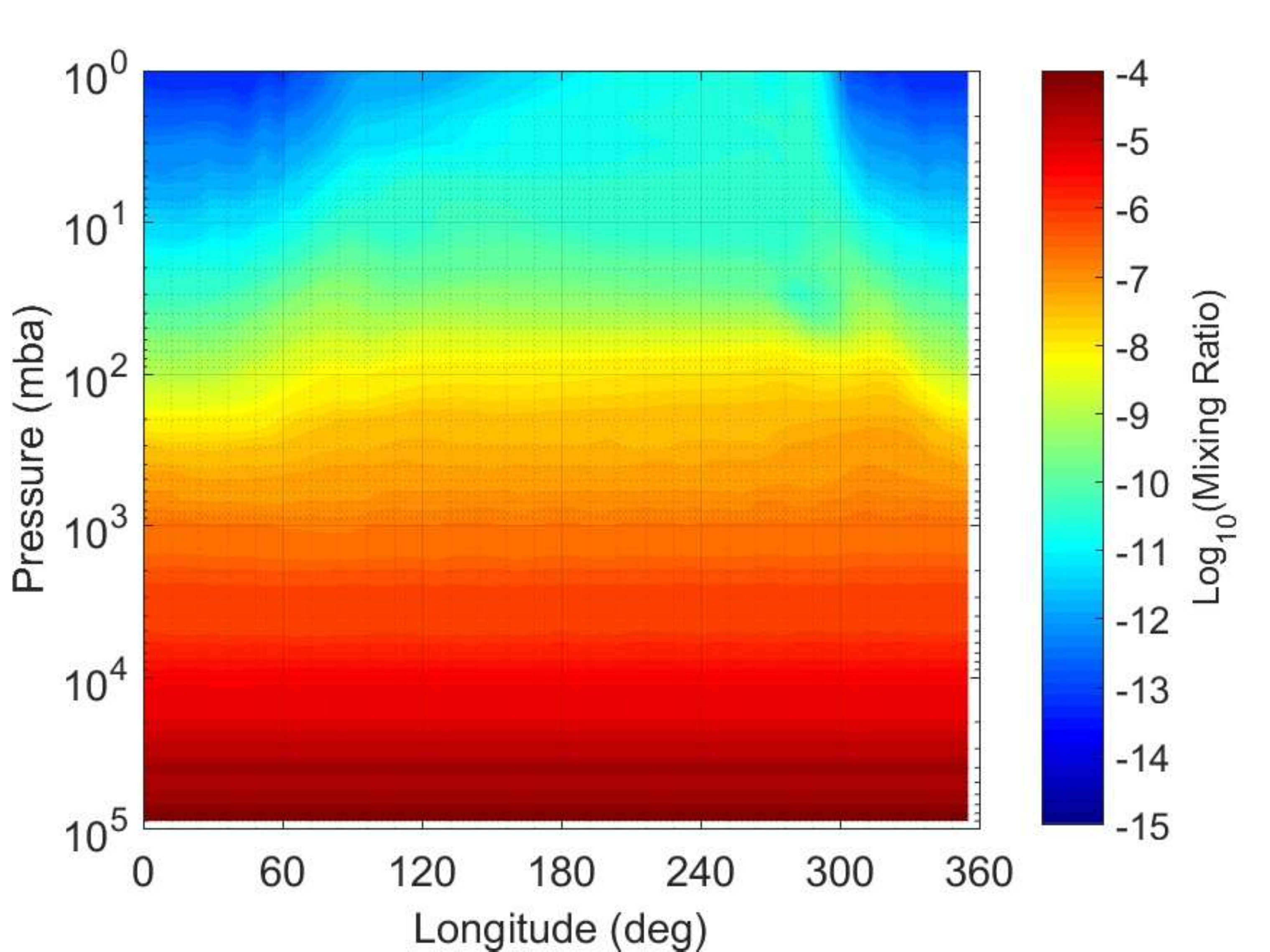}} &
{\includegraphics[width=4.2cm]{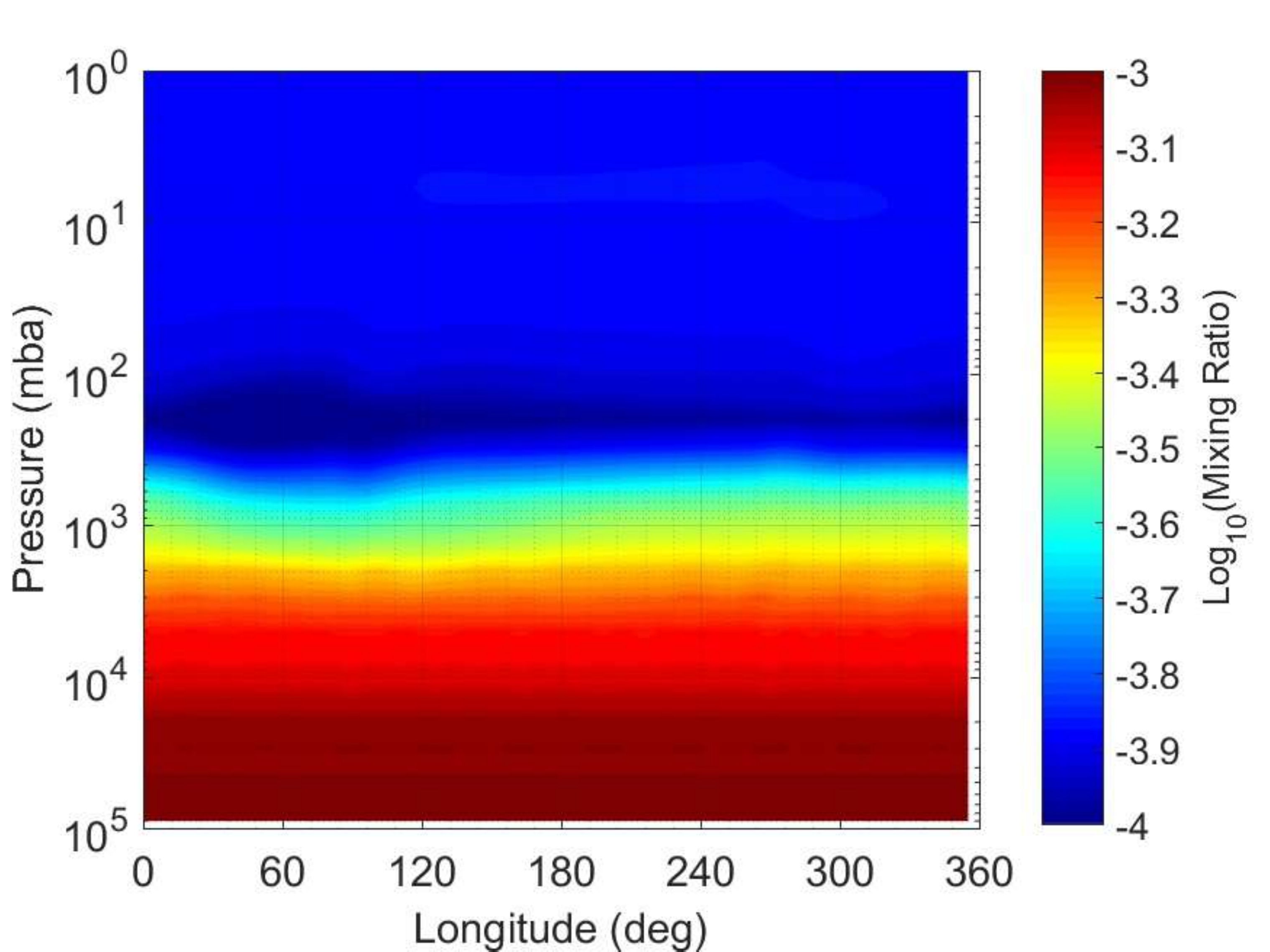}} & 
{\includegraphics[width=4.2cm]{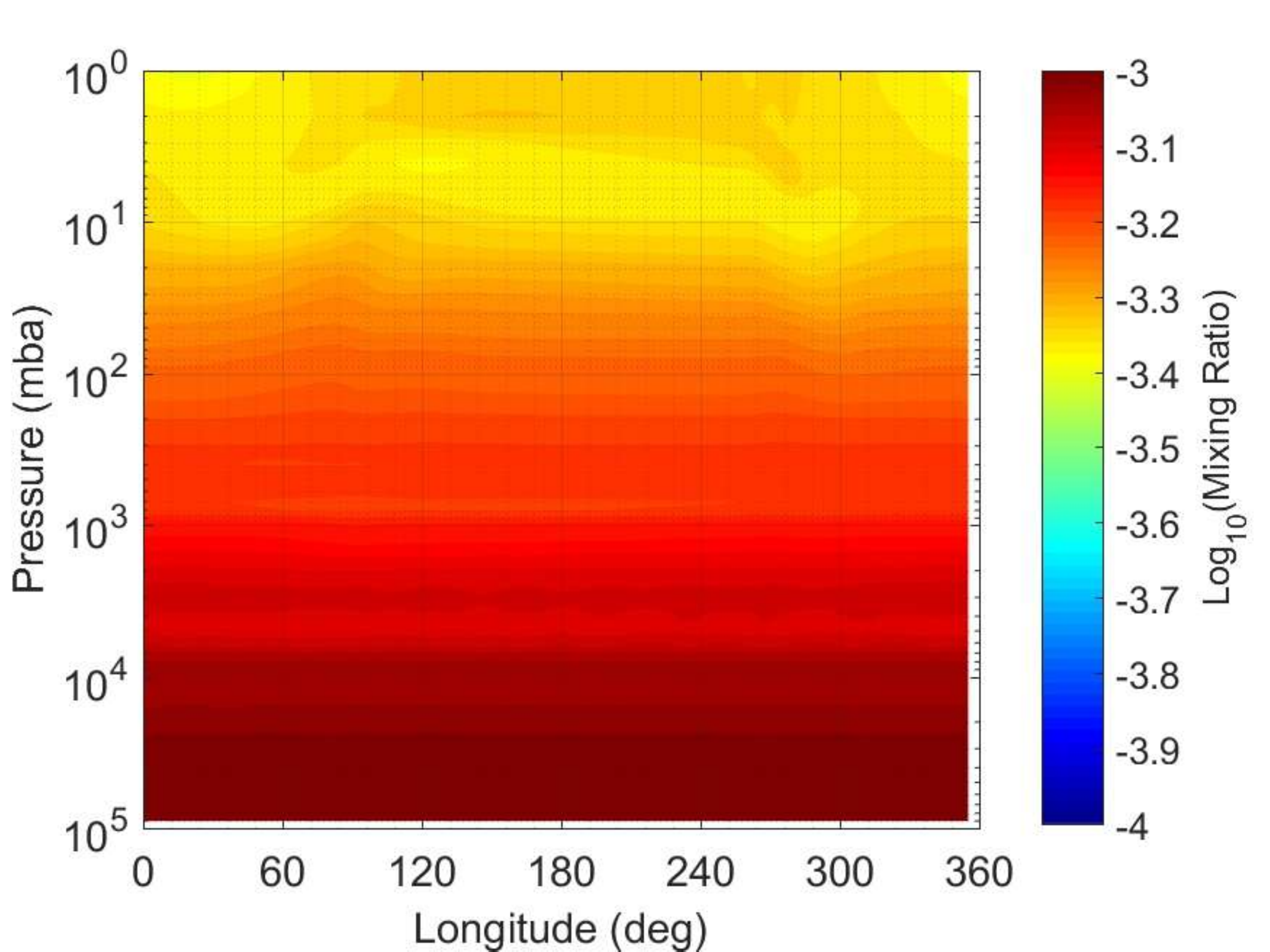}} \\ 
    \vspace{-3cm}\rotatebox{90}{CO} & {\includegraphics[width=4.2cm]{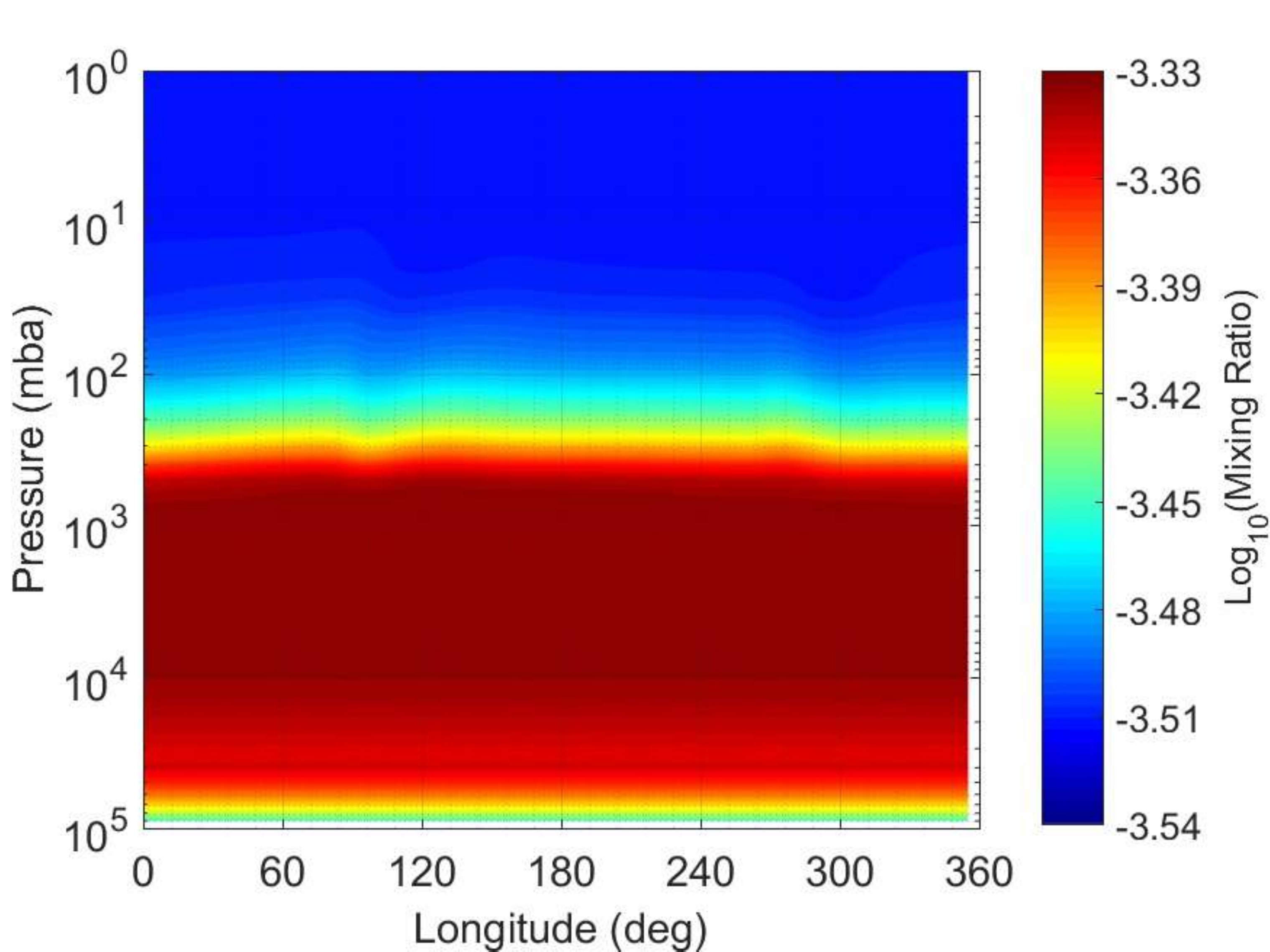}} & {\includegraphics[width=4.2cm]{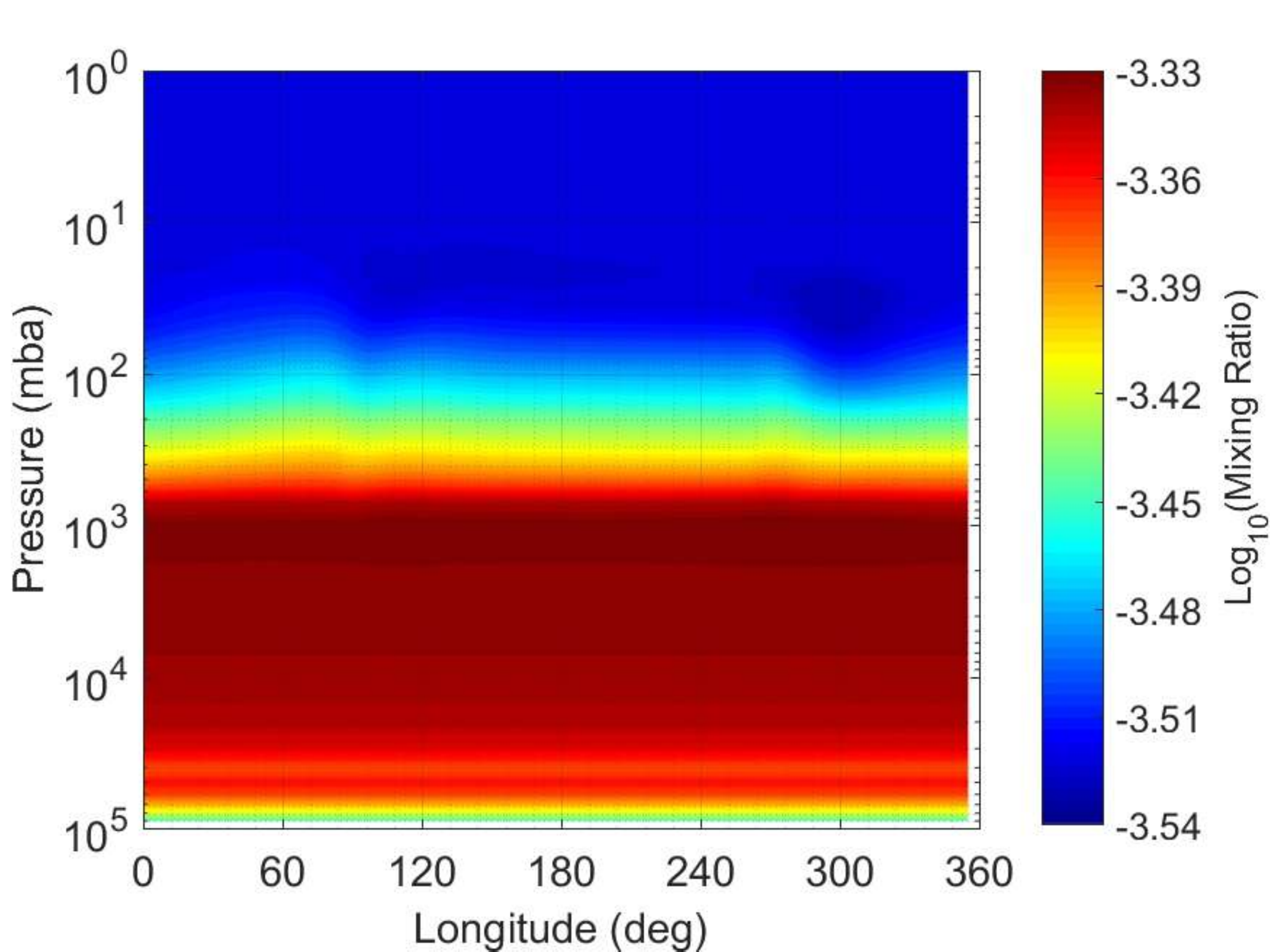}} &
{\includegraphics[width=4.2cm]{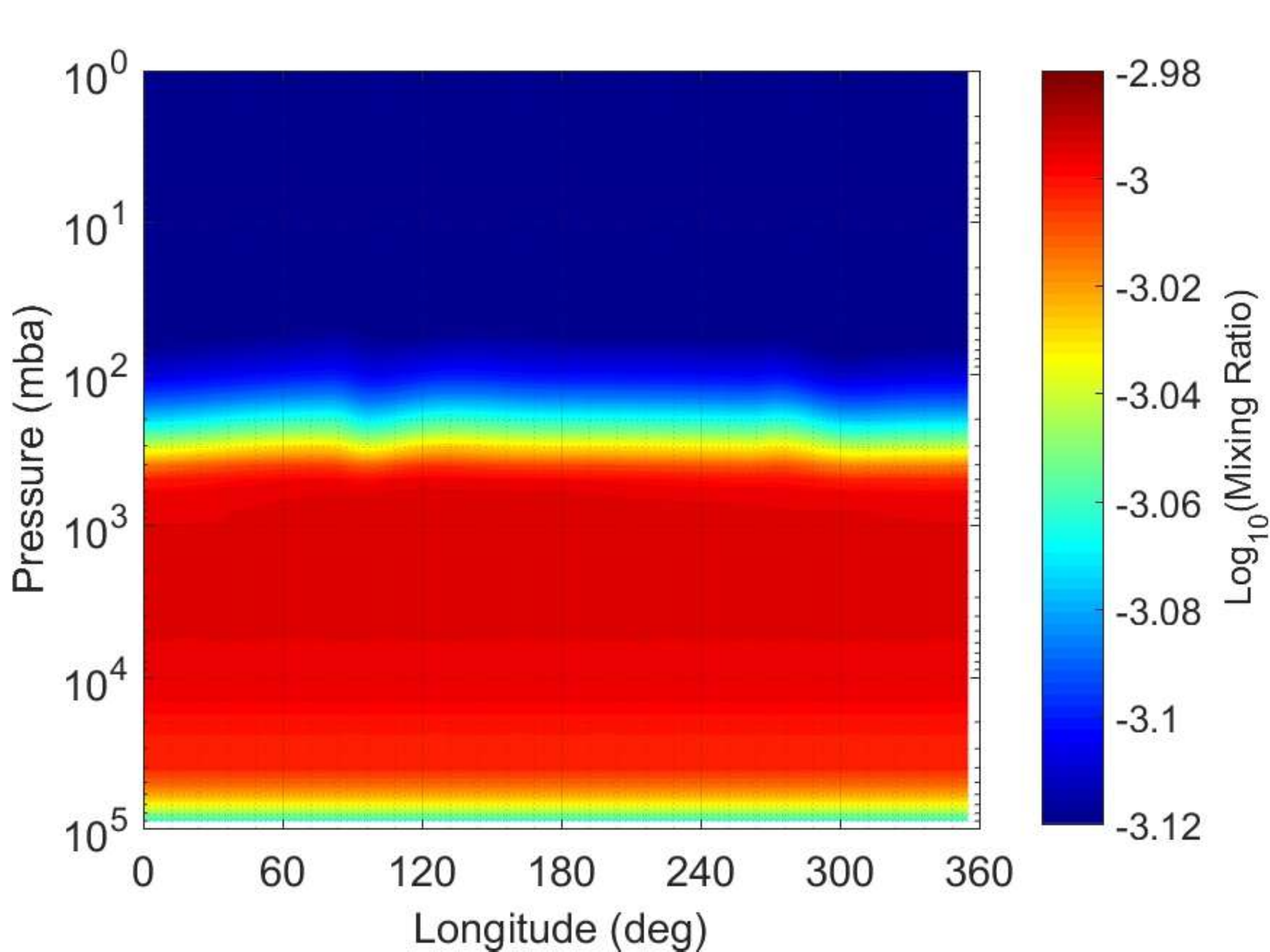}} & 
{\includegraphics[width=4.2cm]{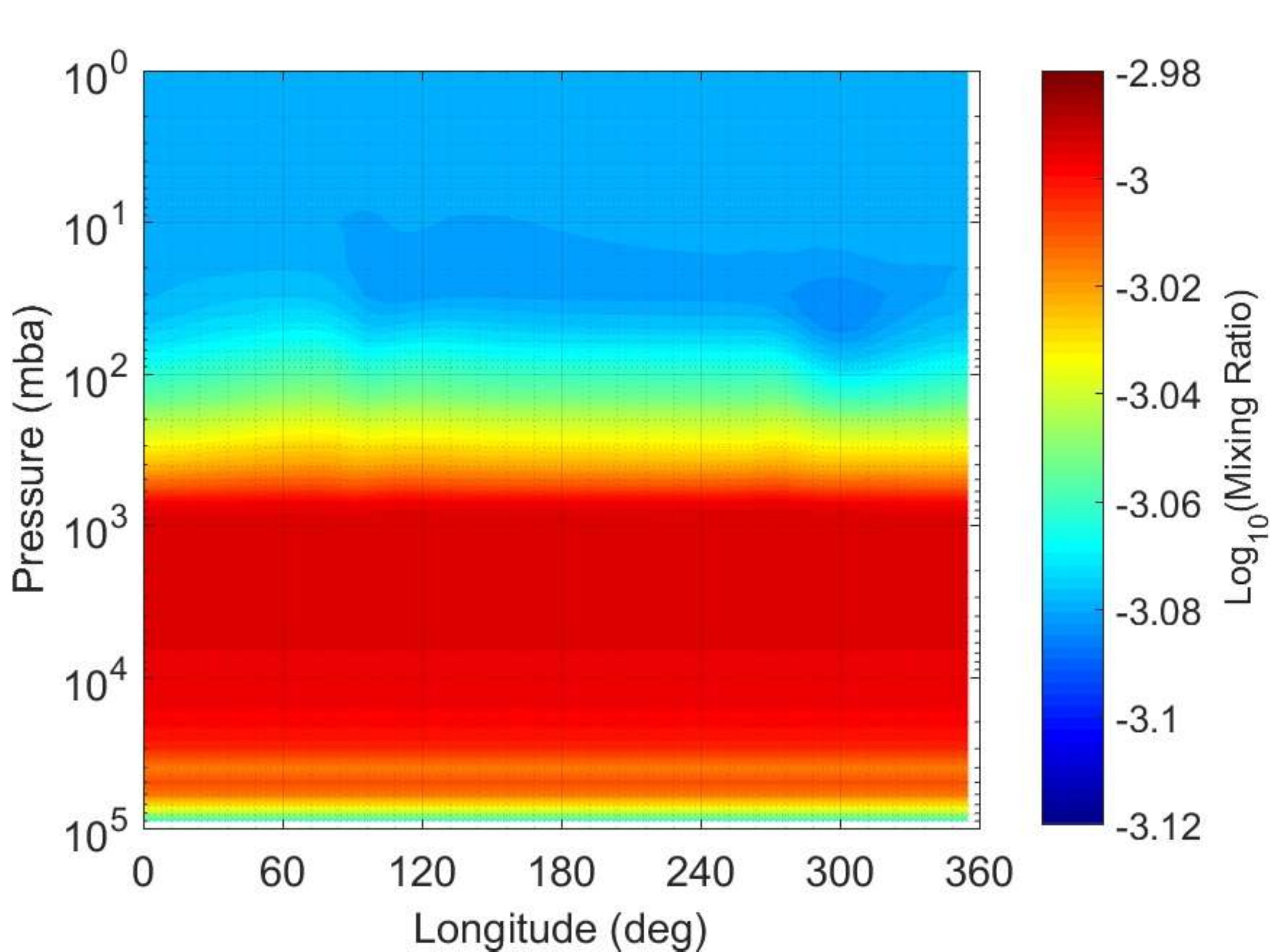}} \\ 
    \vspace{-3cm}\rotatebox{90}{H$_2$O} & {\includegraphics[width=4.2cm]{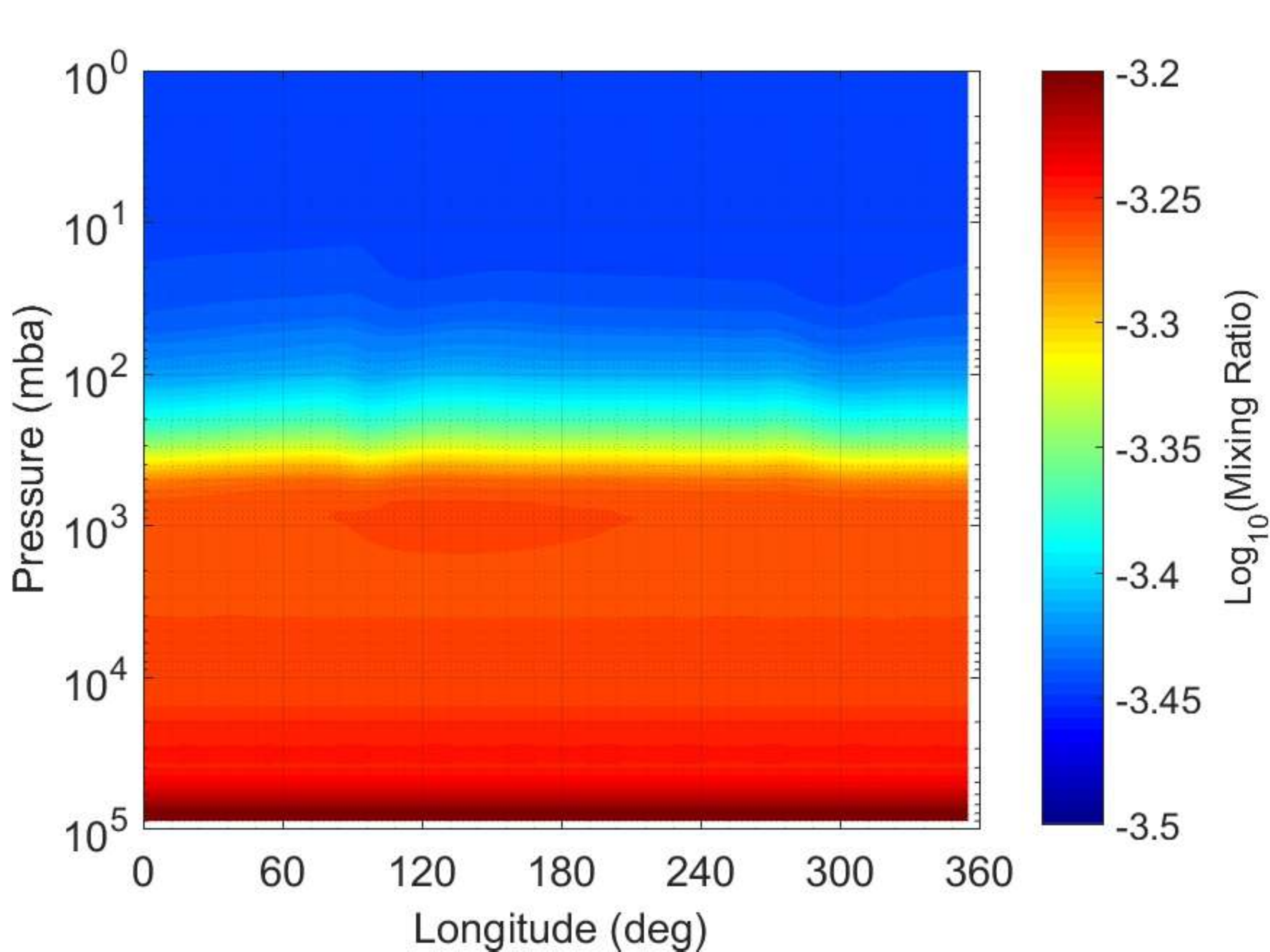}} & {\includegraphics[width=4.2cm]{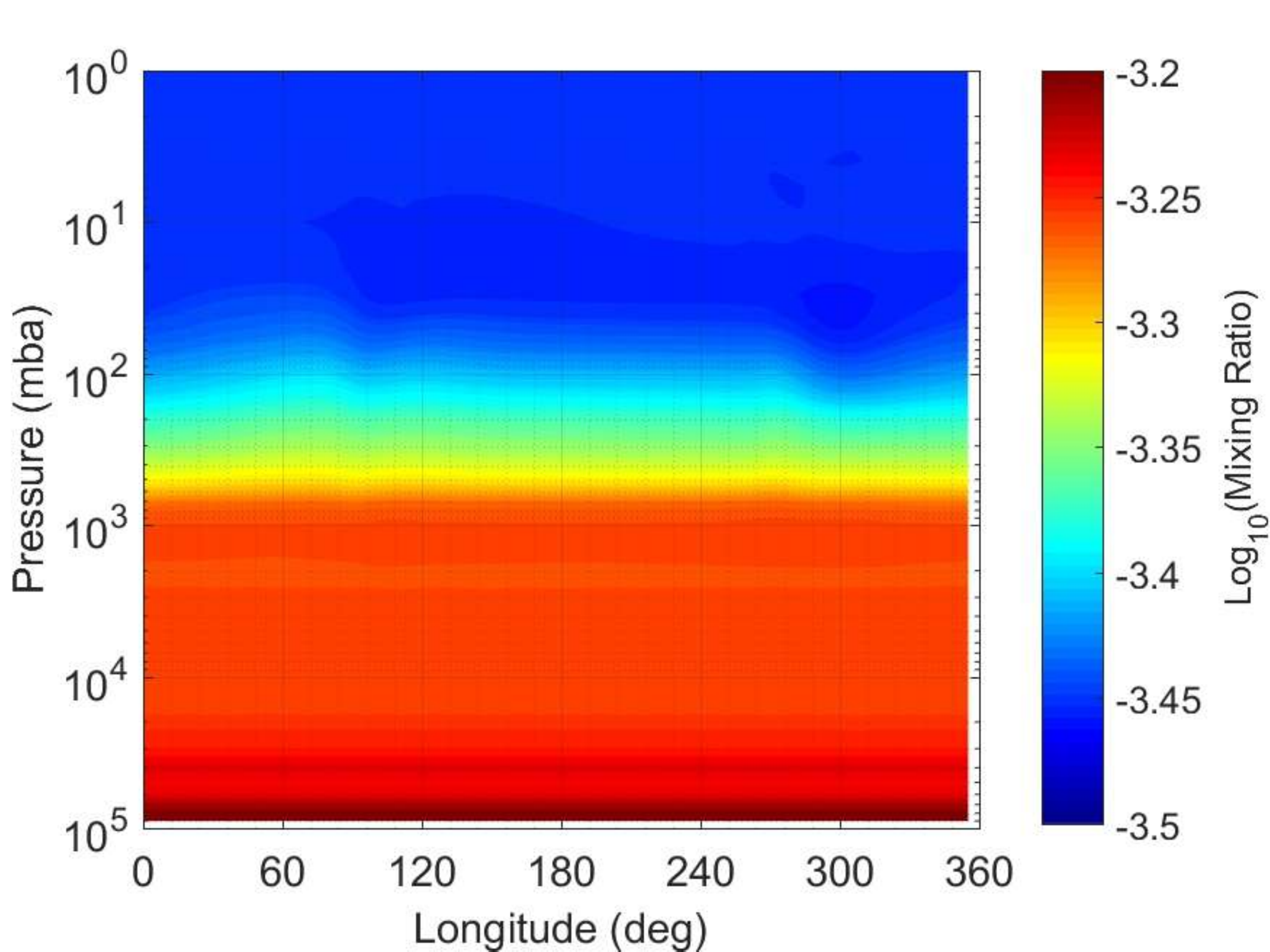}} &
{\includegraphics[width=4.2cm]{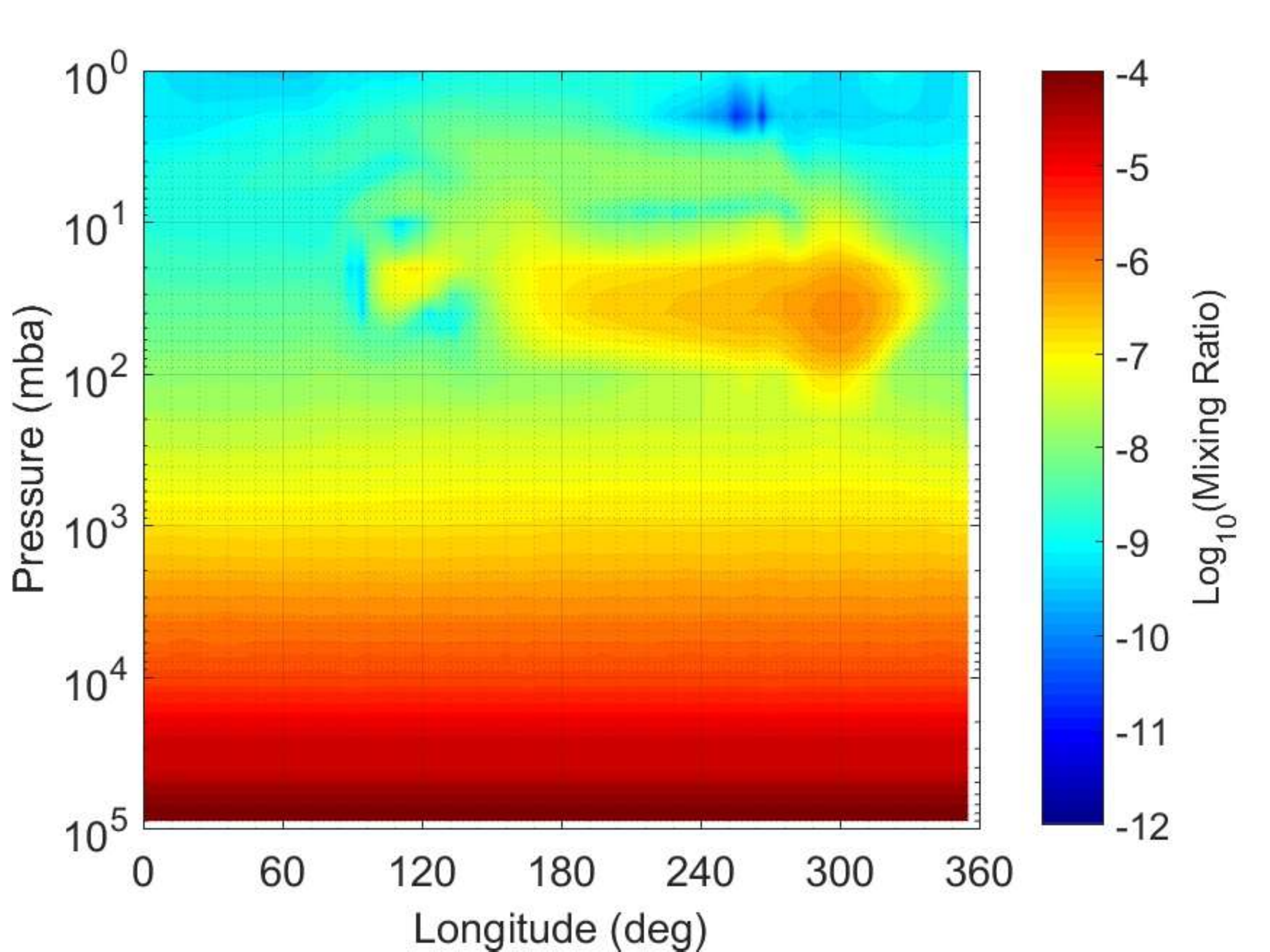}} & 
{\includegraphics[width=4.2cm]{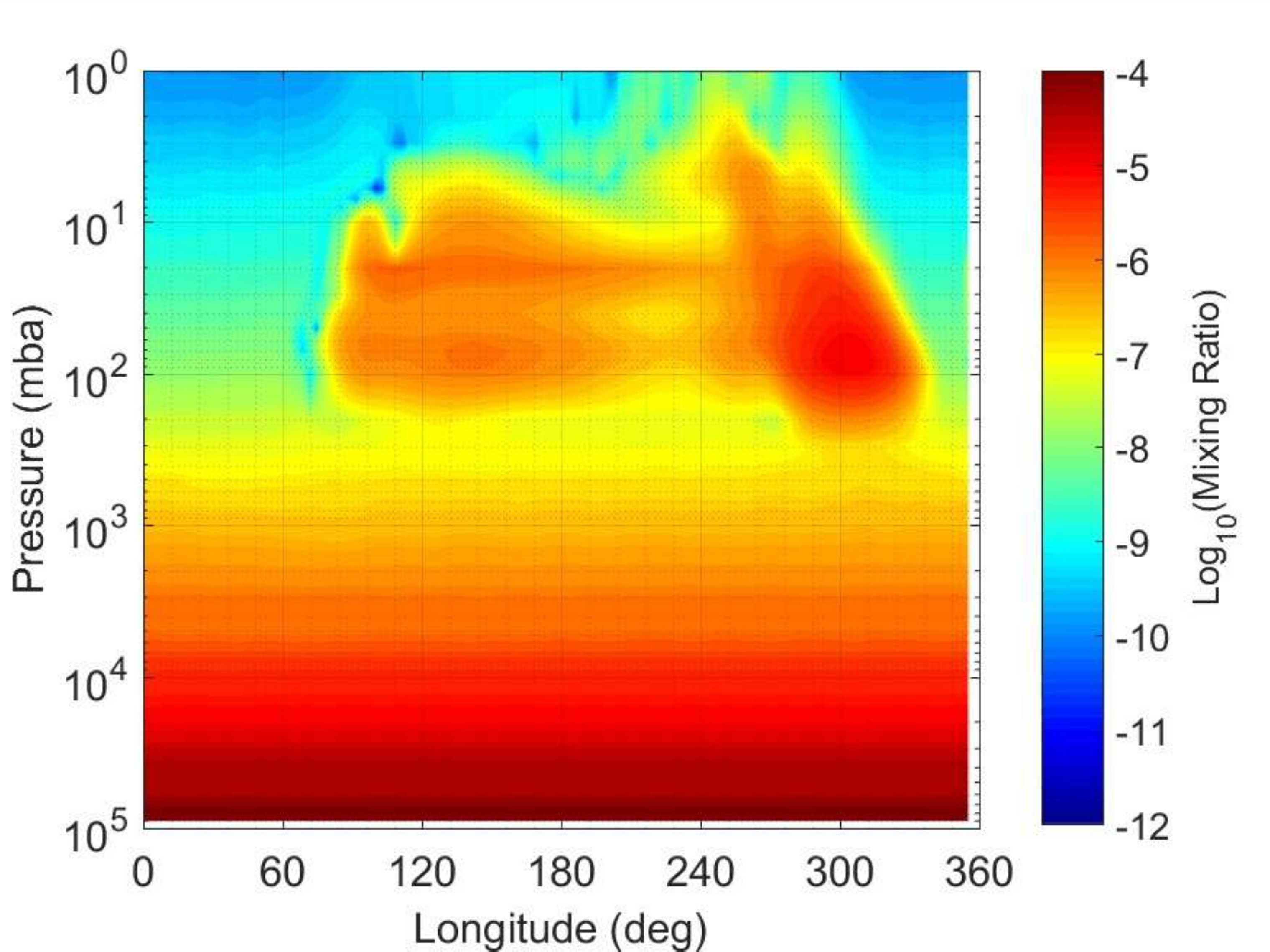}} \\
    \vspace{-3cm}\rotatebox{90}{CO$_2$} & {\includegraphics[width=4.2cm]{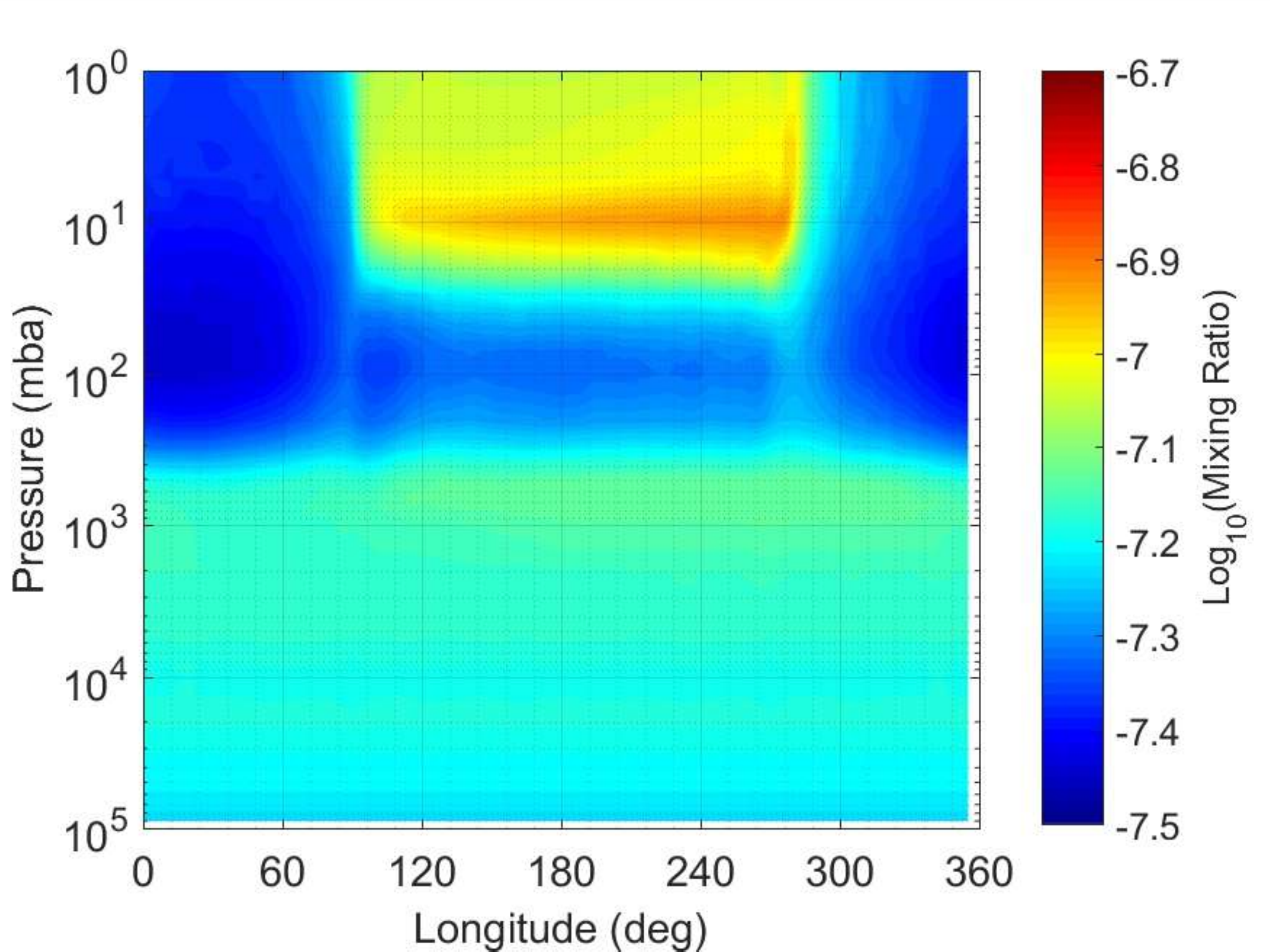}} & {\includegraphics[width=4.2cm]{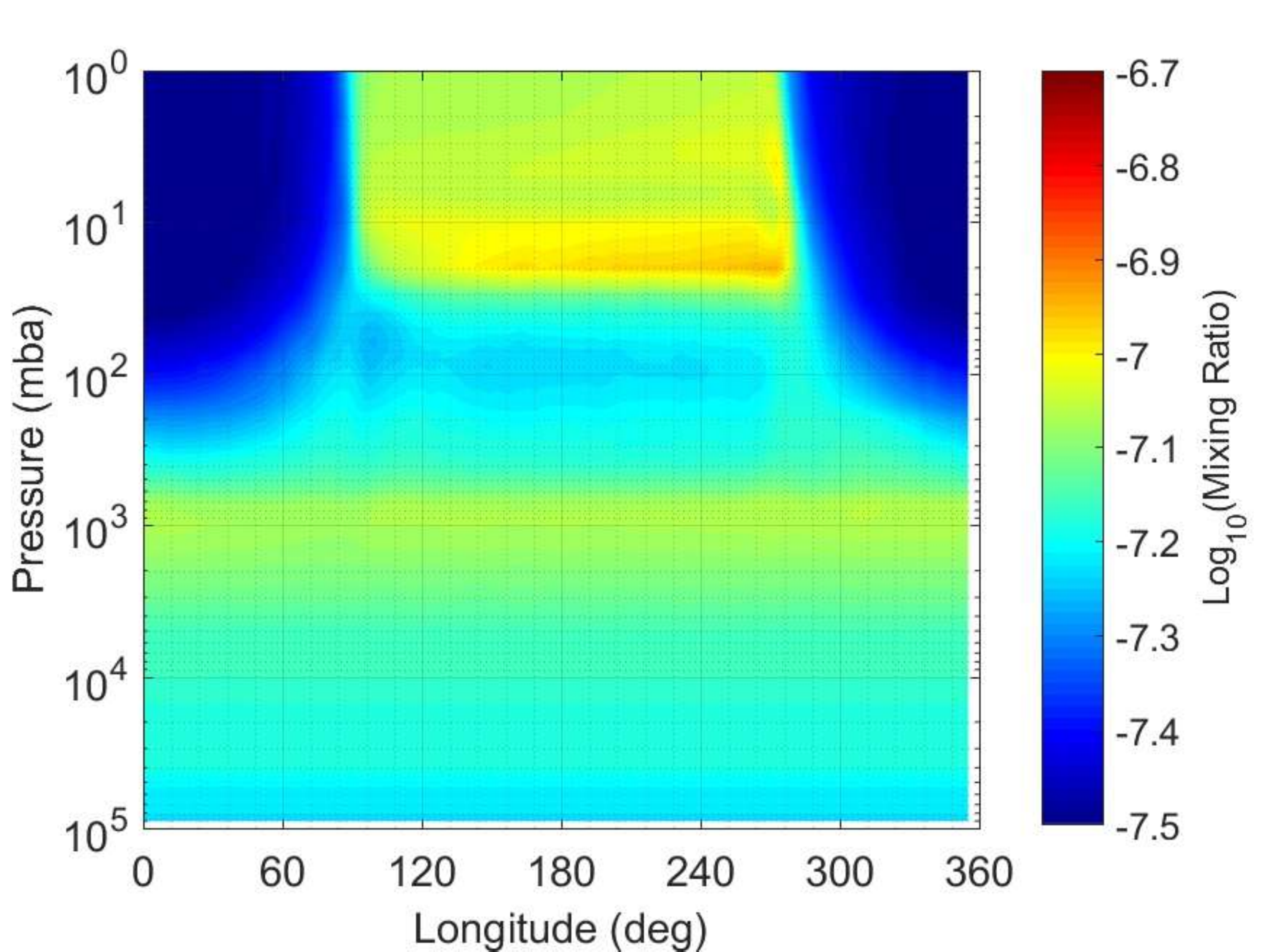}} &
{\includegraphics[width=4.2cm]{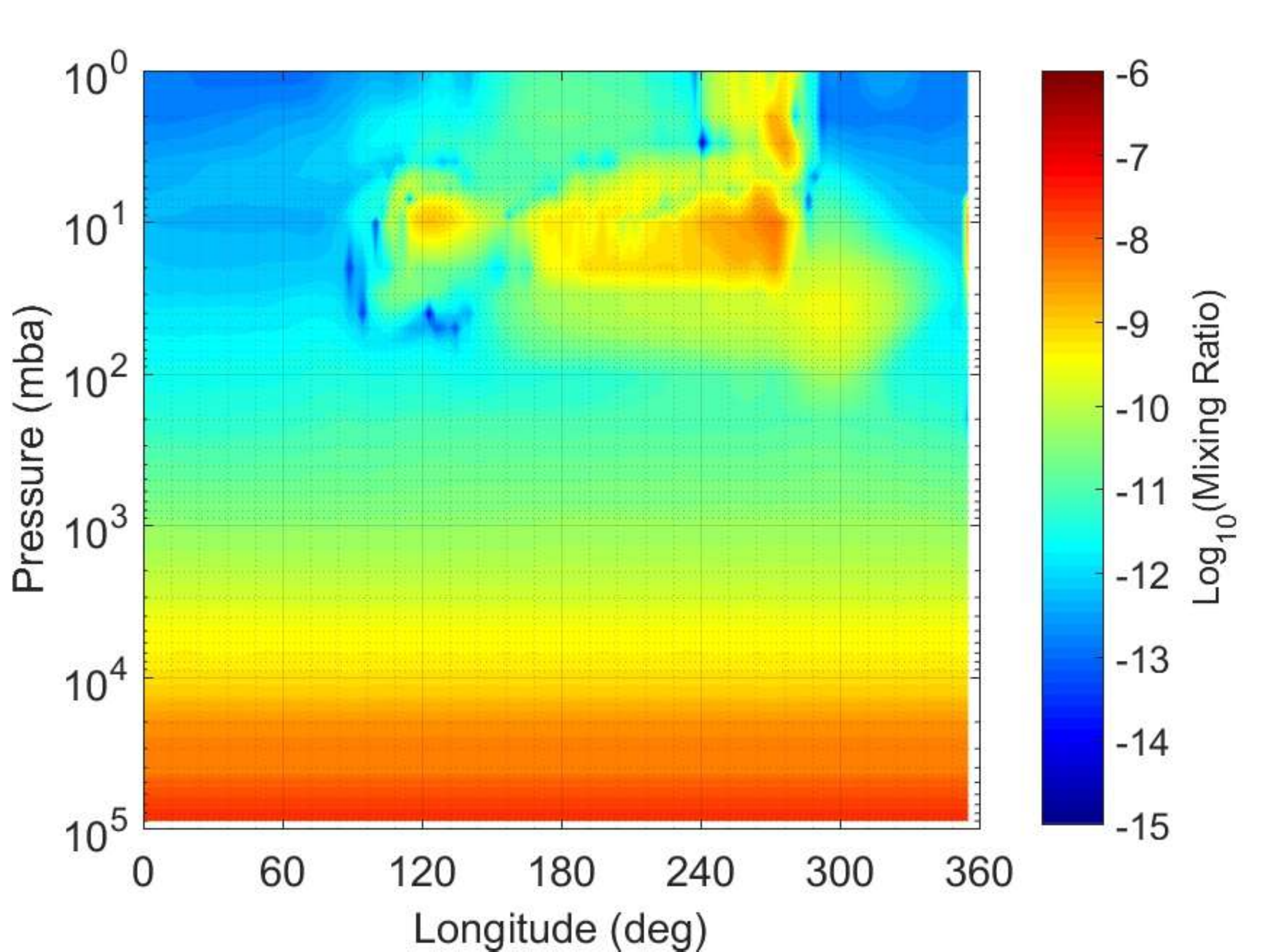}} & 
{\includegraphics[width=4.2cm]{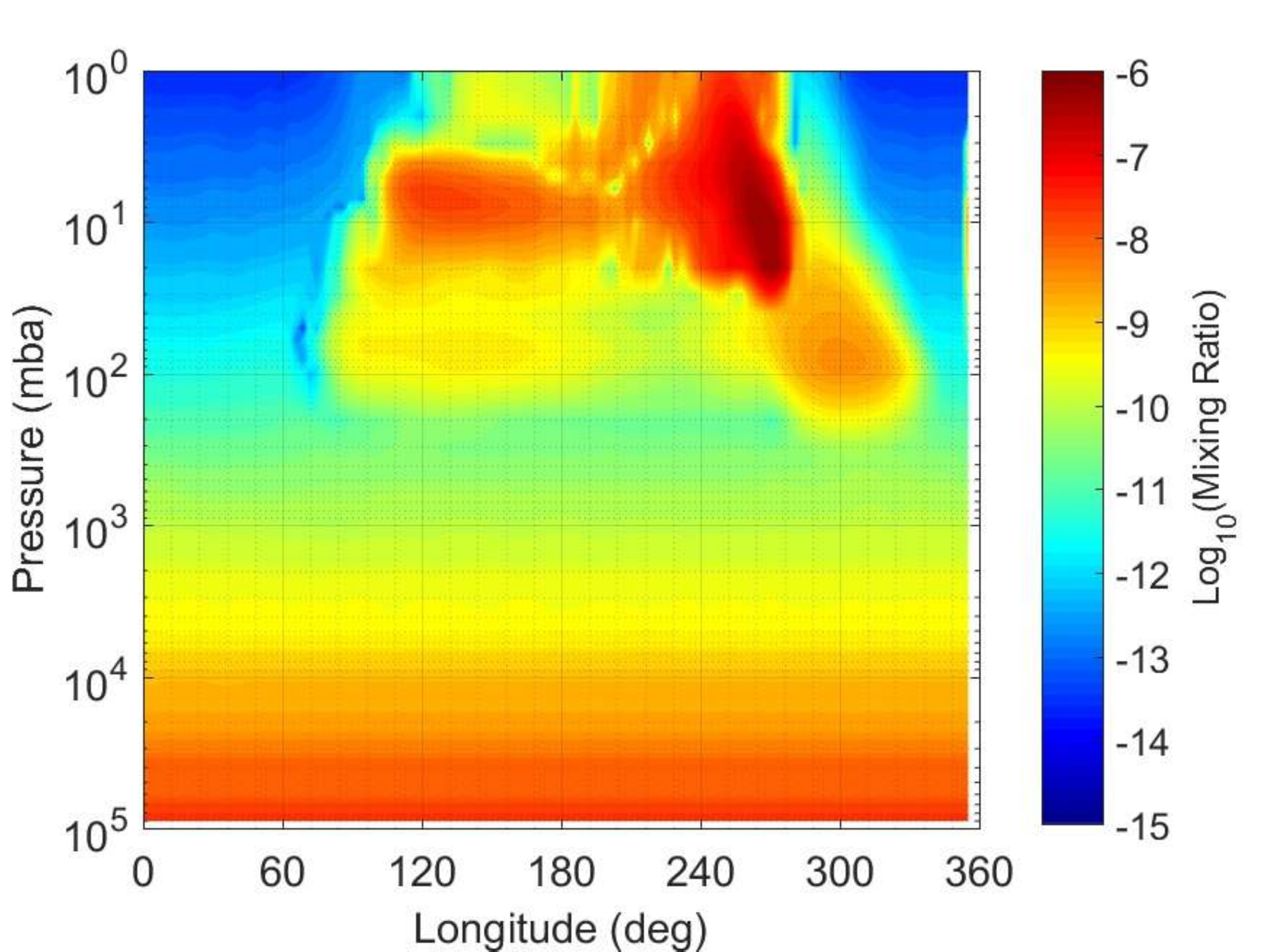}} \\      
  \end{tabular}
  \caption{Distribution of CH$_4$, CO, H$_2$O and CO$_2$ as a function of pressure and longitude from the simulations with different $G$ and $C/O$. The colors represent the volume mixing ratio of the different chemical species. All the results shown in this figure were averaged over the last 100 days of the simulations, and in latitude between 20 and -20 degrees. }
\label{fig:che_ref_0} 
\end{figure*}

\subsection{Spatial chemical distribution}

The distribution of the different chemical species in the atmosphere depends on the competition between chemistry and dynamics. In Fig. \ref{fig:che_ref_v}, the different panels show the vertical profiles of the chemical abundances obtained at different locations: dayside, nightside, morning terminator and evening terminator. As we explain in the figure's caption, the evening terminator is defined as the leading limb in the first transit. The different colors represent the different species: CO$_2$ (red), H$_2$O (blue), CO (magenta) and CH$_4$ (green). The different line styles represent the chemical equilibrium abundances (dotted line) and the results from the dynamics-chemistry coupling (solid line). The four species are in chemical equilibrium below the $\sim$1 bar pressure level. In the upper atmosphere, the abundances start to diverge from chemical equilibrium, in some cases by several orders of magnitude. The magnitude of the departures from chemical equilibrium depend largely on the dynamical mixing, but also on the differences in production rates between the dayside and nightside. The largest departures from equilibrium in the $C/O = 0.5$ case are from CH$_4$ and CO$_2$. In the case of CO$_2$, the large departure is mainly associated to the fact that the CO$_2$ is being forced towards its pseudo equilibrium as shown in Eq. \ref{eq:eq_che_pseudo}. The CO$_2$ departure from the equilibrium is also large in the $C/O = 2$ case. A large departure from the equilibrium is also seen for H$_2$O. It is worth mentioning that with $C/O = 2$, H$_2$O and CH$_4$ swap roles of being the major component in equilibrium (\citealt{2012Madhusudhan} and \citealt{2017Tsai}). The presence of a thermal inversion does not have a large impact on the results. There is no large variability in these plots along the longitude which is a consequence of the chemical components being quench horizontally from the hot dayside as we see later in this section.

Fig. \ref{fig:che_disp} shows the dispersion of the chemical species in regions close to the equator. In the experiments with $C/O = 0.5$, one value could represent the abundances of CO and H$_2$O across the upper atmosphere. When $C/O$ is increased to 2, large variability is obtained for CO$_2$ and H$_2$O. In this carbon rich atmosphere, CO and CH$_4$  have very little dispersion, as being the major species, their equilibrium abundances vary little with altitude. 

In Fig. \ref{fig:che_ref_0}, the different panels show the chemical pressure-longitude distribution. If the chemical reactions dominate over the dynamics, the distribution of the different chemical species will be closely correlated with the temperature distribution. Note that in our simulation and plots, noon is at 0$^o$ longitude. As seen in Fig. \ref{fig:ref_results_u_temp}, the temperature below roughly 1 bar starts to become uniform in longitude, and together with the typical short chemical timescales at large pressures, results in a chemical distribution in the deep atmosphere that does not change much with longitude. This result is true for the four chemical species explored. Above 1 bar the results start becoming more interesting as previously discussed for Fig. \ref{fig:che_ref_v}. In the simulations with solar abundances there are no large differences between simulations with and without thermal inversion in the dayside of the upper atmosphere. The main differences are associated with the larger day-night contrast in the case with thermal inversion which also enhances the day-night contrast in the distribution of CH$_4$ and CO$_2$. For these two species the warmer dayside reduces their abundance in the dayside of the atmosphere. The uniform distribution in longitude for H$_2$O and CO in the upper atmosphere is tied to the strong horizontal mixing (horizontal quenching) driven by the strong equatorial jet. When $C/O$ is increased to 2, the abundance of CH$_4$ in the upper atmosphere increases. The abundance of CO is also slightly increased when compared to the results from the $C/O = 0.5$ experiment, where the warmer dayside favours a slight increase in CO abundance in the upper atmosphere. The same conclusions on horizontal quenching on CO, was obtained by \cite{2012Agundez}. Compared to the solar abundance experiments, H$_2$O naturally decreases its abundance for cases where $C/O$ is increased. In the upper atmosphere (above the pressure level 100 mbar) the higher concentrations of H$_2$O and CO$_2$ are in the nightside. In both cases the concentration is pushed eastwards (in the same direction of the broad jet stream), which results in higher concentrations in regions around the morning terminator. 

\begin{figure*}
  \begin{tabular}{m{0.1cm} c c c c}
     & G = 0.5 $\&$ C/O = 0.5 & G = 2.0 $\&$ C/O = 0.5 & G = 0.5 $\&$ C/O = 2 & G = 2 $\&$ C/O = 2 \\
    \vspace{-3cm}\rotatebox{90}{CH$_4$} & {\includegraphics[width=4.2cm]{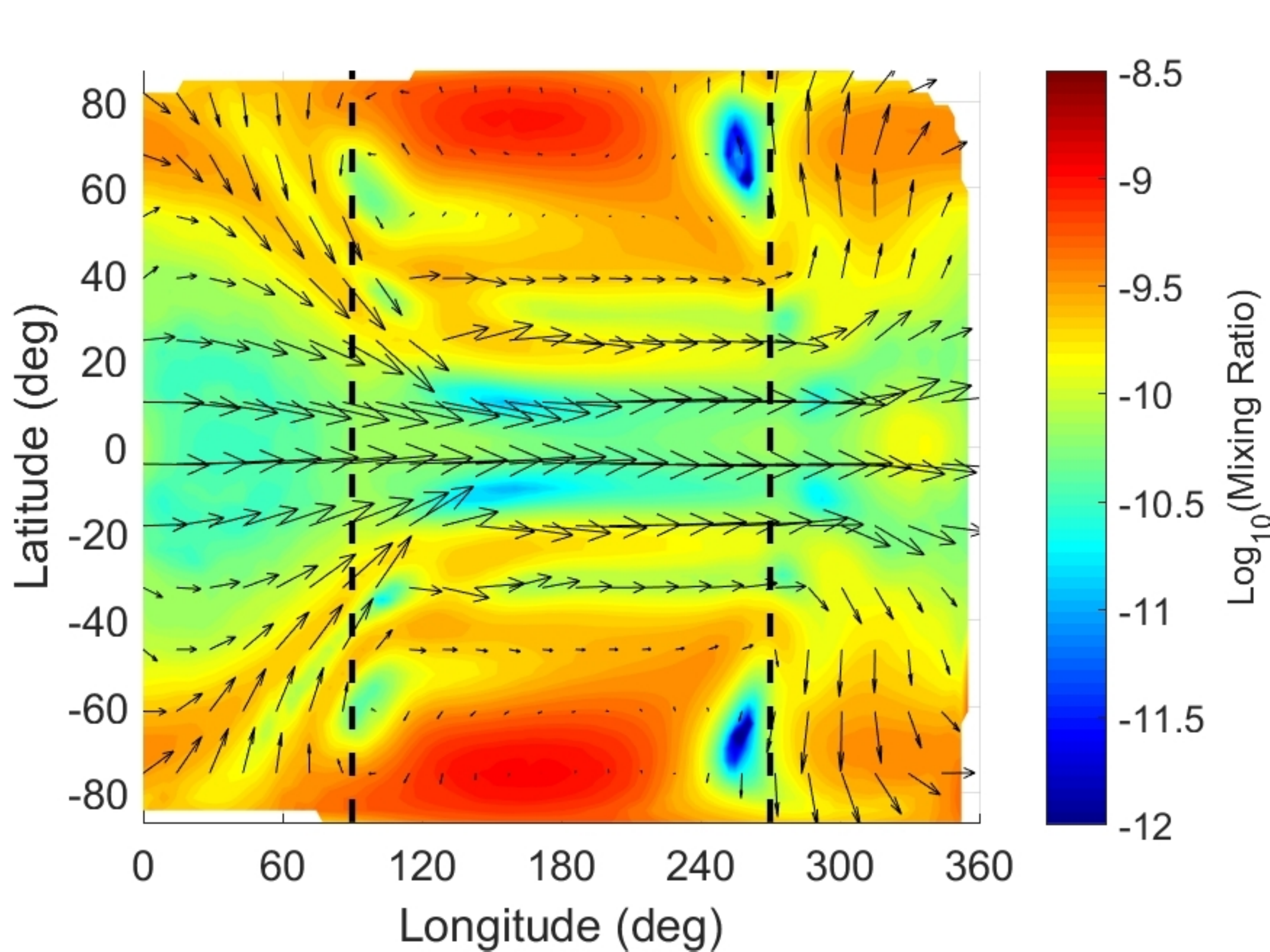}} & {\includegraphics[width=4.2cm]{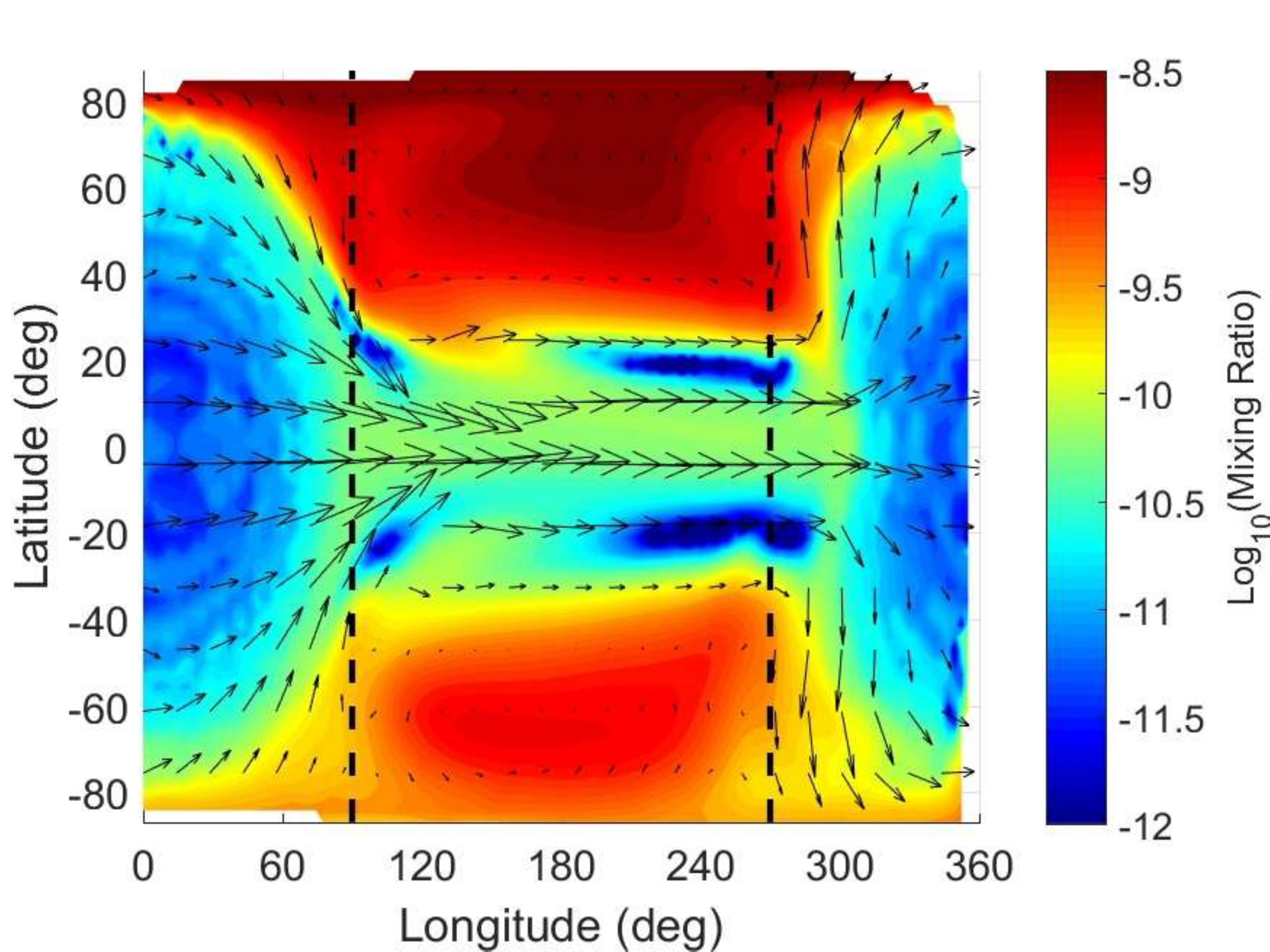}} &
{\includegraphics[width=4.2cm]{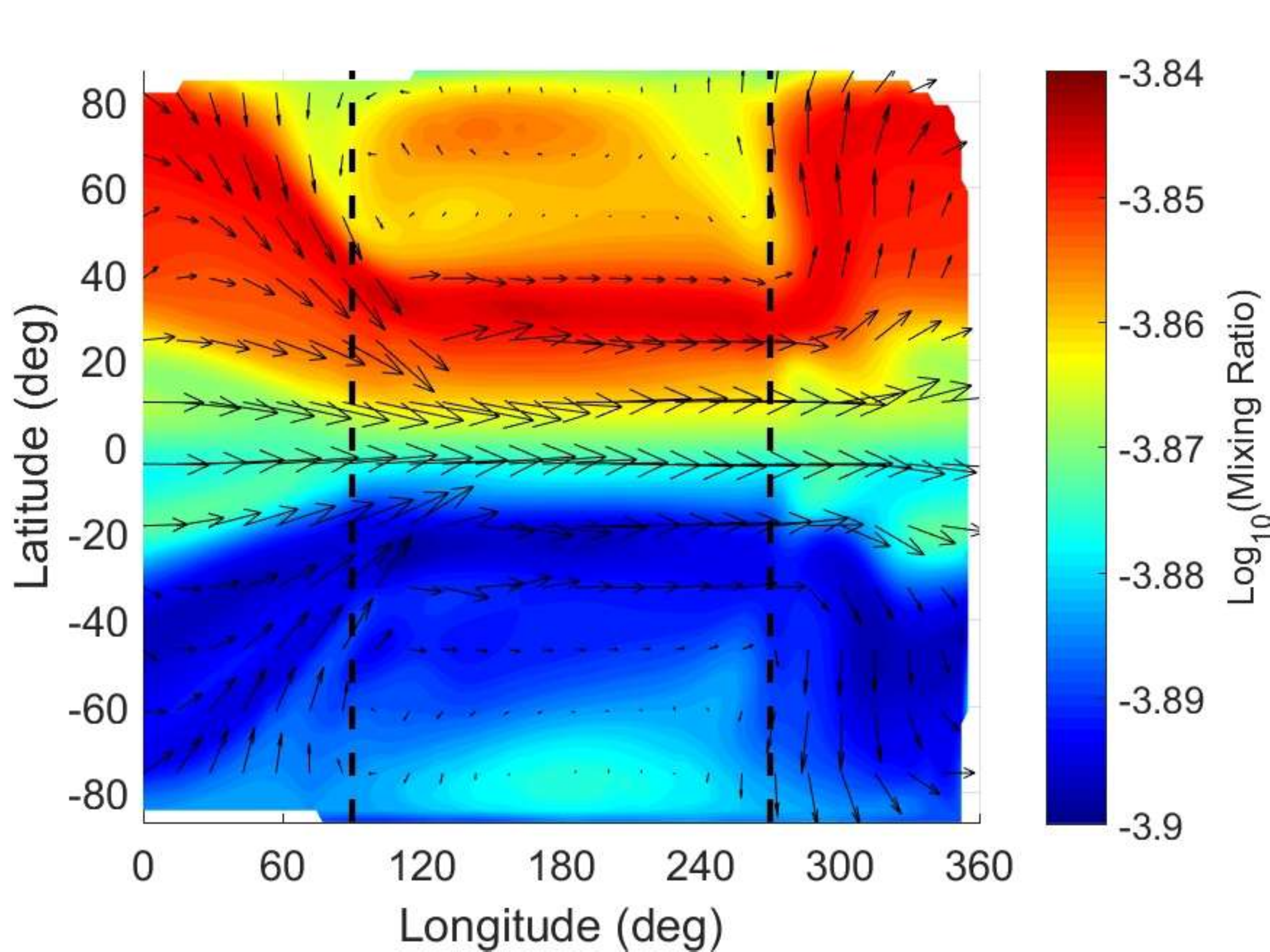}} & 
{\includegraphics[width=4.2cm]{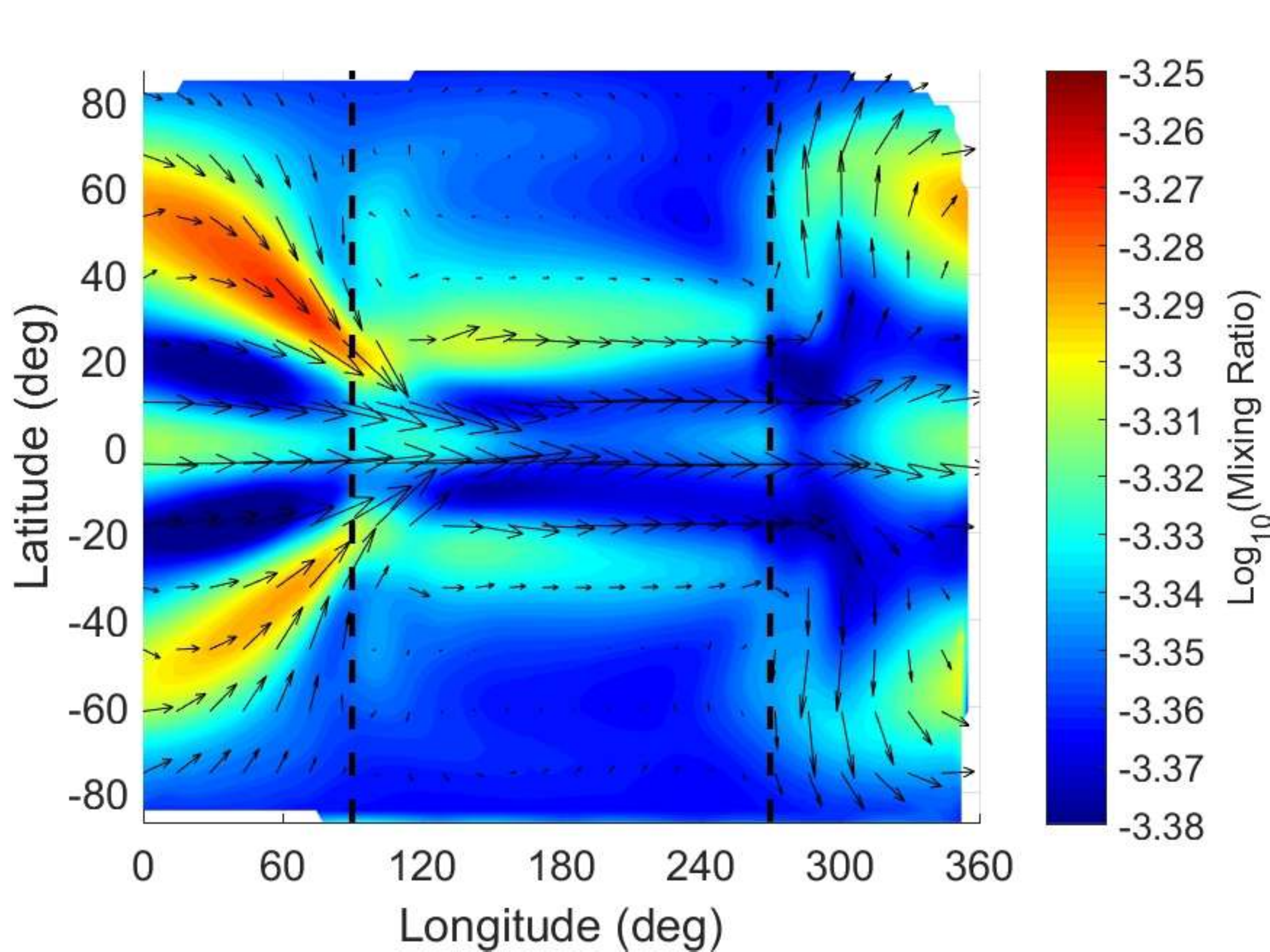}} \\ 
    \vspace{-3cm}\rotatebox{90}{CO} & {\includegraphics[width=4.2cm]{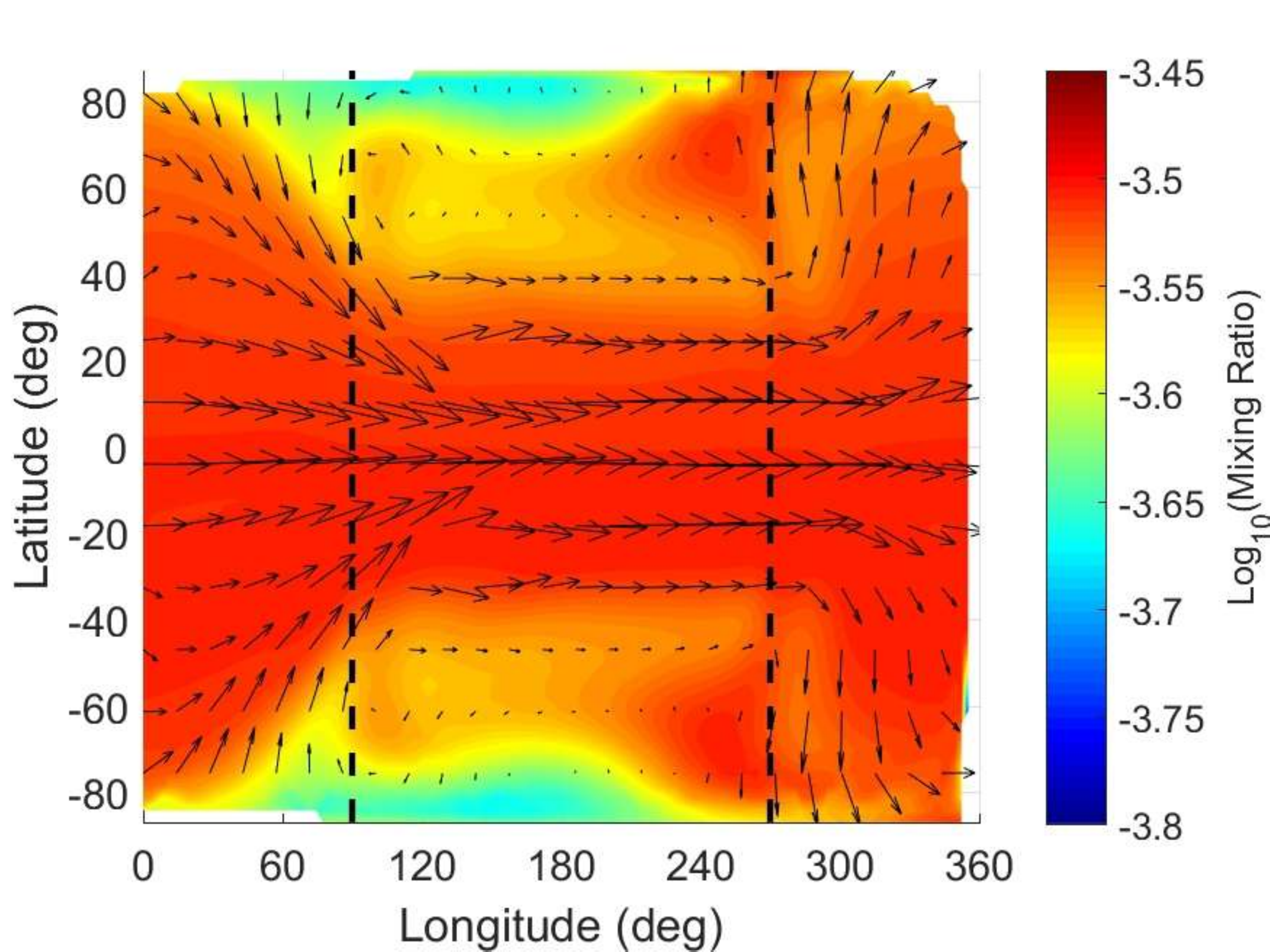}} & {\includegraphics[width=4.2cm]{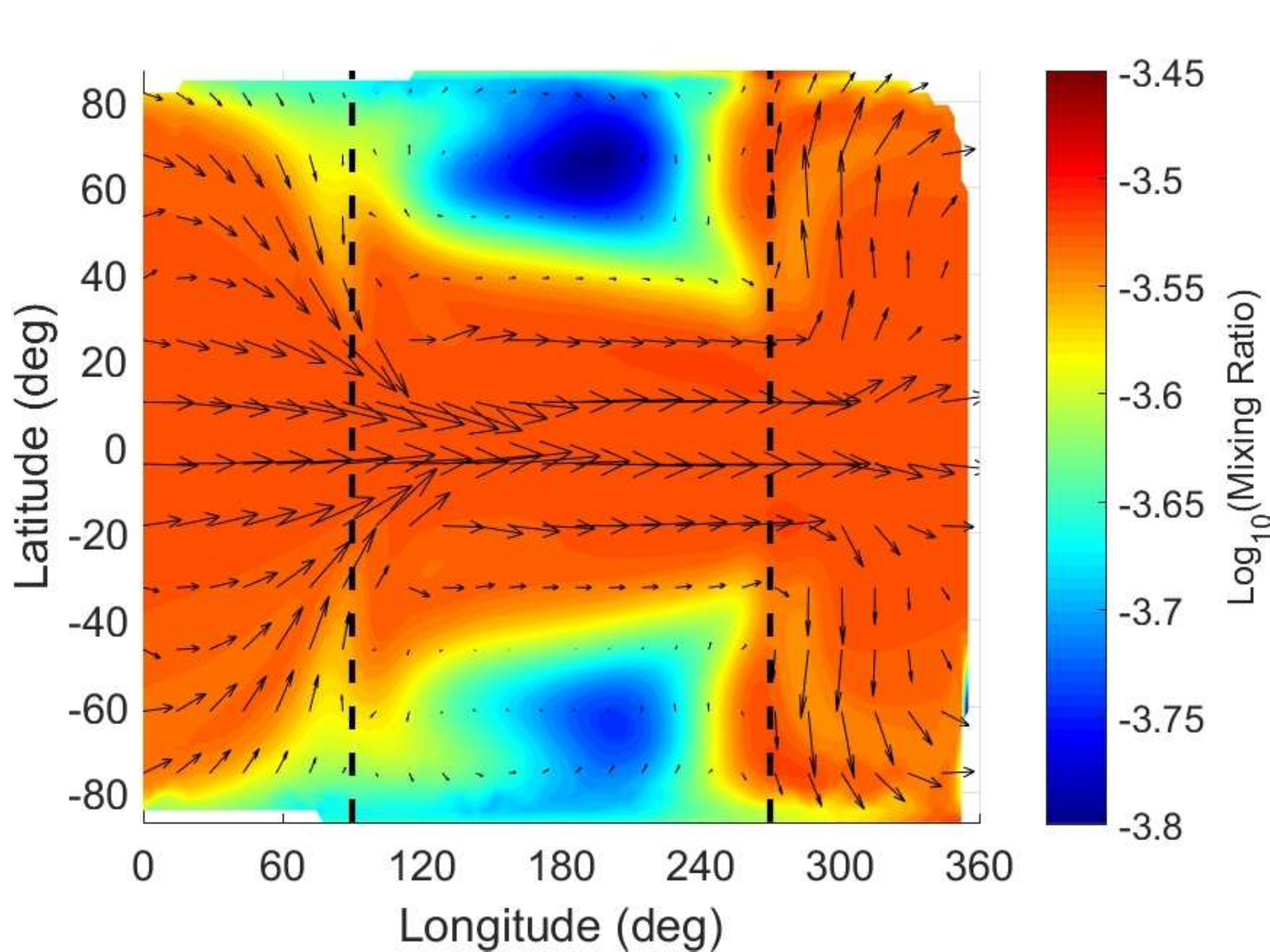}} &
{\includegraphics[width=4.2cm]{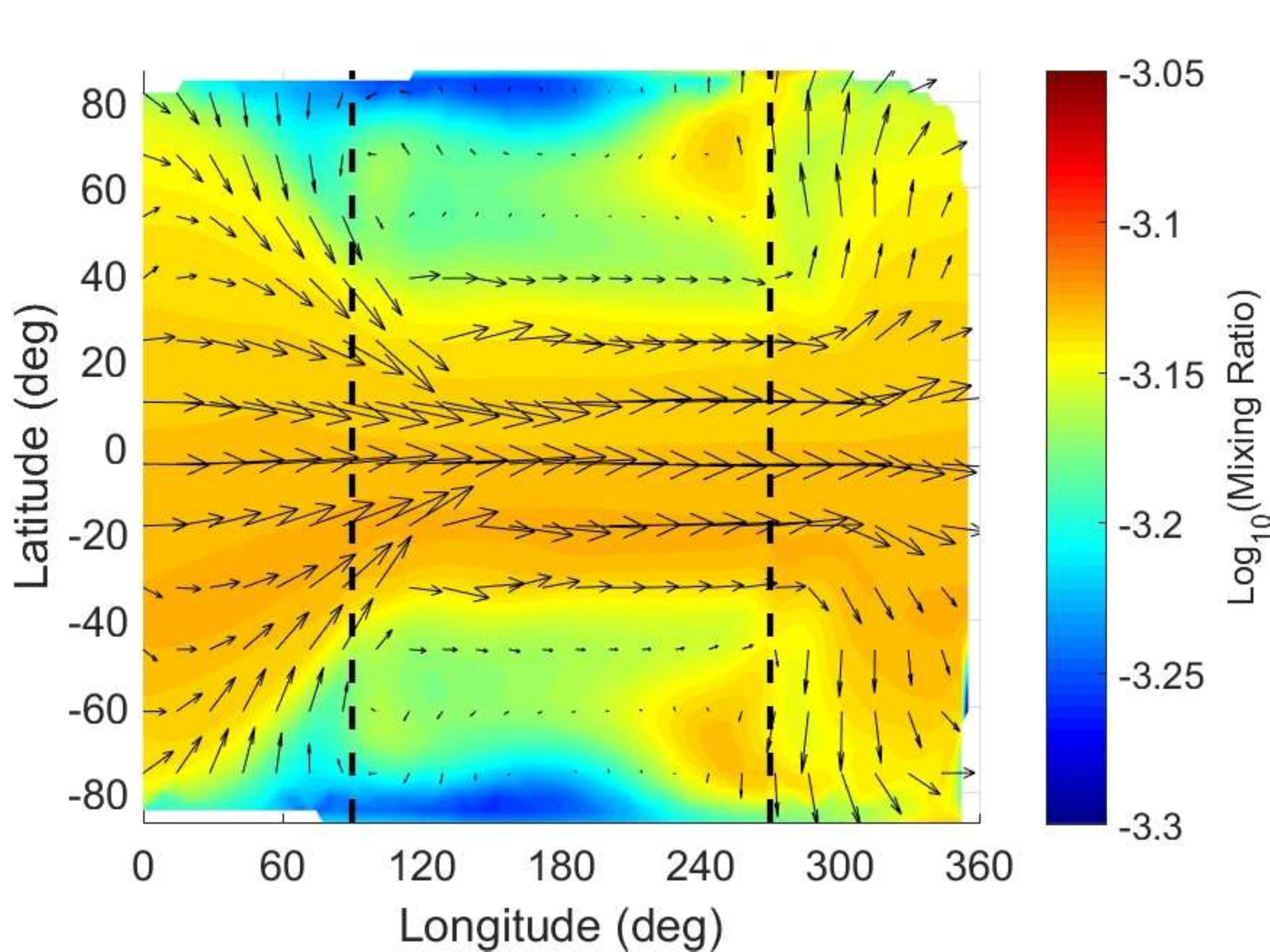}} & 
{\includegraphics[width=4.2cm]{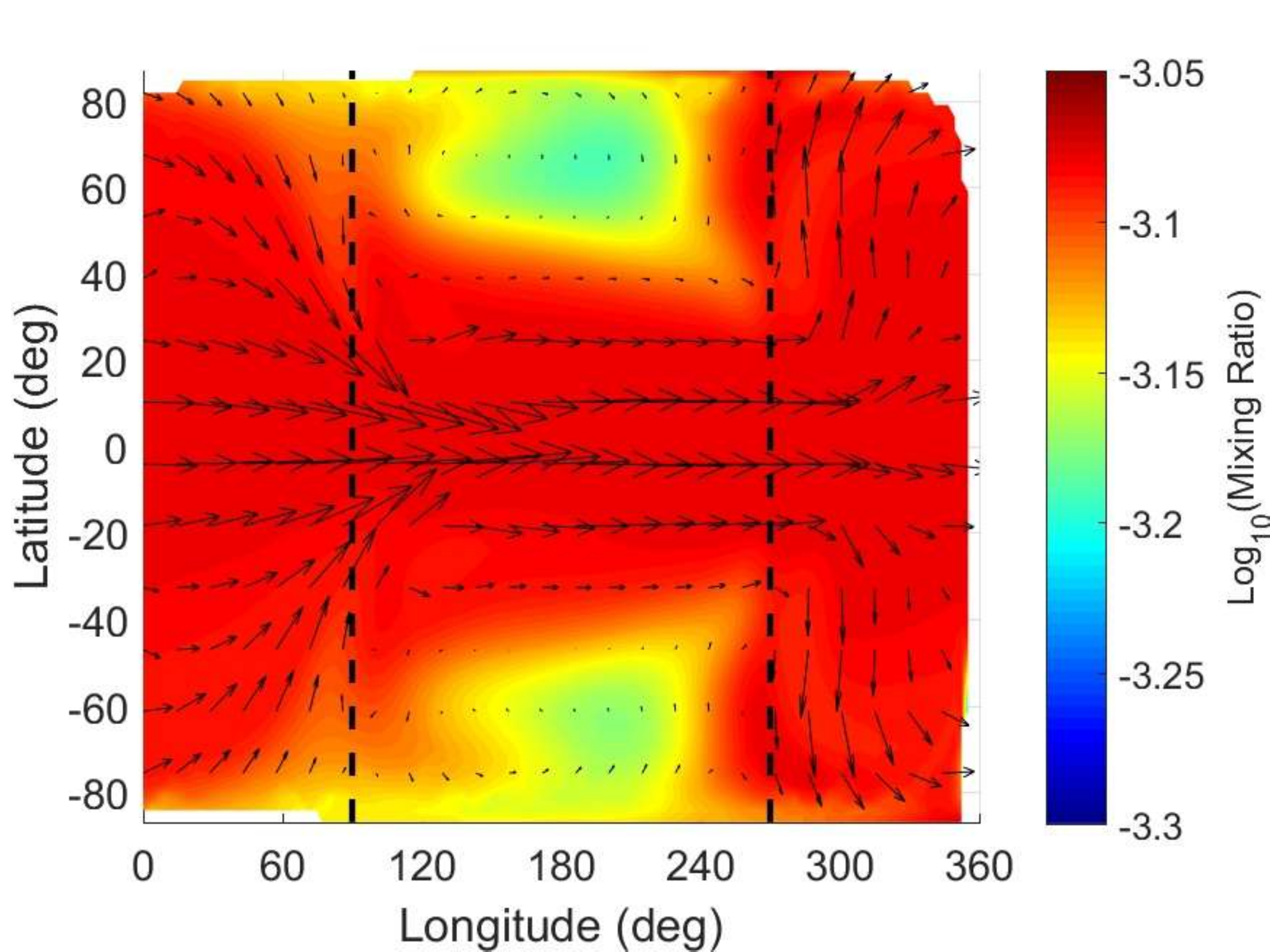}} \\ 
    \vspace{-3cm}\rotatebox{90}{H$_2$O} & {\includegraphics[width=4.2cm]{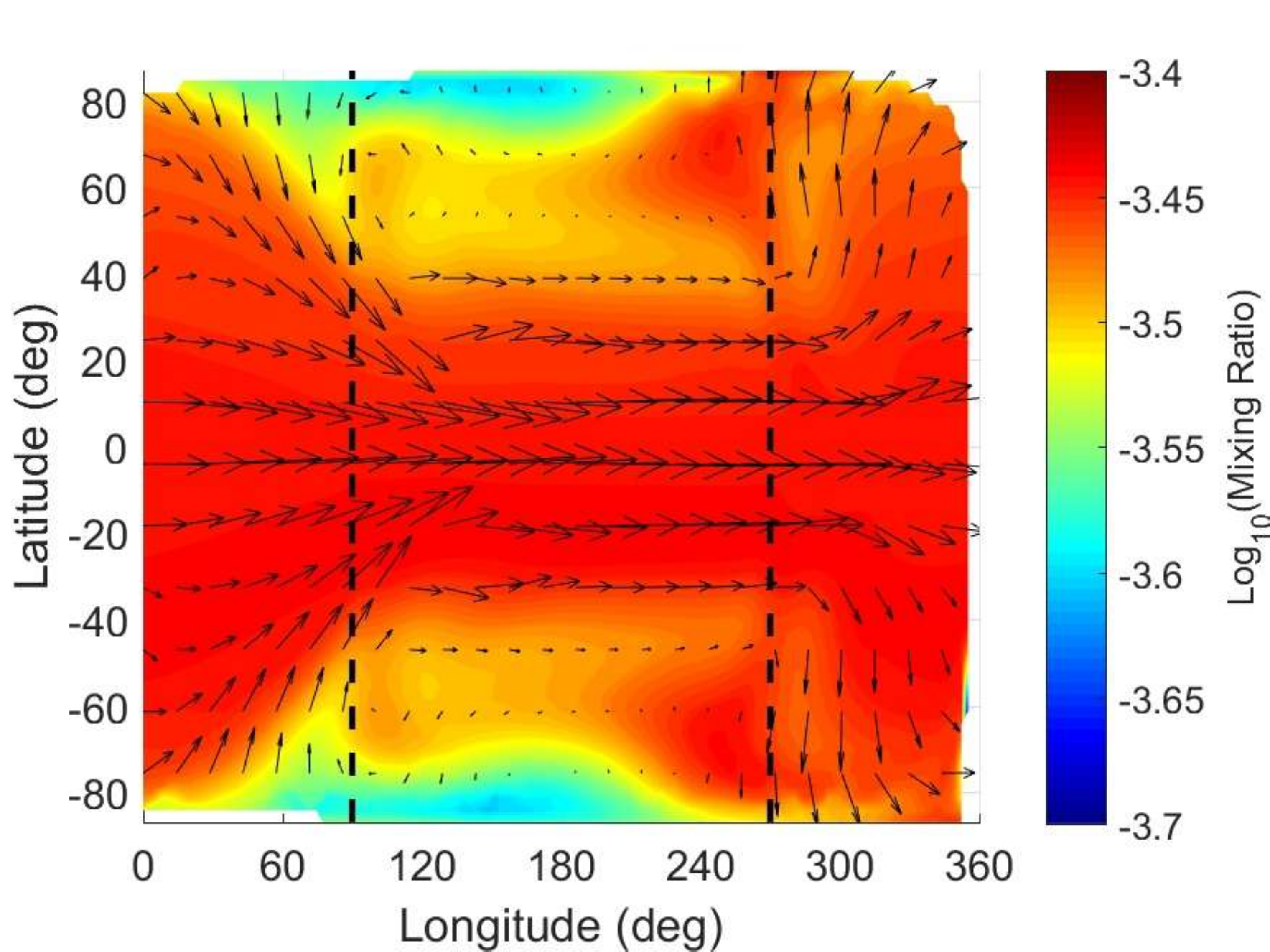}} & {\includegraphics[width=4.2cm]{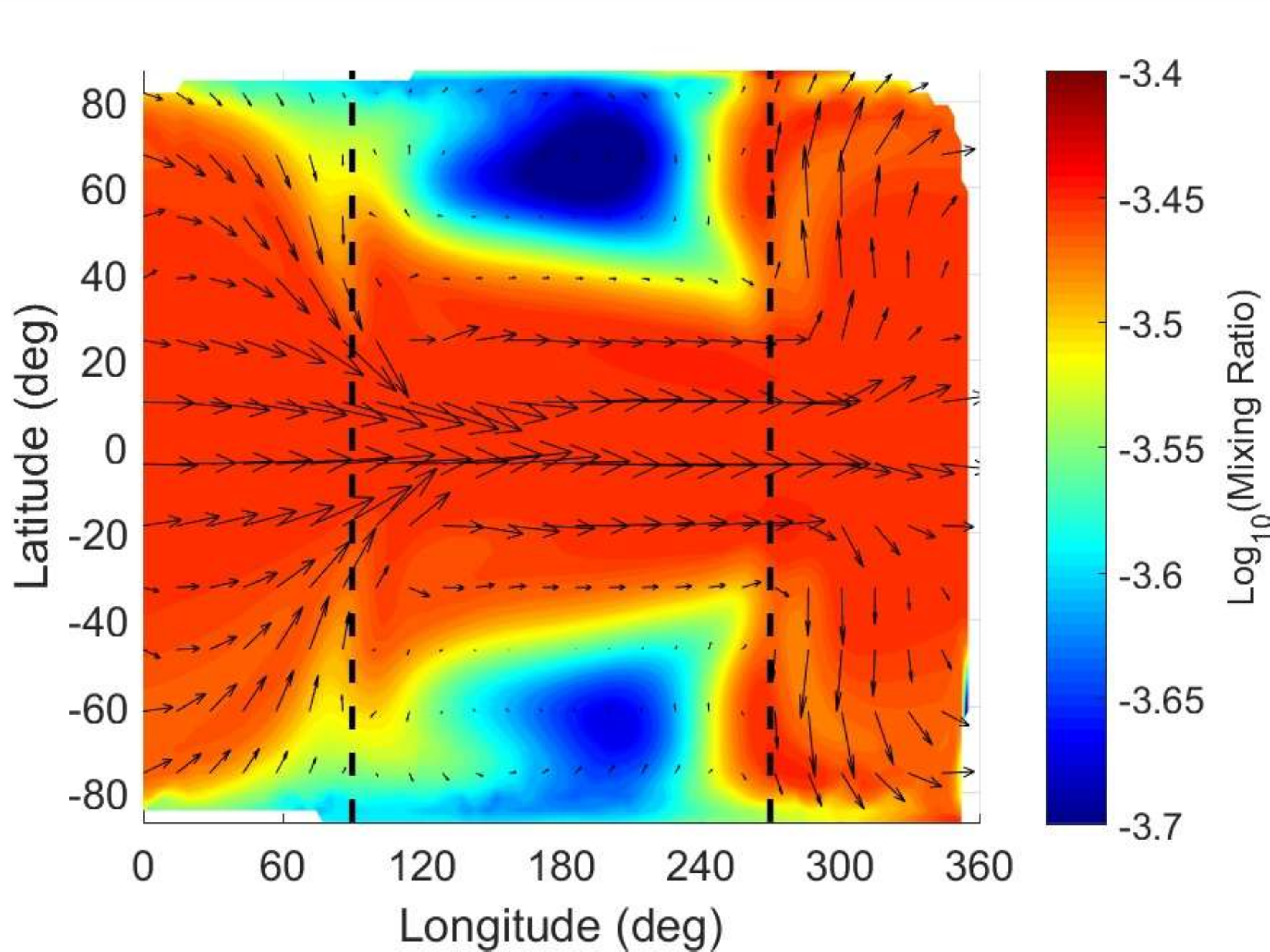}} &
{\includegraphics[width=4.2cm]{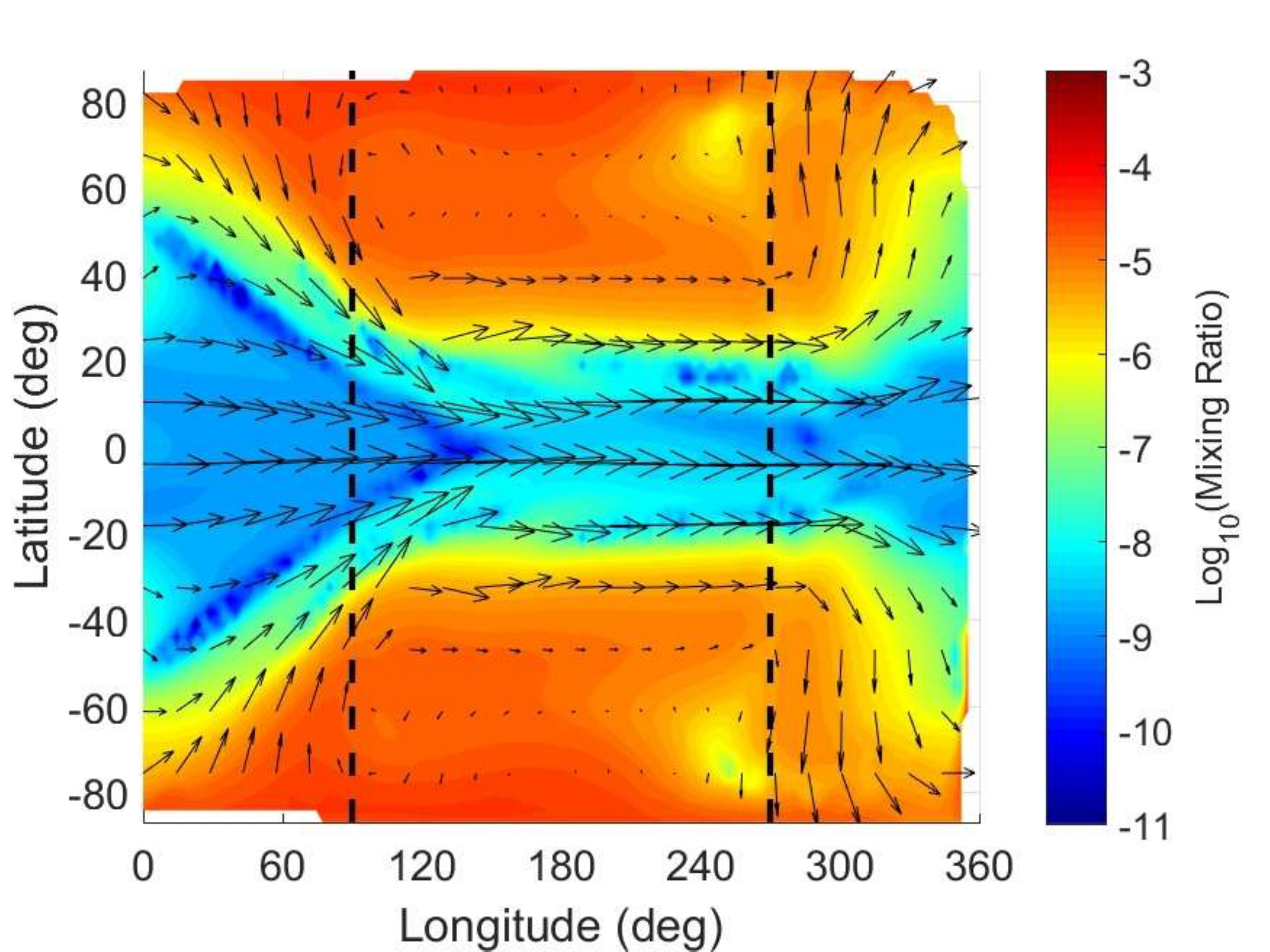}} & 
{\includegraphics[width=4.2cm]{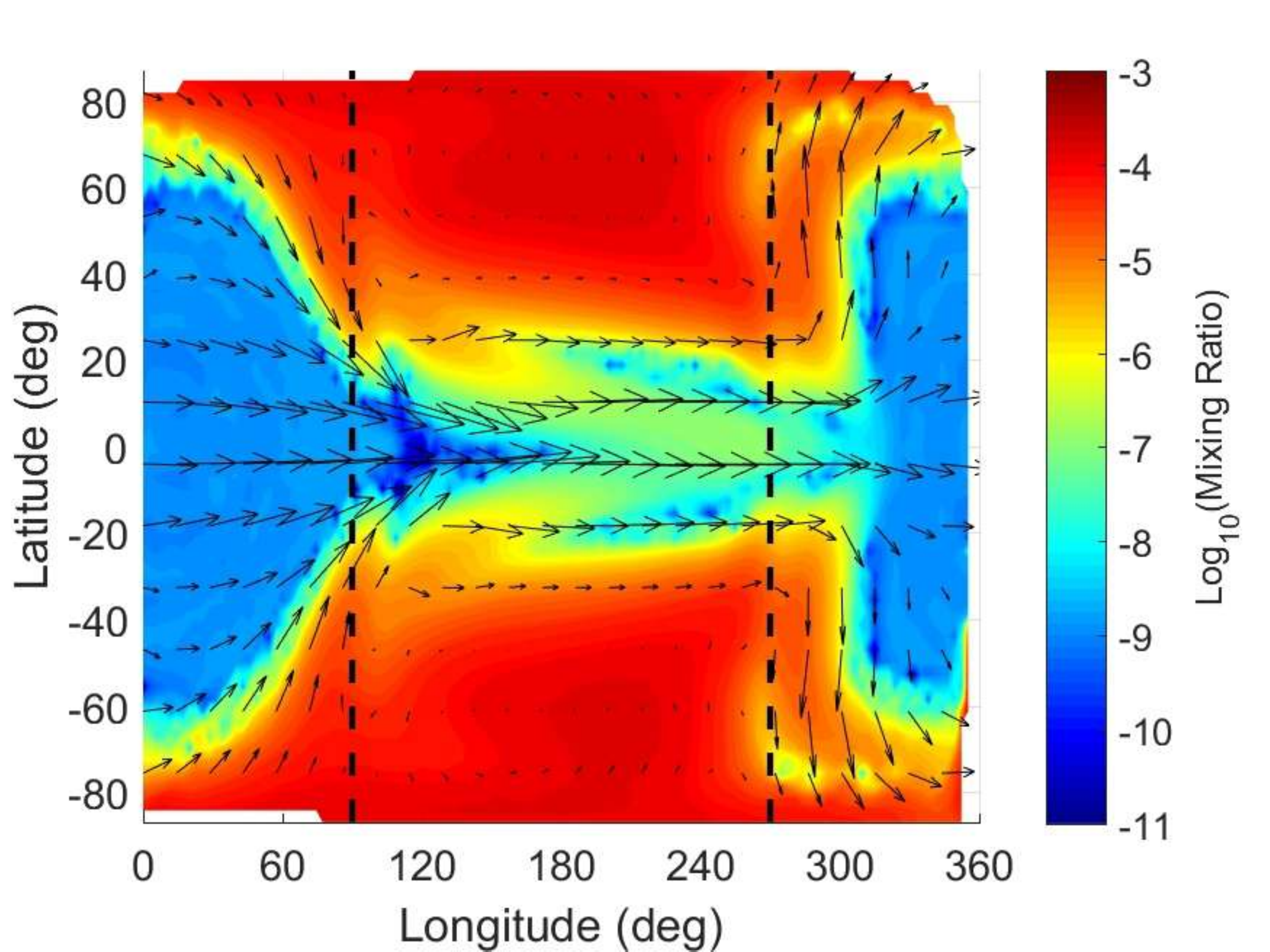}} \\
    \vspace{-3cm}\rotatebox{90}{CO$_2$} & {\includegraphics[width=4.2cm]{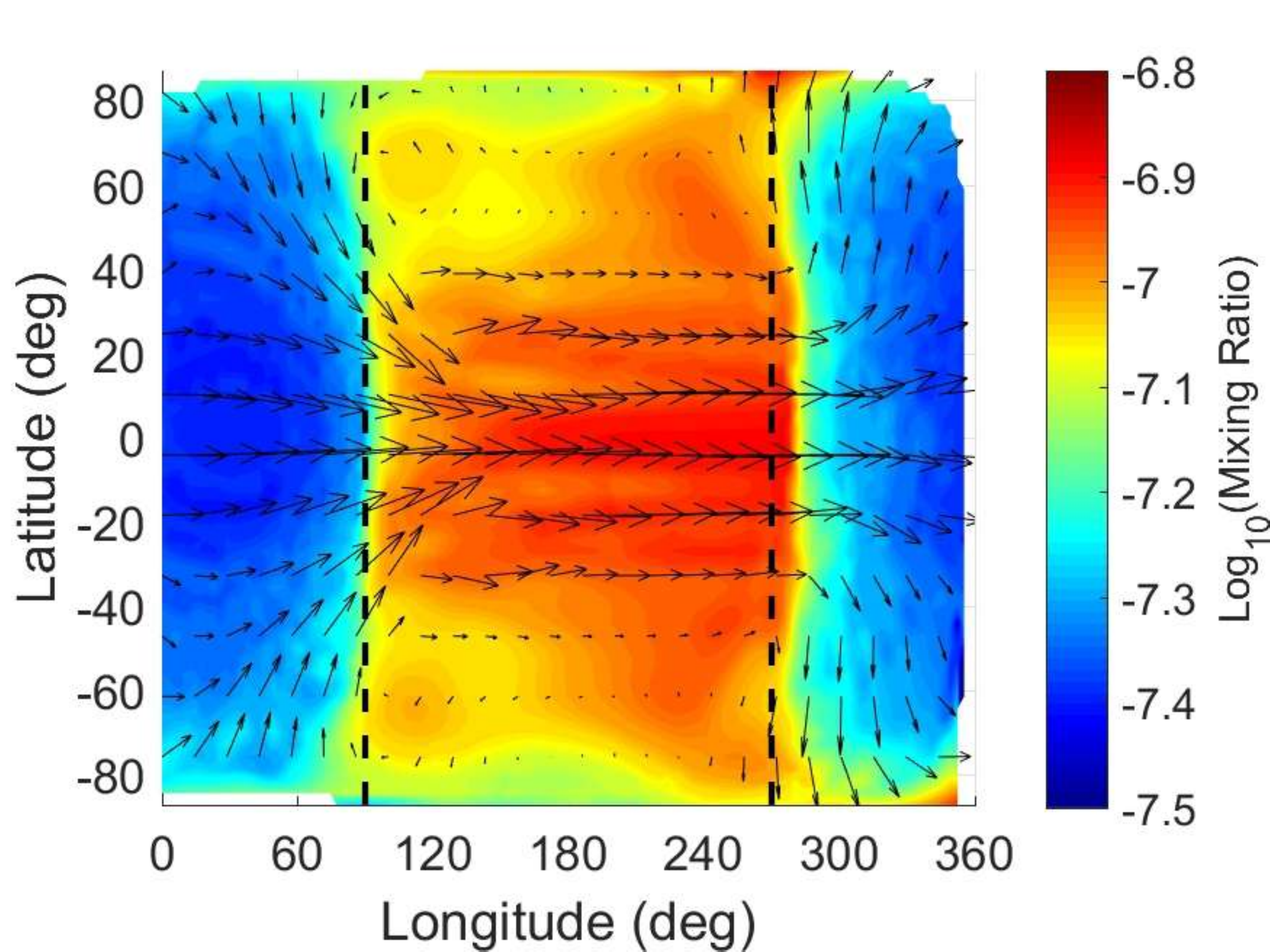}} & {\includegraphics[width=4.2cm]{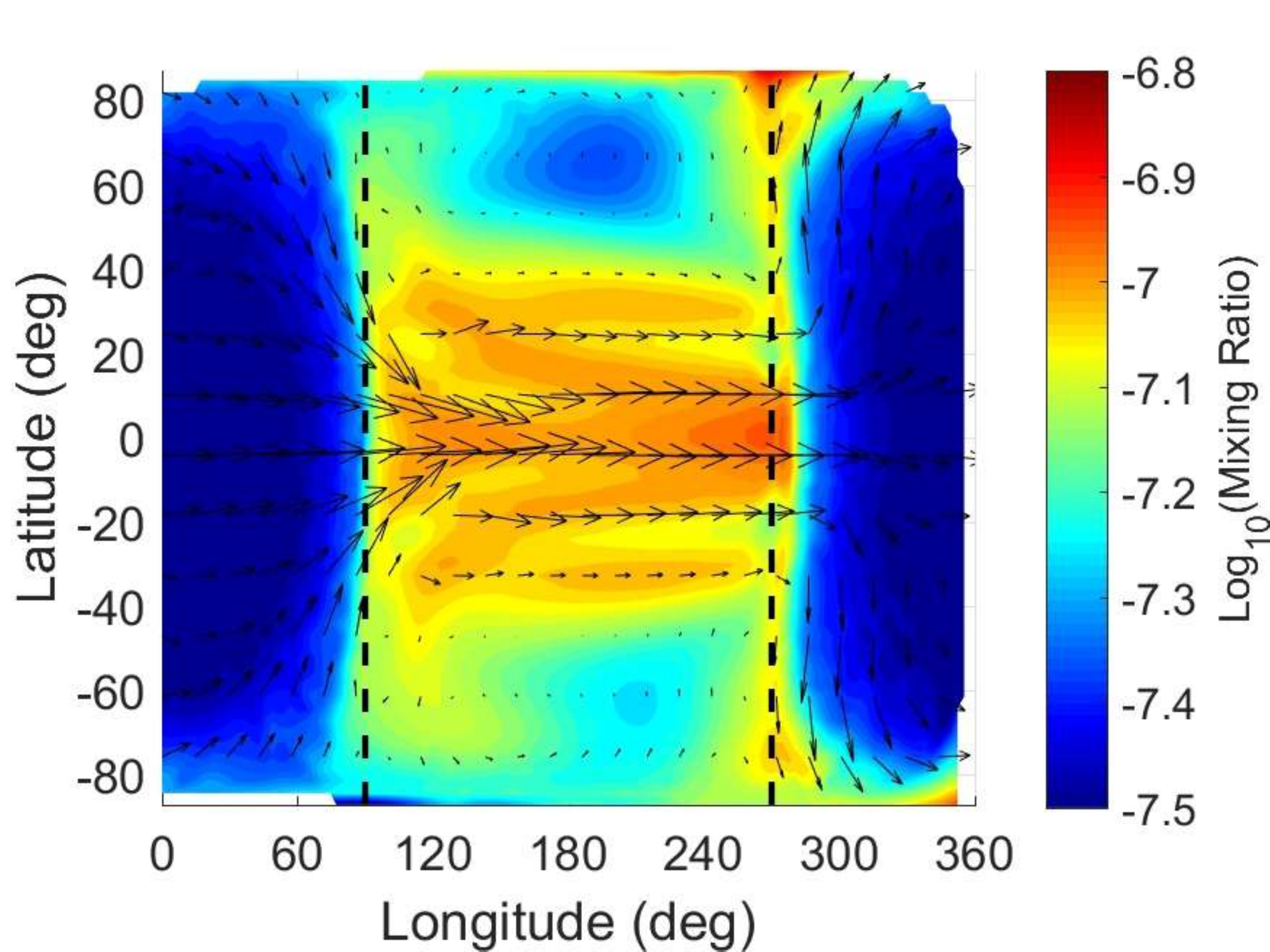}} &
{\includegraphics[width=4.2cm]{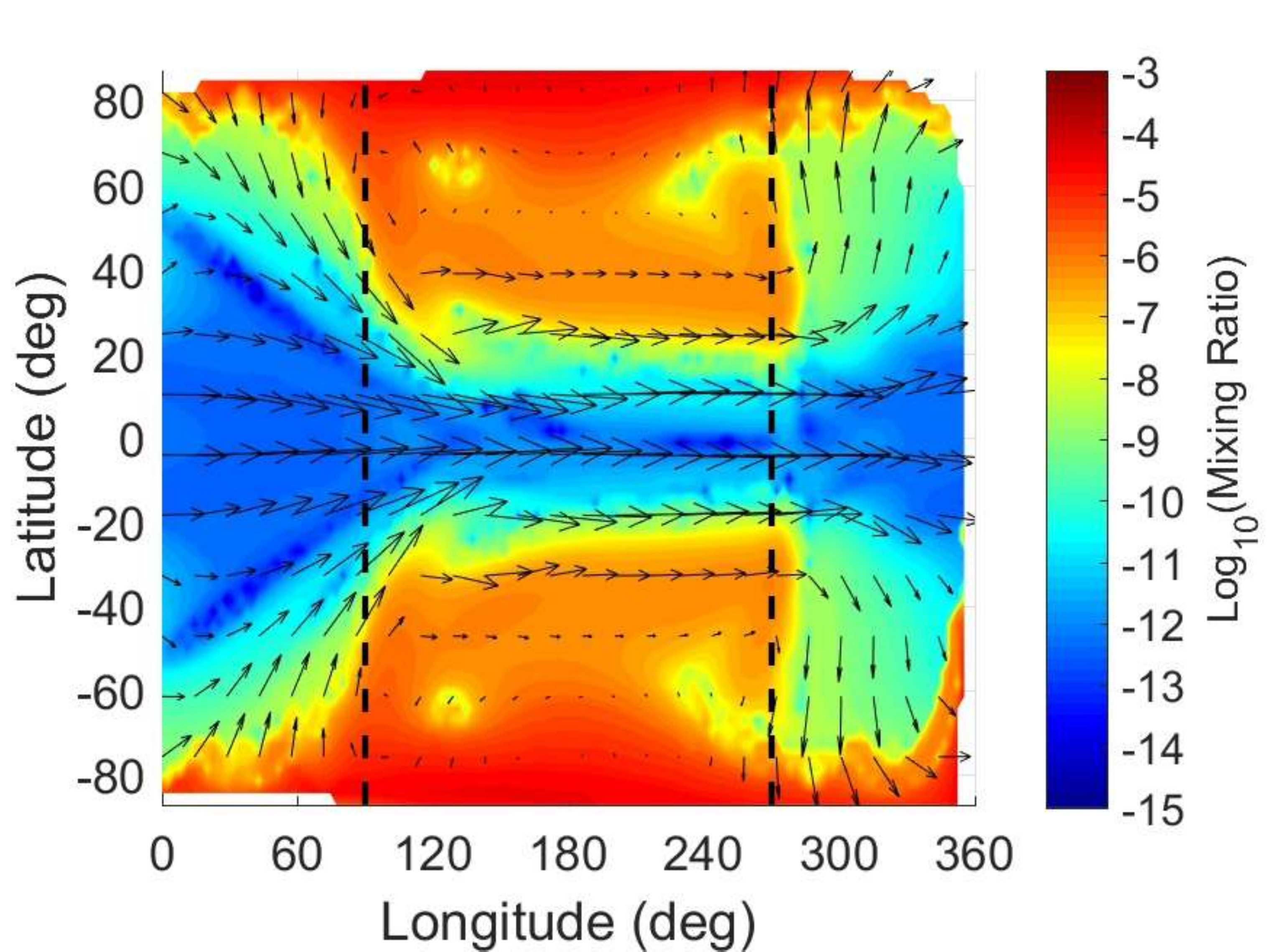}} & 
{\includegraphics[width=4.2cm]{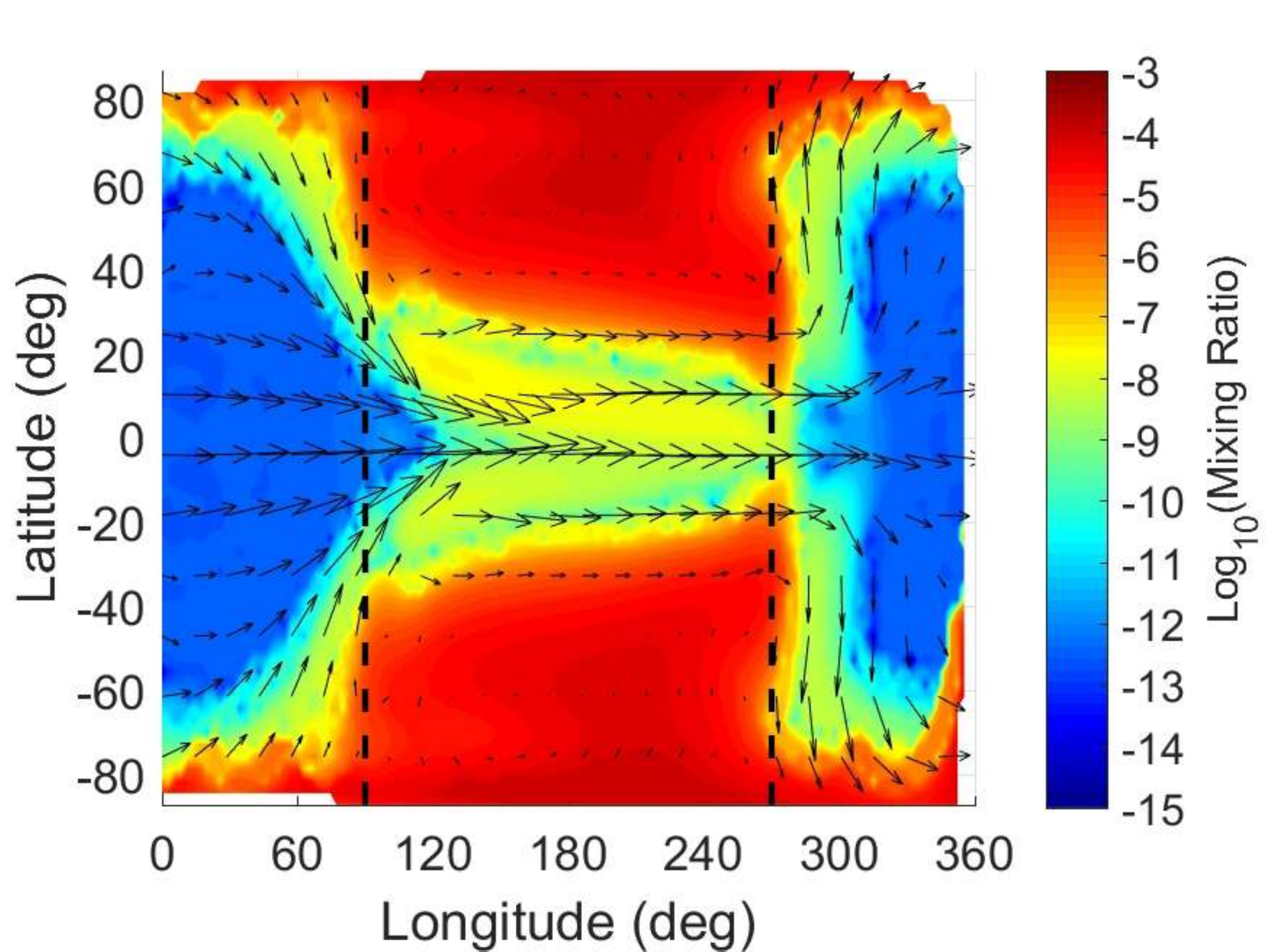}} \\      
  \end{tabular}
  \caption{Distribution of CH$_4$, CO, H$_2$O and CO$_2$ at 10 mbar as a function of latitude and longitude from the simulations with different $G$ and $C/O$. The colors represent the volume mixing ratio of the different chemical species. All the results shown in this figure were averaged over the last 100 days of the simulations.}
\label{fig:che_ref_h} 
\end{figure*}

In Fig. \ref{fig:che_ref_h}, we show the chemical horizontal distribution at 10 mbar, which is roughly the pressure level where the jet reaches maximum strength and is also the photosphere in the nightside of the planet. In general, at low latitudes the broad jet is very efficient mixing the chemical components along the longitude in the atmosphere, except for CO$_2$ in 3 experiments ($G = 0.5$ and $C/O = 0.5$; $G = 2$ and $C/O = 0.5$; $G = 2$ and $C/O = 2$). In the nightside, the different maps show a well mixed atmosphere at low latitudes but each polar vortex in each hemisphere tend to a have distinct chemical distribution. This feature is associated with the typical low temperatures obtained in this region and the low dynamical mixing. However, a very long chemical timescale can lead to the chemical species being well mixed over all latitudes and longitudes as is, for example, CH$_4$ for $C/O$ = 2. It is not the goal of this paper to explore in detail the permeability of the polar vortex area, but this region could be surrounded by a steep Ertel Potential Vorticity (PV) gradient, which can behave as a ``transport barrier'' for tracers (\citealt{1985Hoskins}). In these areas of strong PV gradient, the Rossby waves restoring mechanism suppresses Rossby wave breaking, which results in parcels taking the form of reversible wavelike motions (\citealt{1995Mcintyre}). However, the permeability of this transport barrier is reduced by the erosion of atmospheric wave breaking at all scales and due to vertical transport. A more detailed exploration of the transport into the polar vortex could be  carried out in the future by computing the effective diffusivity on an isentropic level of an atmospheric tracer map (e.g. \citealt{2000Haynes1}; \citealt{2000Haynes2}). 

For C/O equals 2 and $G = 0.5$, the latitude-longitude map shows an antisymmetric distribution of CH$_4$. The antisymmetric distribution is associated to: the quasi-uniform distribution of CH$_4$, the initial conditions which allowed more CH$_4$ to be contained in one hemispheres than in the other and the low permeability crossing the equatorial region due to the presence of the strong equatorial jet.

\subsection{Timescales}
\label{subsec_timescales}

\begin{figure*}
  \begin{tabular}{m{0.1cm} c c c c}
     & G = 0.5 $\&$ C/O = 0.5 & G = 2.0 $\&$ C/O = 0.5 & G = 0.5 $\&$ C/O = 2 & G = 2 $\&$ C/O = 2 \\
    \vspace{-3cm}\rotatebox{90}{Dayside} & {\includegraphics[width=4.2cm]{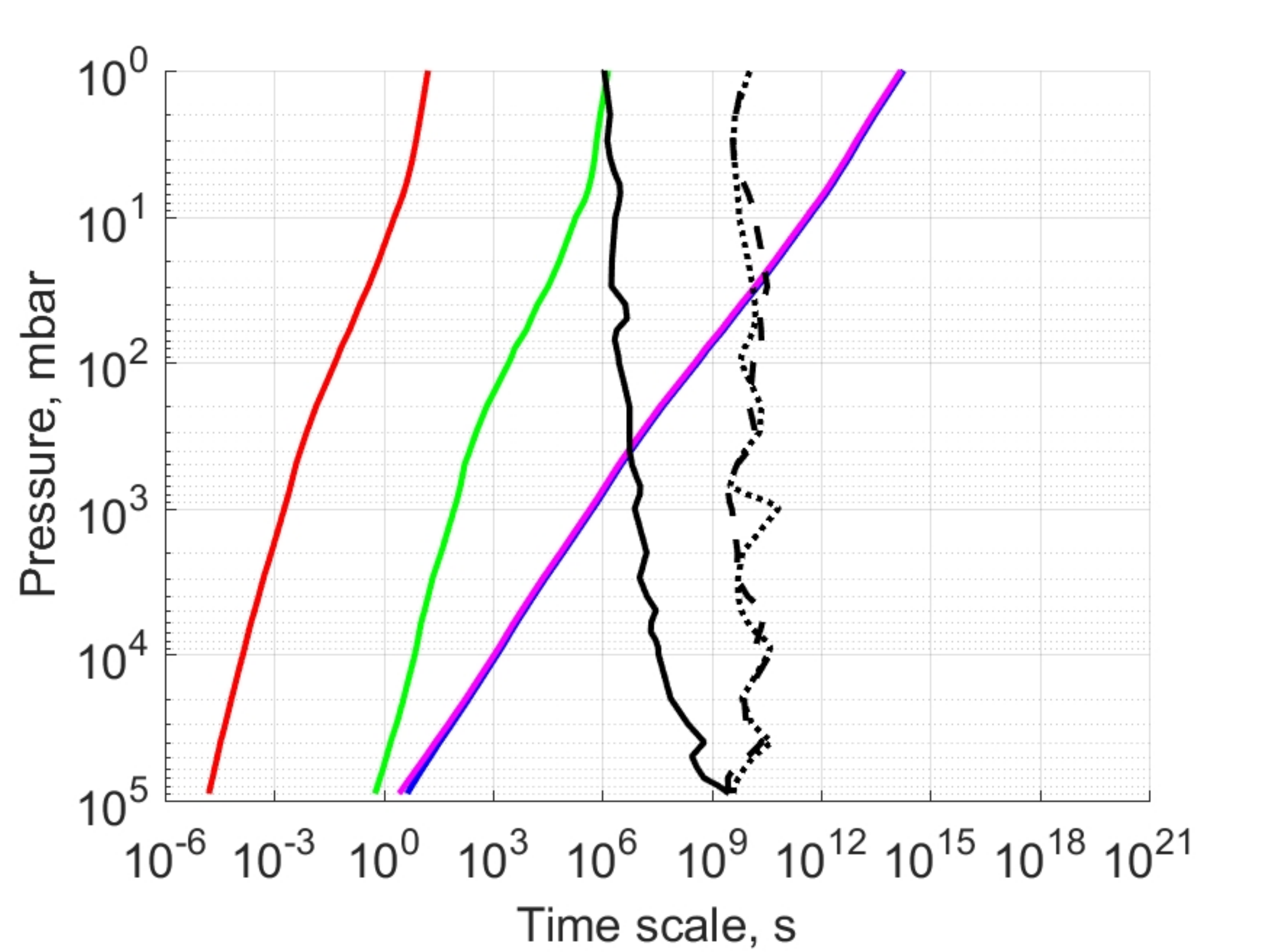}} & {\includegraphics[width=4.2cm]{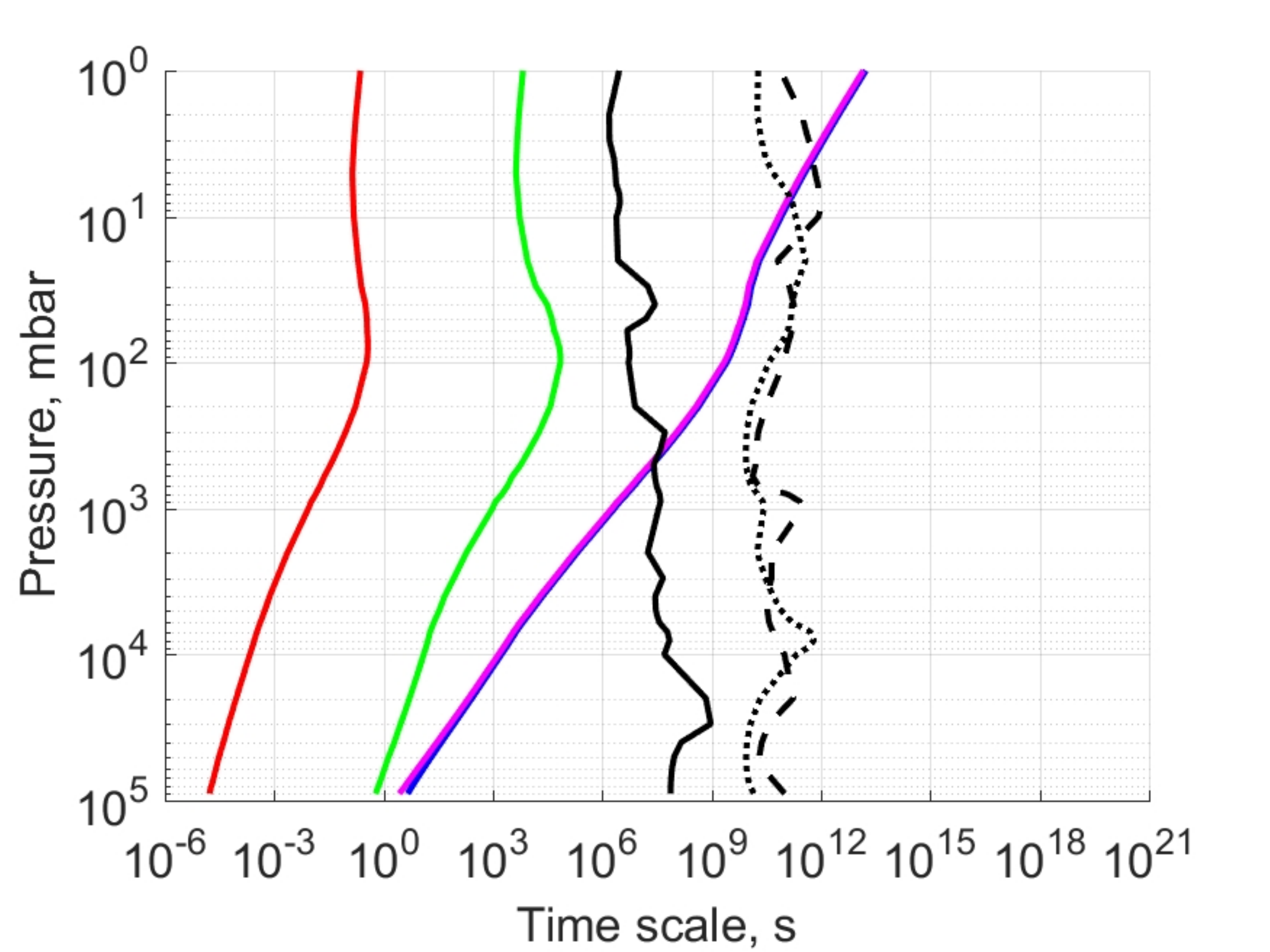}} &
{\includegraphics[width=4.2cm]{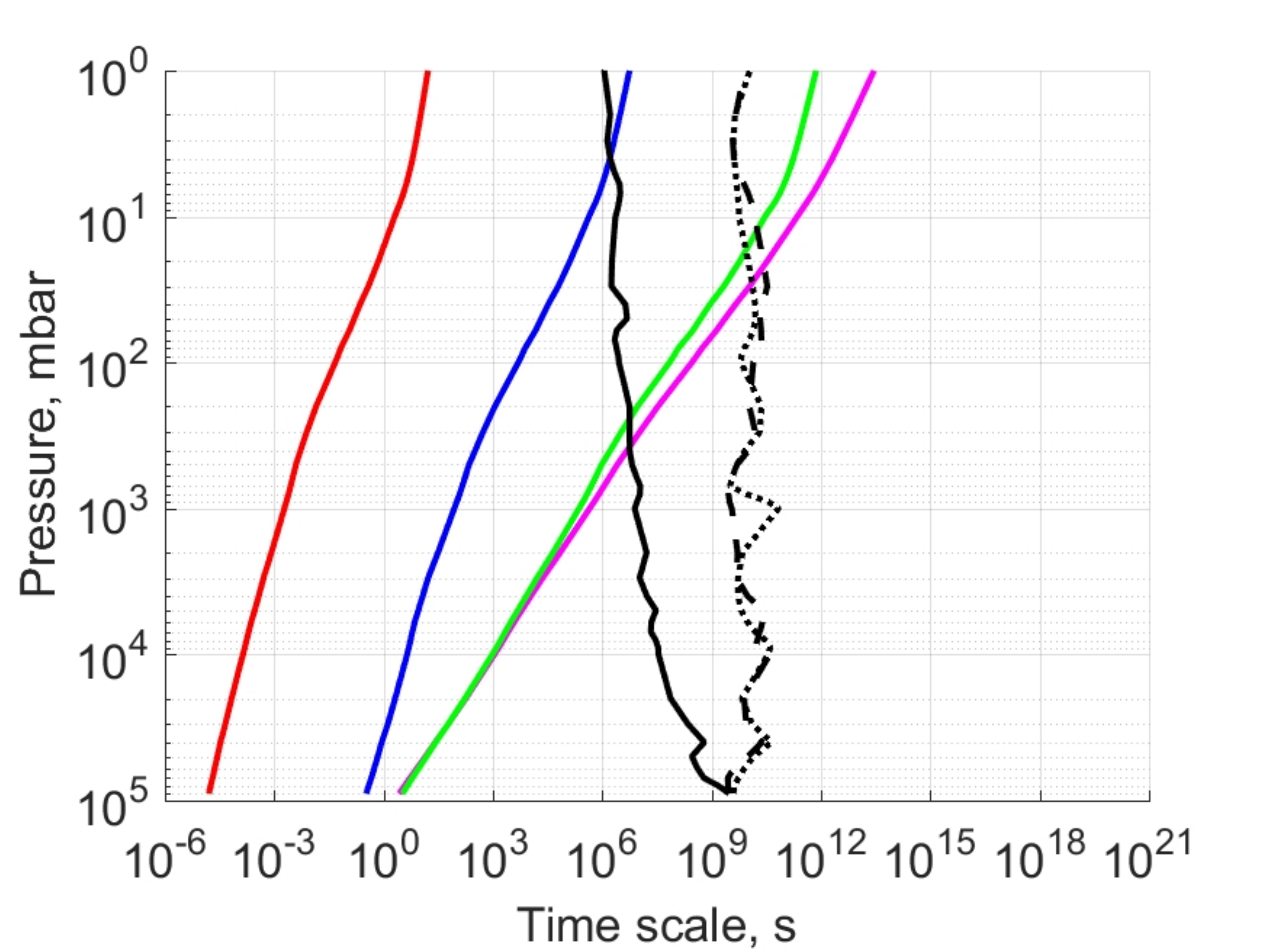}} & 
{\includegraphics[width=4.2cm]{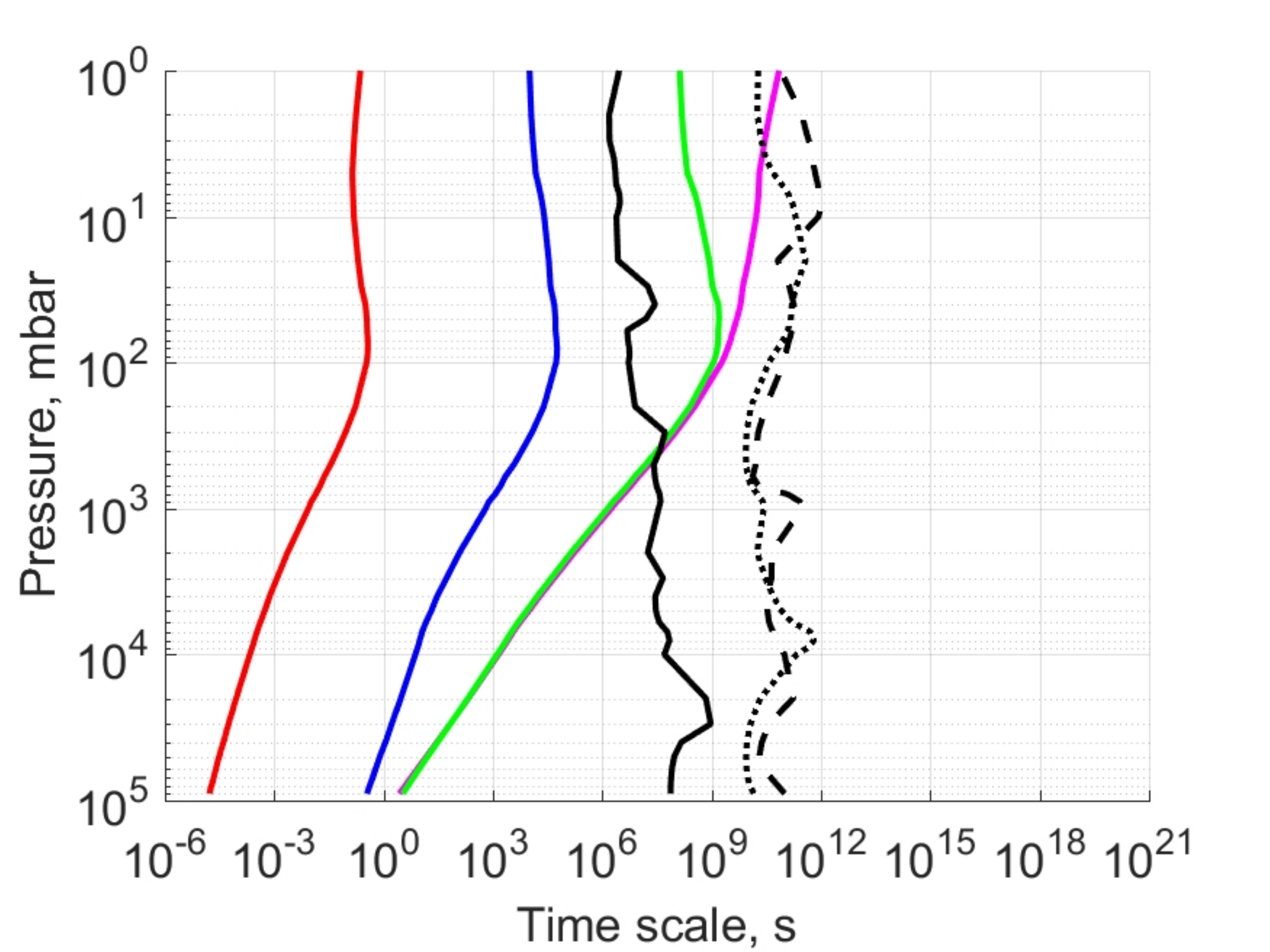}} \\ 
    \vspace{-3cm}\rotatebox{90}{Nightside} & {\includegraphics[width=4.2cm]{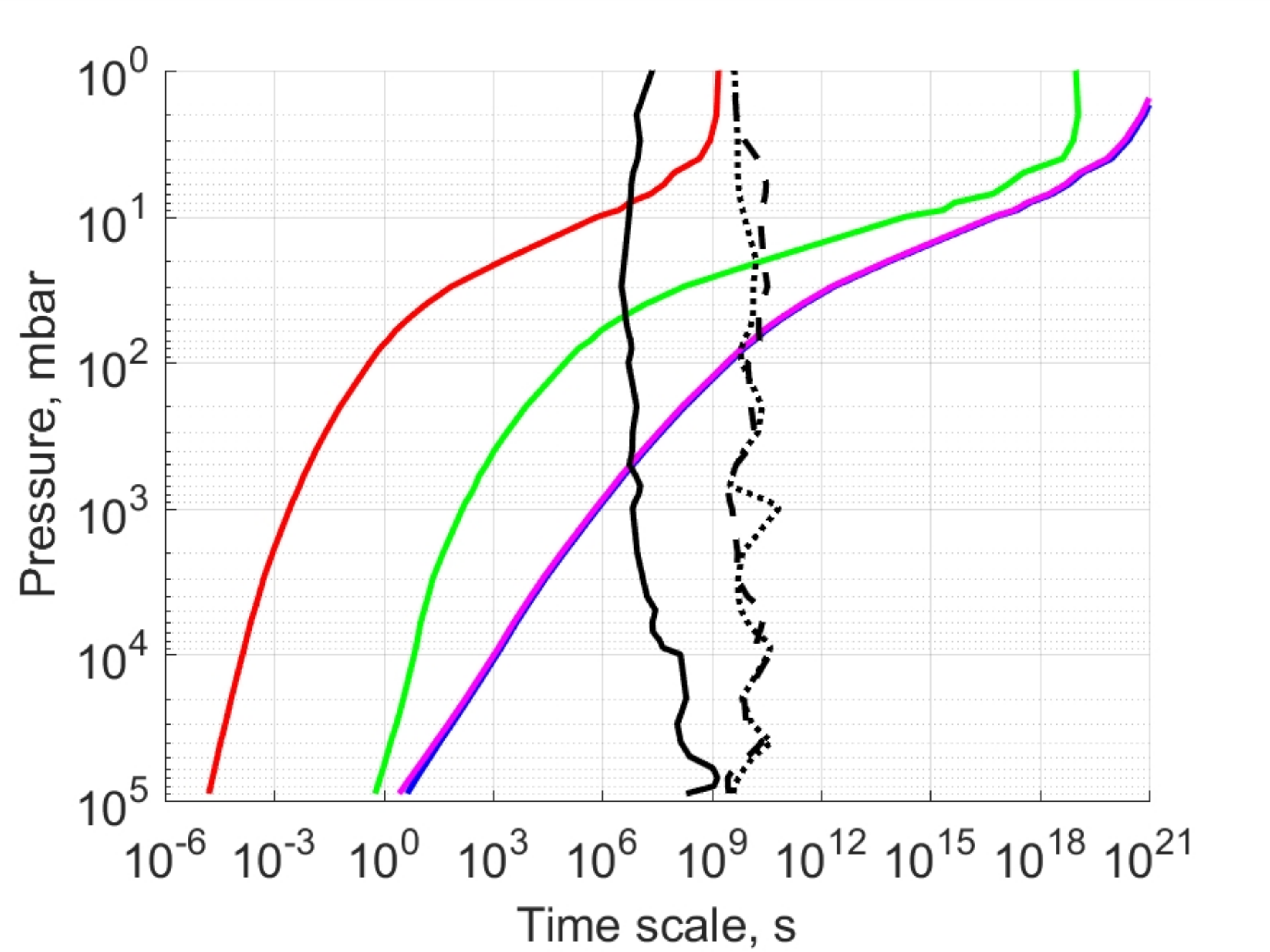}} & {\includegraphics[width=4.2cm]{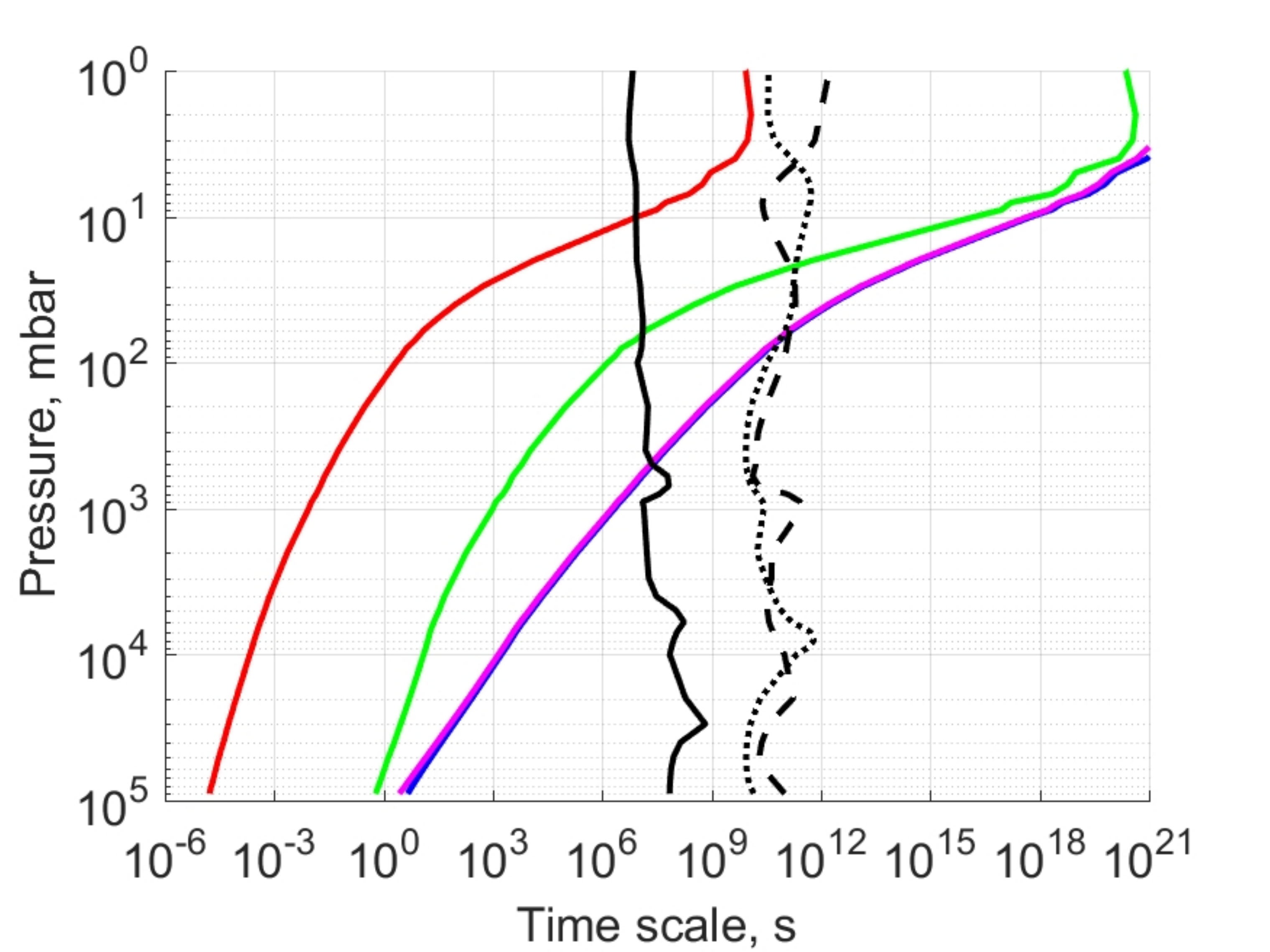}} &
{\includegraphics[width=4.2cm]{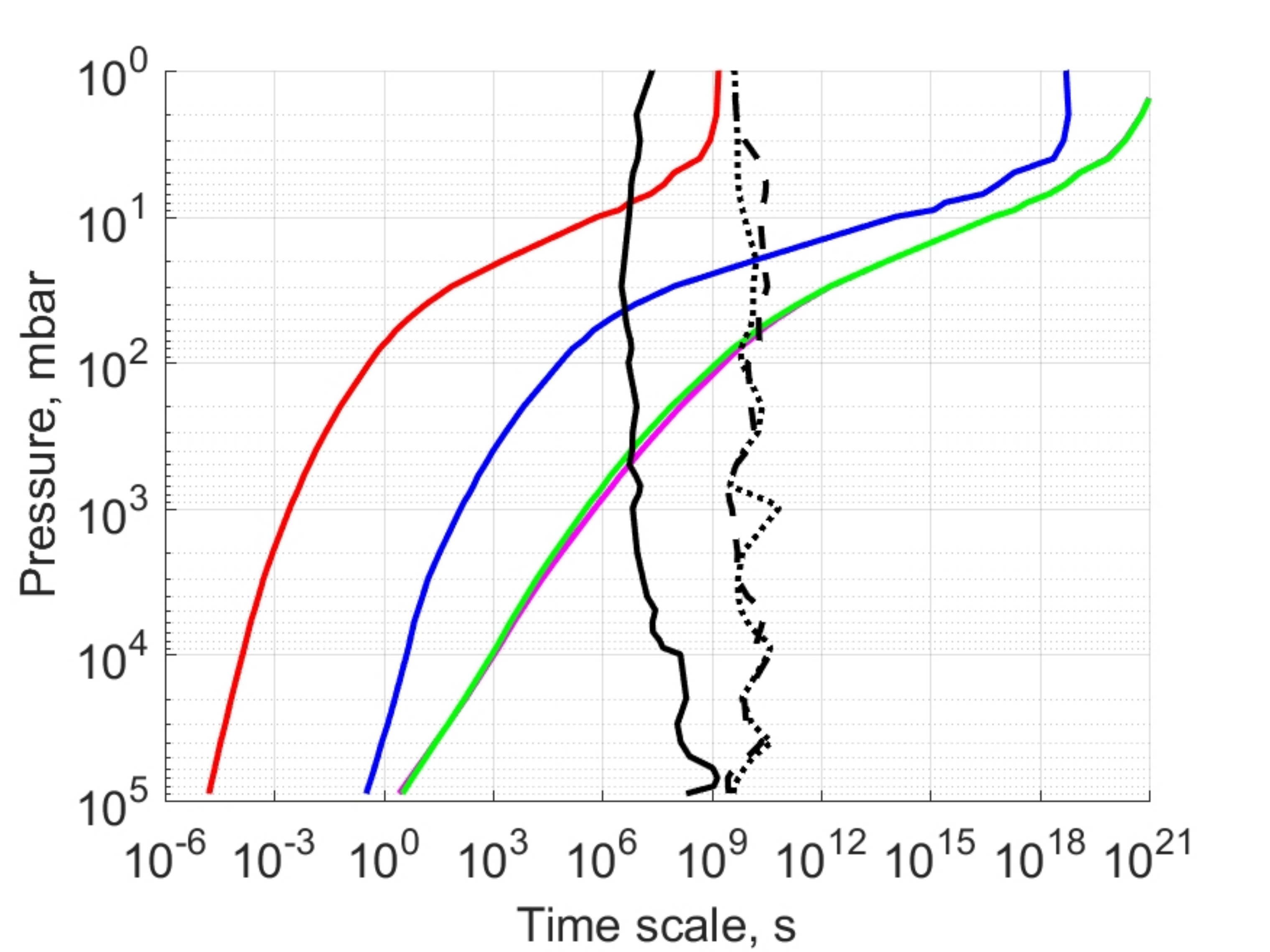}} & 
{\includegraphics[width=4.2cm]{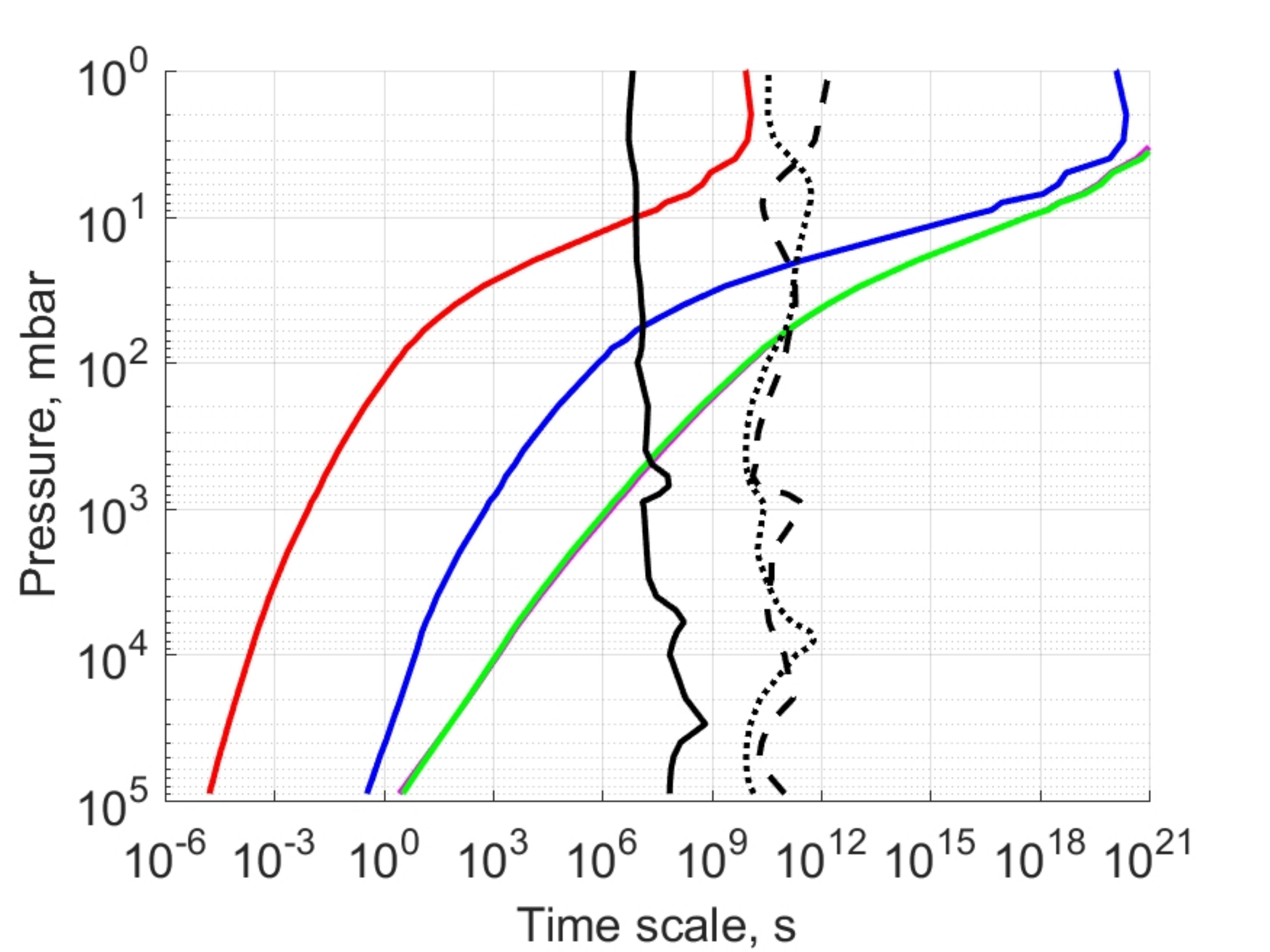}} \\ 
    \vspace{-3cm}\rotatebox{90}{Morning terminator} & {\includegraphics[width=4.2cm]{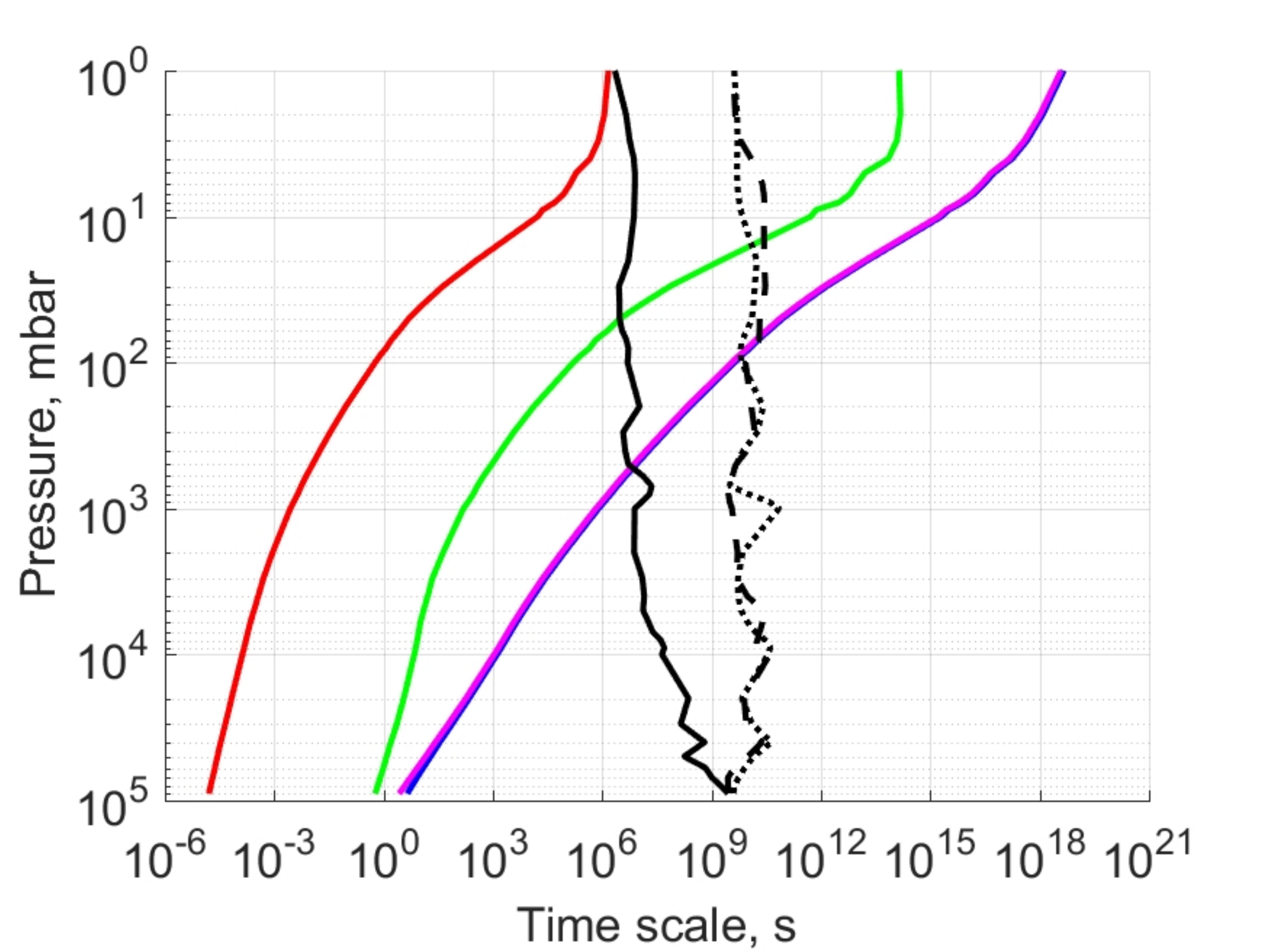}} & {\includegraphics[width=4.2cm]{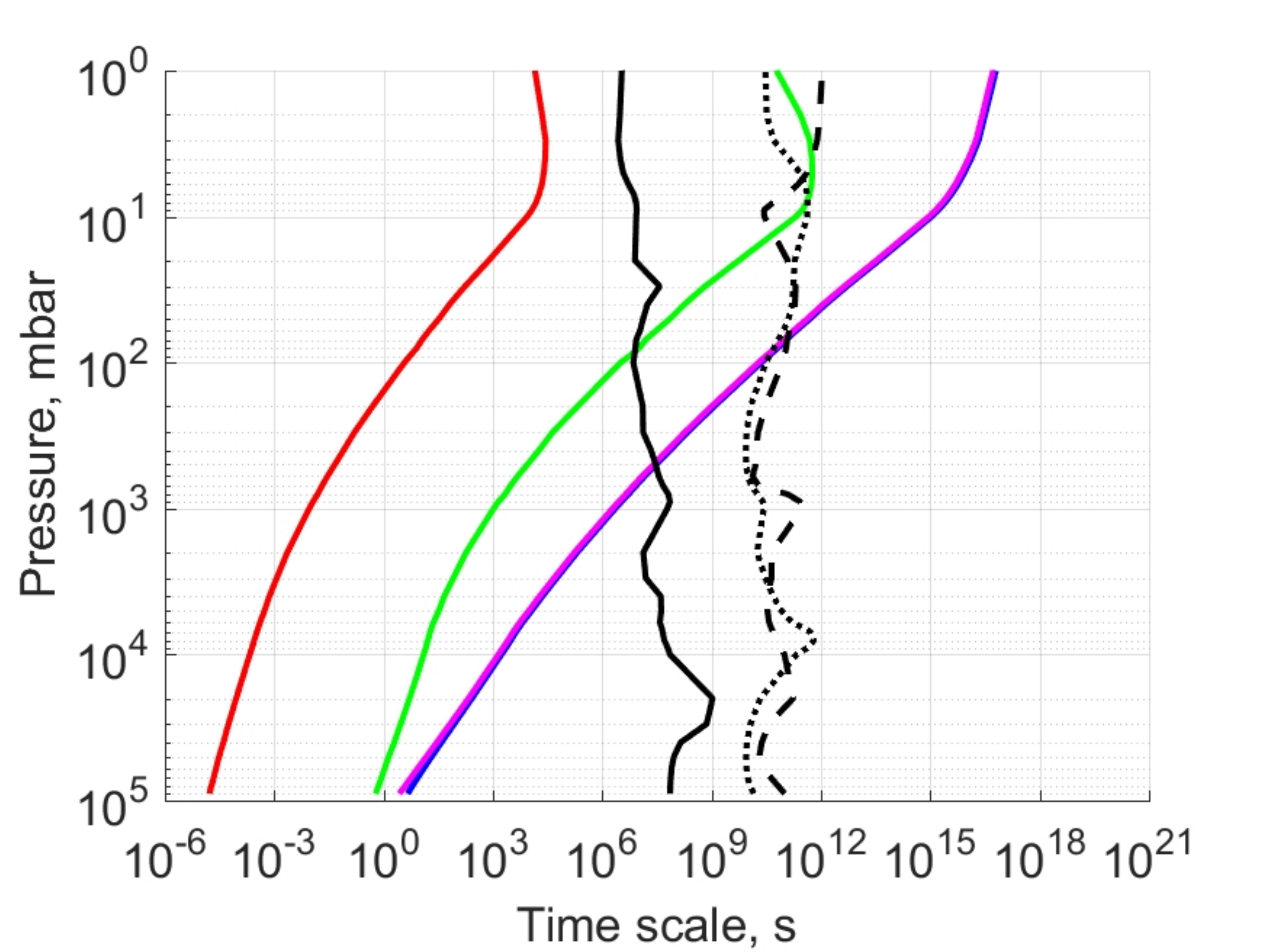}} &
{\includegraphics[width=4.2cm]{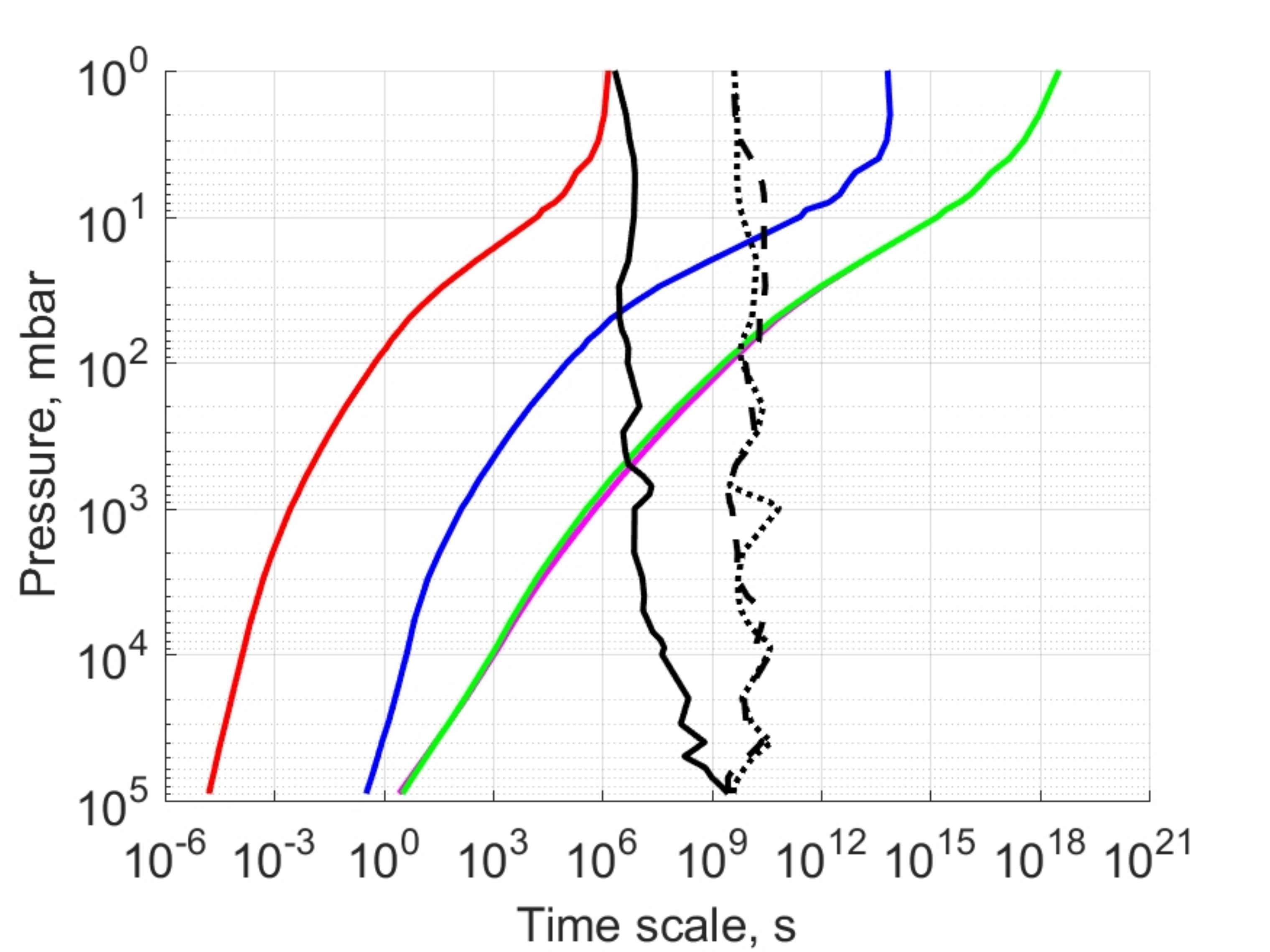}} & 
{\includegraphics[width=4.2cm]{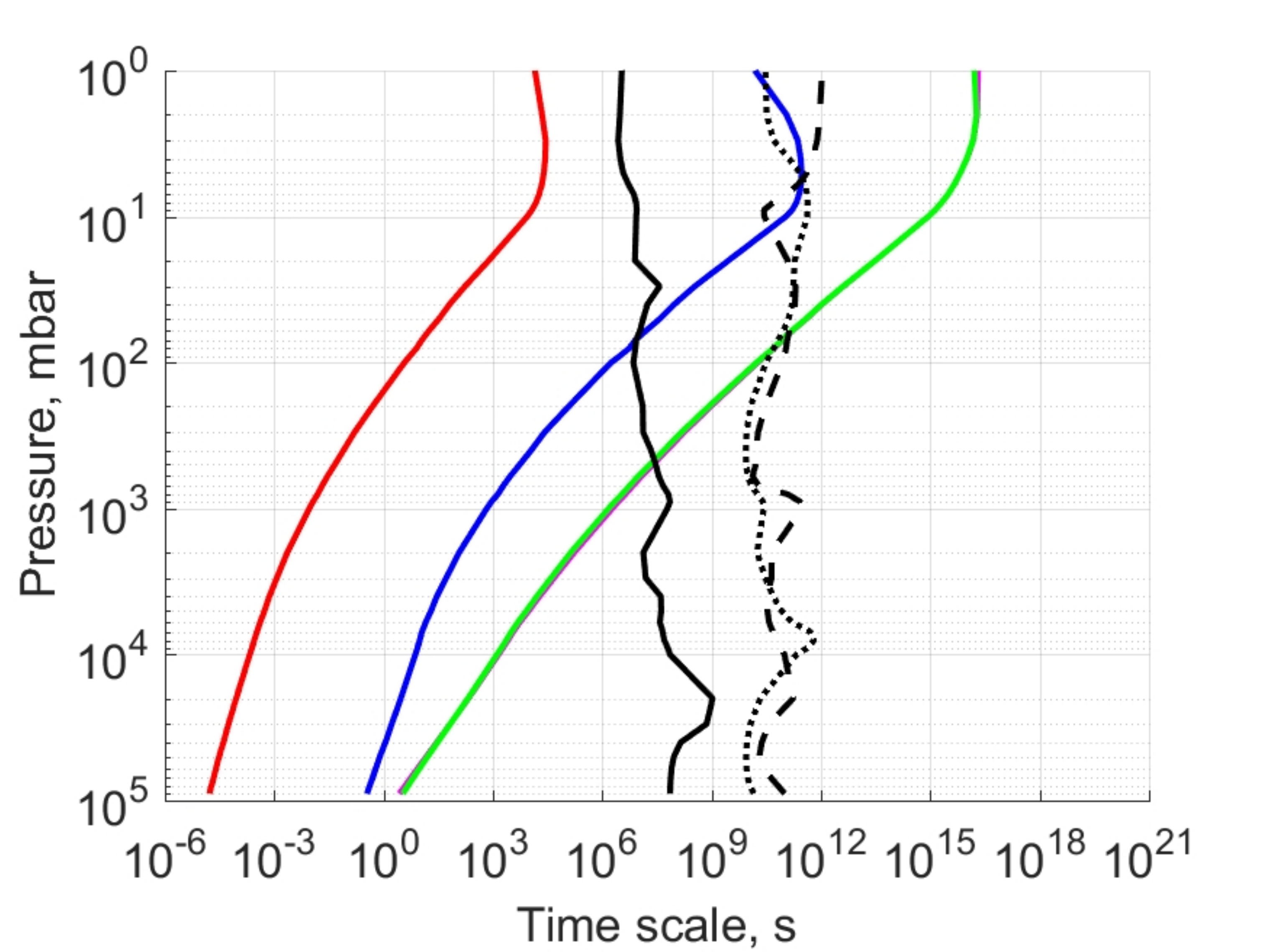}} \\ 
    \vspace{-3cm}\rotatebox{90}{Evening terminator} & {\includegraphics[width=4.2cm]{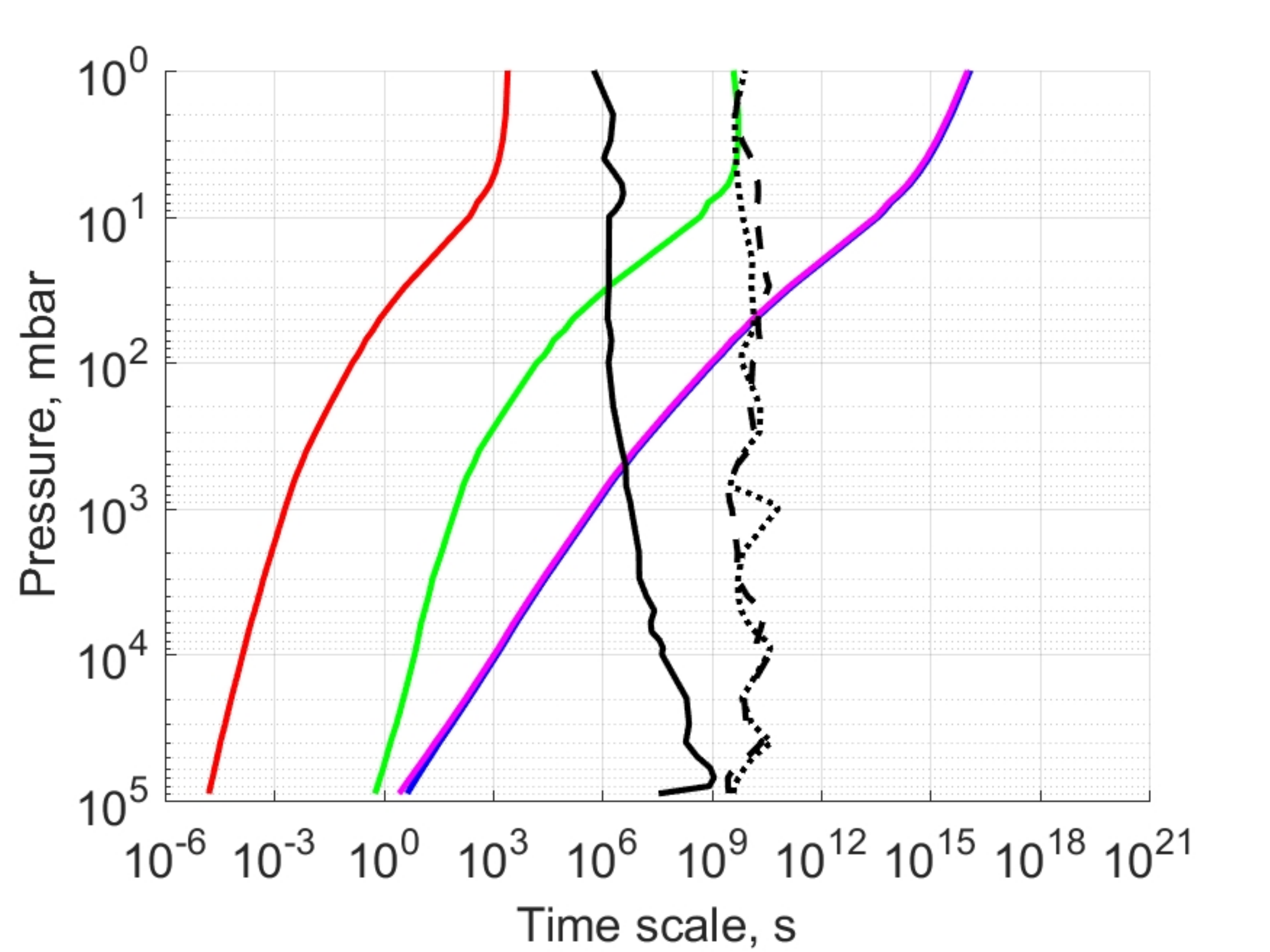}} & {\includegraphics[width=3.8cm]{Times_EveningT_cheG1_SOLAR}} &
{\includegraphics[width=4.2cm]{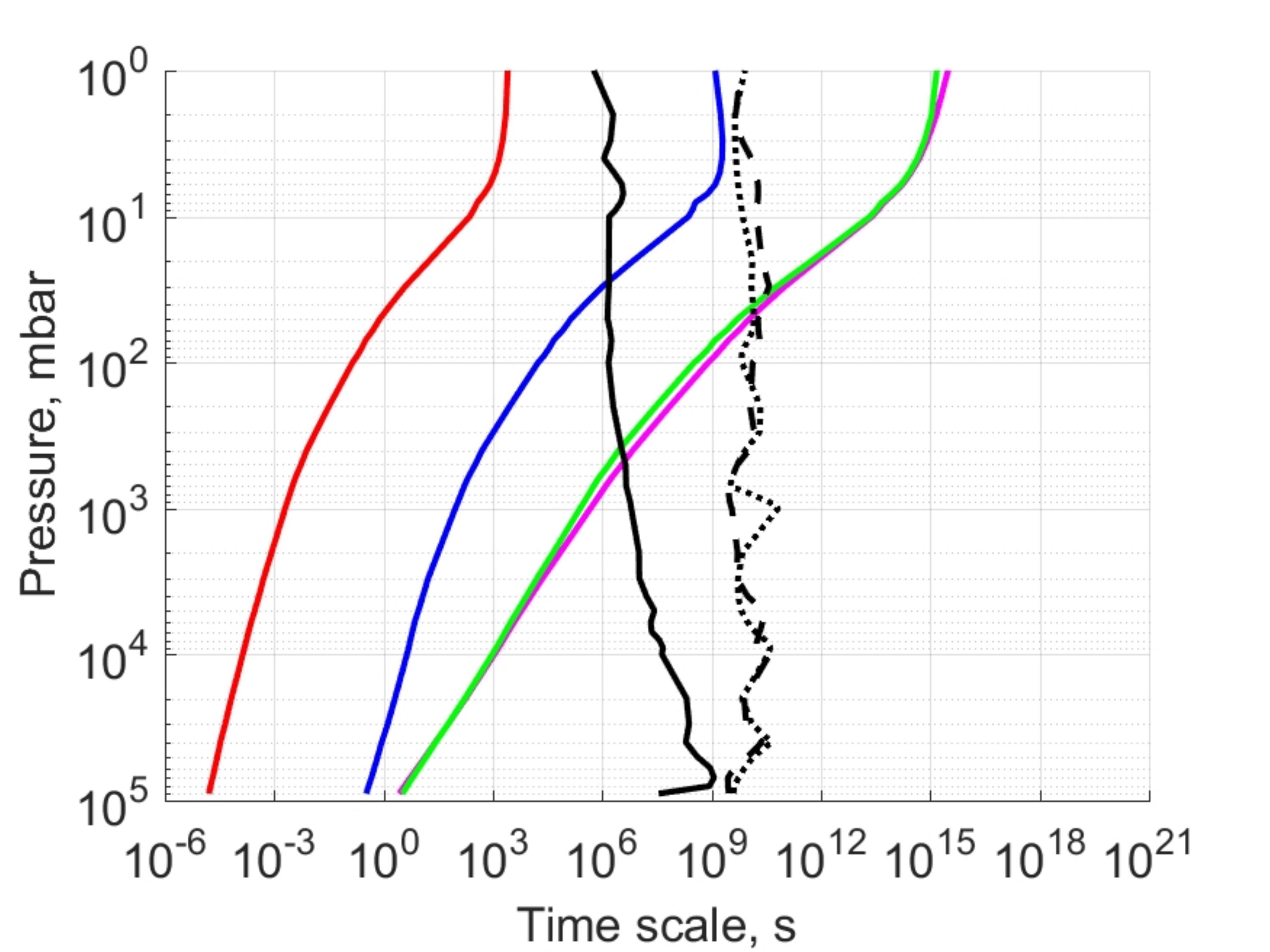}} & 
{\includegraphics[width=4.2cm]{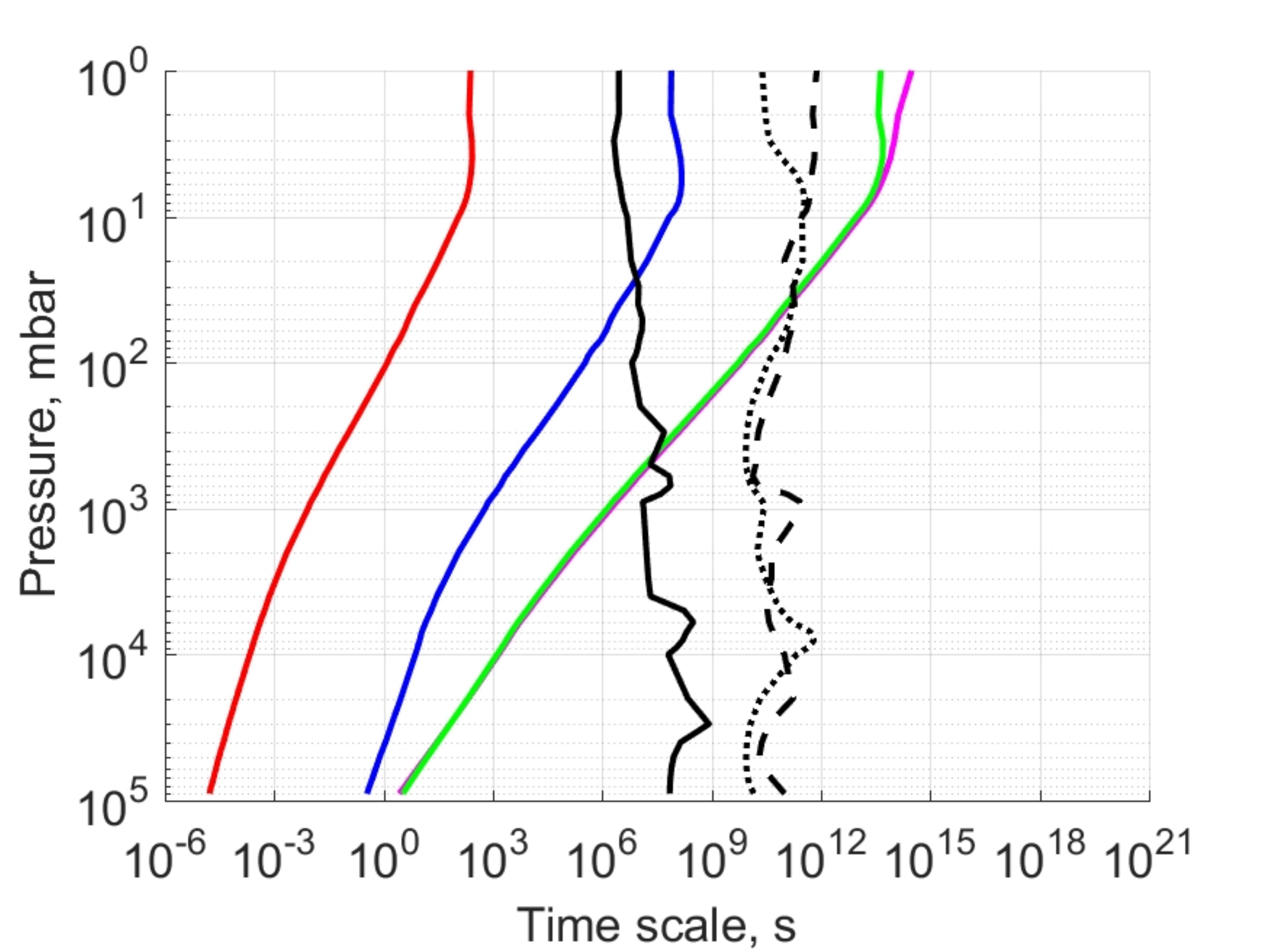}} \\ 
  \end{tabular}
  \caption{This figure shows the chemical and dynamical time scales from the simulations with different $G$ and $C/O$ and regions as defined previously in Fig. \ref{fig:che_ref_v}. The colors represent the chemical species: CO$_2$ (red), H$_2$O (blue), CO (magenta) and CH$_4$ (green). The black lines are the dynamical timescales at different directions: longitude (solid line), latitude (dotted line) and vertical (dashed line). All the results shown in this figure were averaged over the last 100 days of the simulations.}
  \label{fig:che_ref_time}     
\end{figure*}

As it is explained in the introduction, the chemical distribution can be understood in terms of chemical and dynamical timescales. In this section, we compare the chemical timescales computed in \cite{2018Tsai}  (represent how fast the species react toward equilibrium) to the dynamical timescales (represent the efficiency of the atmospheric circulation mixing tracers in the atmosphere). The dynamical timecales are defined as a characteristic spatial scale divided by a mean wind speed:
\begin{equation}
\label{eq:tauu}
\tau_u = \frac{2\pi R_{Planet}}{\bar{u}}
\end{equation}
\begin{equation}
\label{eq:tauv}
\tau_v = \frac{\pi R_{Planet}}{\bar{v}}
\end{equation}
\begin{equation}
\label{eq:tauw}
\tau_w = \frac{\bar{H_a}}{\bar{w}}.
\end{equation}

Eqs. \ref{eq:tauu}, \ref{eq:tauv} and \ref{eq:tauw} represent the dynamical timescales in different directions. Eq. \ref{eq:tauu}, refers to mixing in the longitudinal direction, Eq. \ref{eq:tauv} in the meridional direction and Eq. \ref{eq:tauw} in the vertical direction. The numerator in these three equations is the characteristic spatial scale that represents the scale of the dominant pattern in the atmospheric circulation. The characteristic spatial scale in the longitudinal and meridional directions was chosen to be the planet perimeter and half of the planet perimeter respectively, and the atmospheric scale hight in the vertical direction. On the denominator we included the latitude, longitude and time averaged wind speed ($u$ - longitude; $v$ - latitude; $w$ - vertical). The spatial averaging is based on a root-mean-square method. These timescales are rough estimates of the real values and we use them here to obtain a zero-order estimate of the real values to guide us on understanding the main drivers setting the chemical distribution across the atmosphere. 

Fig. \ref{fig:che_ref_time}, shows a comparison between the chemical and the different dynamical timescales. The different panels in this figure contain a large amount of information about the competition between the chemical and dynamical timescales at different locations for the 4 different experiments. We will just focus on the main features. The major component of the dynamical mixing in the atmosphere of WASP-43b is according to our timescales the longitudinal component of the wind field. This result is associated with the formation of the strong equatorial jet that transports tracers eastwards. In the nightside of the planet the other two components may become more efficient than the chemical timescale but still a couple of order of magnitude weaker than the longitudinal mixing.

Fig. \ref{fig:che_ref_time} shows that the chemical timescales for CO$_2$ are the shortest of the other 4 species, and the only region where it is not shorter than the dynamical timescale is above the pressure level 10 mbar in the nightside. The short chemical timescale of CO$_2$ explains the tendency for the formation of a clear day-night contrast. The differences between a scenario with or without thermal inversion are very small. The presence of a thermal inversion results in a slight increase of the chemical distribution day-night contrast, unless the chemical species are very well mixed in the atmosphere, as is the case for CH$_4$ when $C/O$ is 2. For $C/O$ = 0.5, CH$_4$ is the second chemical species with the shortest timescale. The distribution of CH$_4$ in the upper atmosphere in this scenario is controlled by the conditions on the dayside. CH$_4$ maintains in chemical equilibrium on the dayside due to the high temperature. This abundance is then transported horizontally and quenched outside the dayside. For $C/O$ = 2.0, in the case with no thermal inversion, we obtain similar results than the previous two scenarios, where the chemical abundance of CH$_4$ is set by the dayside production and quenched in the longitudinal direction, but for $C/O$ = 2.0 the vertical mixing has a stronger contribution vertically homogenizing CH$_4$. When $G$ is increased to 2 (keeping $C/O$ = 2), the chemical timescales of some species become shorter than the vertical dynamical timescale, and it quenches vertically from pressures around 100 mbar. Our vertical dynamical timescale is overestimated, and does not capture the vertical quenching point suggested by the vertical distribution of CH$_4$. A more detailed analysis of the mass transport by mean circulation, eddies and diffusion would reveal the exact nature and efficiency of the mixing, but this detailed analysis is not the main goal of our paper. CO and H$_2$O for $C/O$ = 0.5, have similar distributions. As we also saw before in Fig. \ref{fig:che_disp}, the dispersion of CO and H$_2$O is very small compared to the other molecules. The chemical timescales of these two species are very similar at $C/O$ = 0.5, and both are quenched horizontally and vertically above 100 mbar, which agrees well with the results from Fig. \ref{fig:che_ref_v}. There are no differences again between the cases with and without thermal inversion. The slight decrease with altitude is associated with the production of CO and H$_2$O that is favoured in the hotter regions in the upper atmosphere, and mixed rapidly towards the nightside, where it is trapped and destroyed near the polar  vortexes, which maintains the abundances close to chemical equilibrium (see the low dynamics in Fig. \ref{fig:che_ref_h}). For $C/O$ = 2, the CO maintains roughly the same distribution than with $C/O$ = 0.5 because the equilibrium chemistry and timescales are similar in the two cases. However, for H$_2$O, the cold polar vortexes are the regions with the largest production of H$_2$O, which then get eroded by the dynamics (equatorward transport) raising the levels of H$_2$O in the upper atmosphere. Another mechanism raising the levels of H$_2$O is the higher production of H$_2$O at low pressures in the nightside of the planet, which is then mixed in the horizontal and vertical directions as seen in Fig. \ref{fig:che_ref_time}. Also it shows that H$_2$O is longitudinally quenched across the atmosphere except in the dayside at low pressures (longitudinal and chemistry timescales are comparable at low pressures in the case $G = 0.5$).  The comparable timescales in the dayside of the upper atmosphere produce a clear day-night contrast (see Fig. \ref{fig:che_ref_h}). Again the differences between the results with and without thermal inversion are minor. 

\section{Phase-curve spectra}
\label{sec:phcrv_spec}
\begin{figure*}
\label{fig:phcv}
\begin{centering}
\subfigure[0.1250 ($\sim$ nightside)]{
\includegraphics[width=0.8\columnwidth]{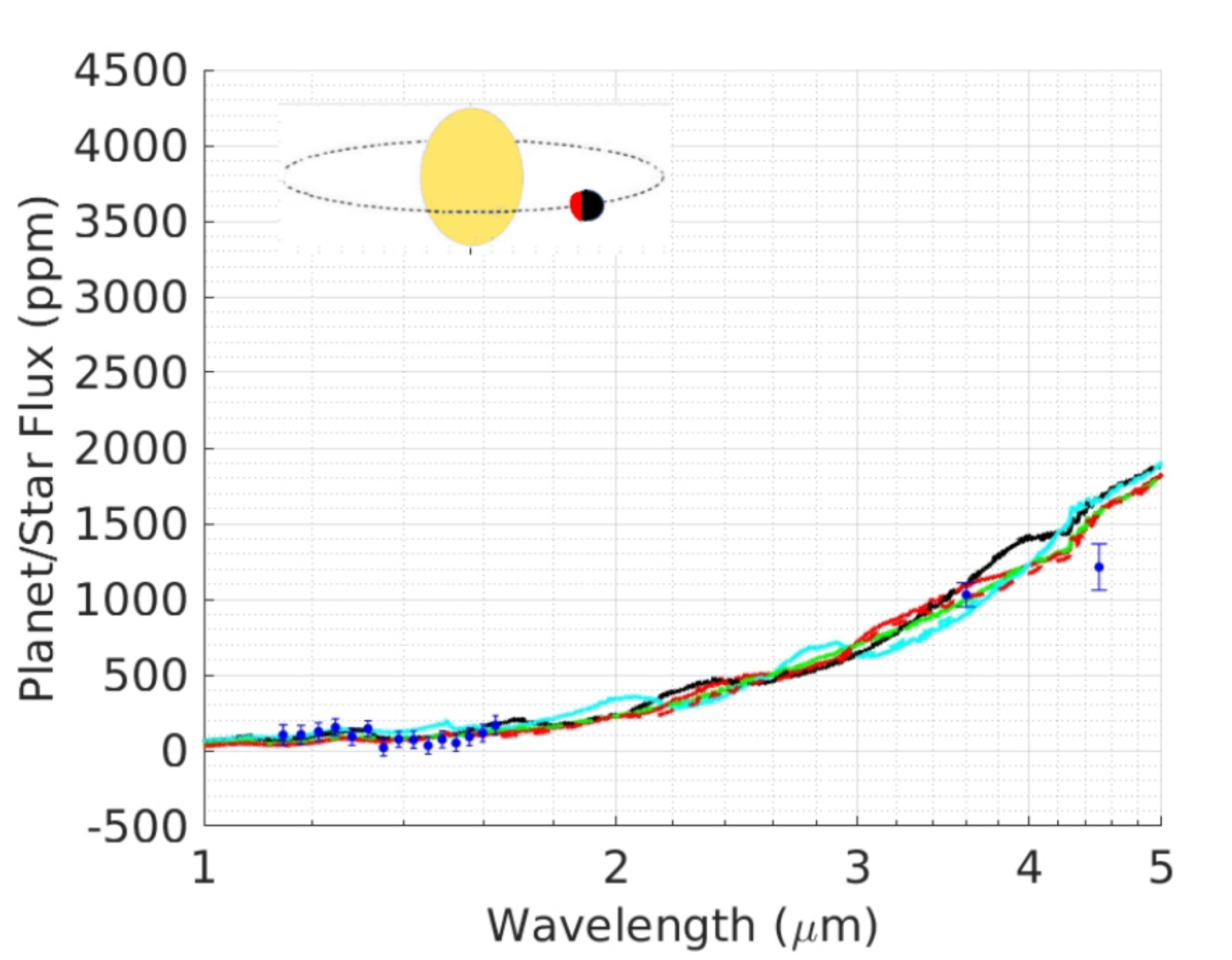}}
\subfigure[0.3125]{
\includegraphics[width=0.8\columnwidth]{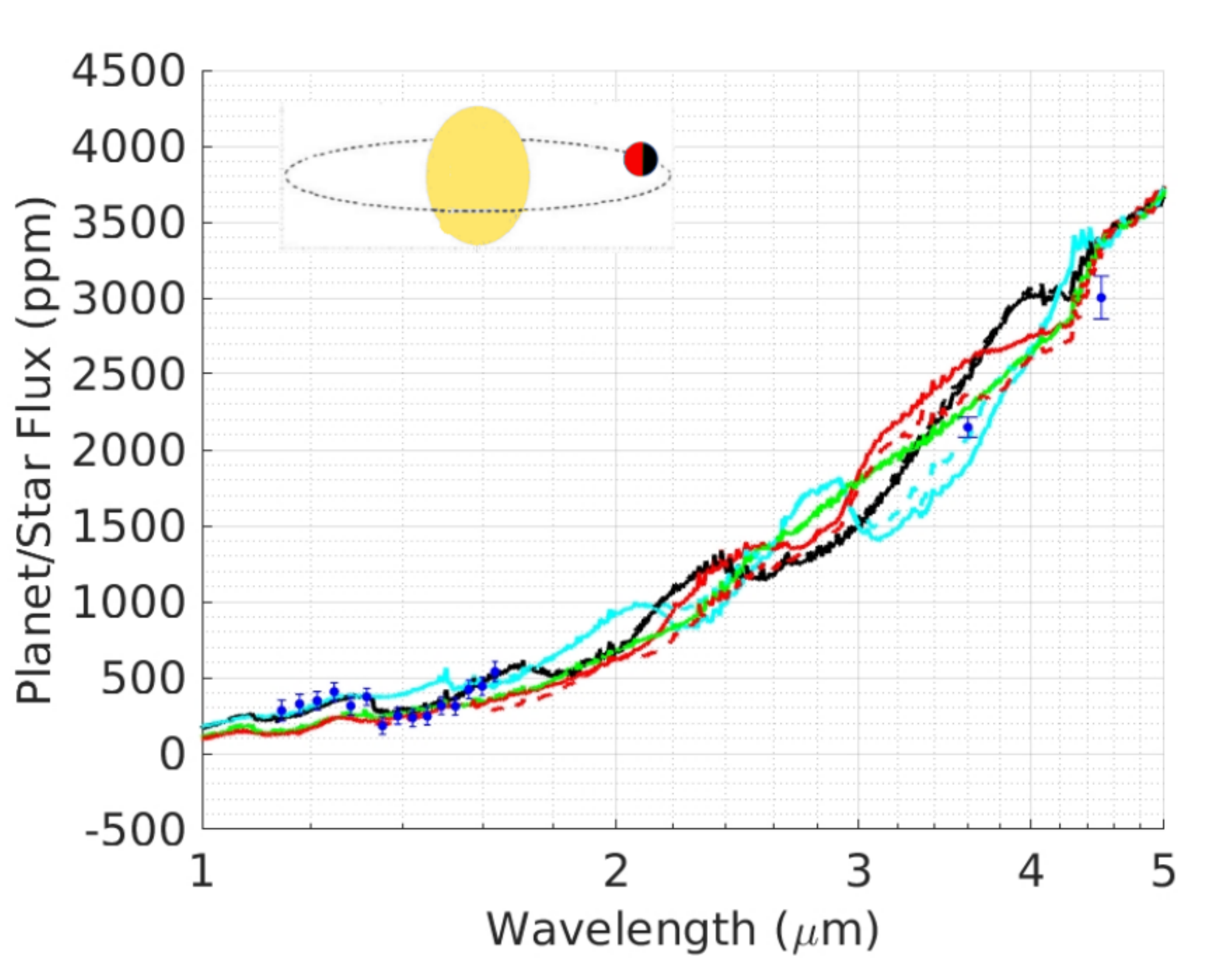}}
\subfigure[0.5000 (dayside)]{
\includegraphics[width=0.8\columnwidth]{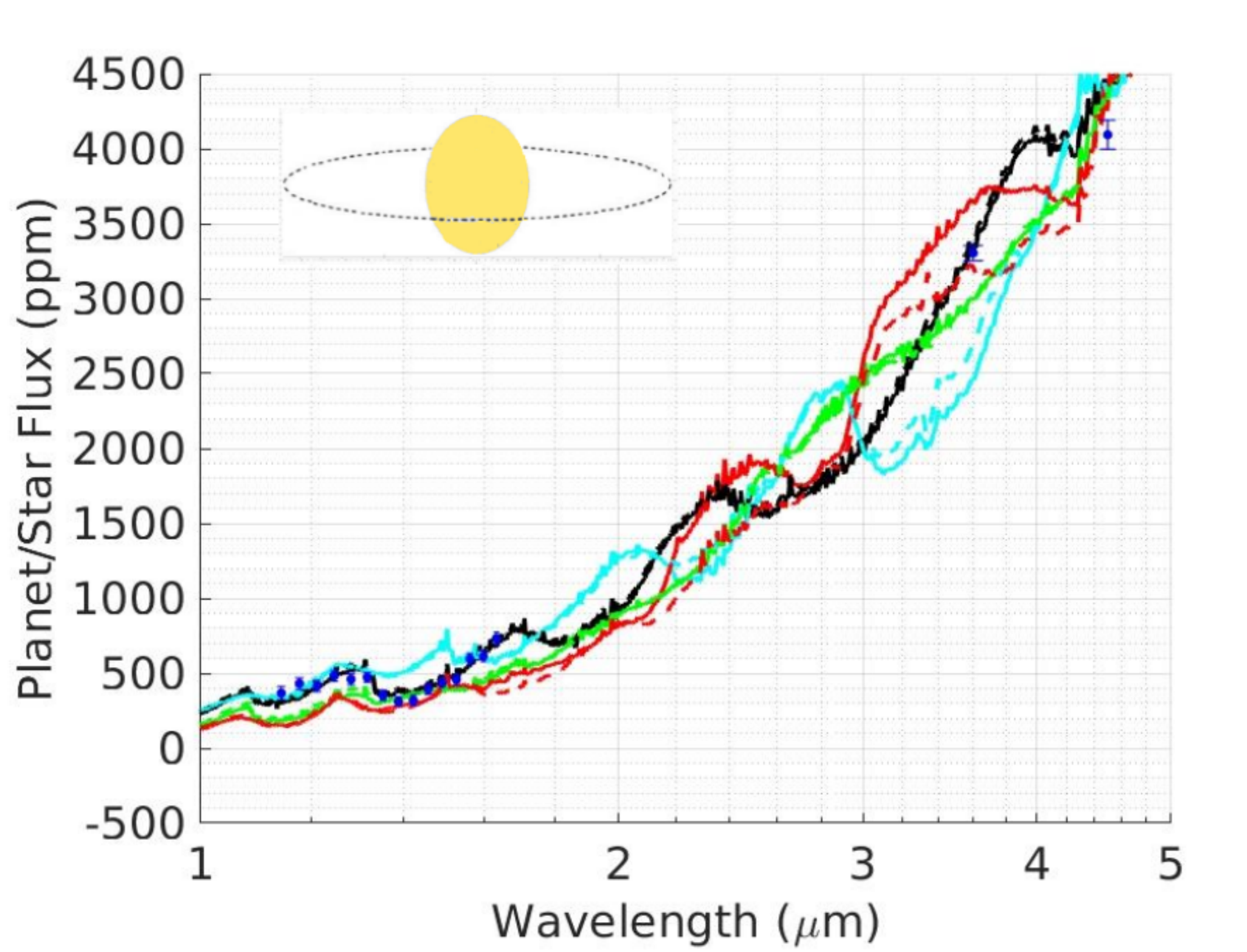}}
\subfigure[0.5625]{
\includegraphics[width=0.8\columnwidth]{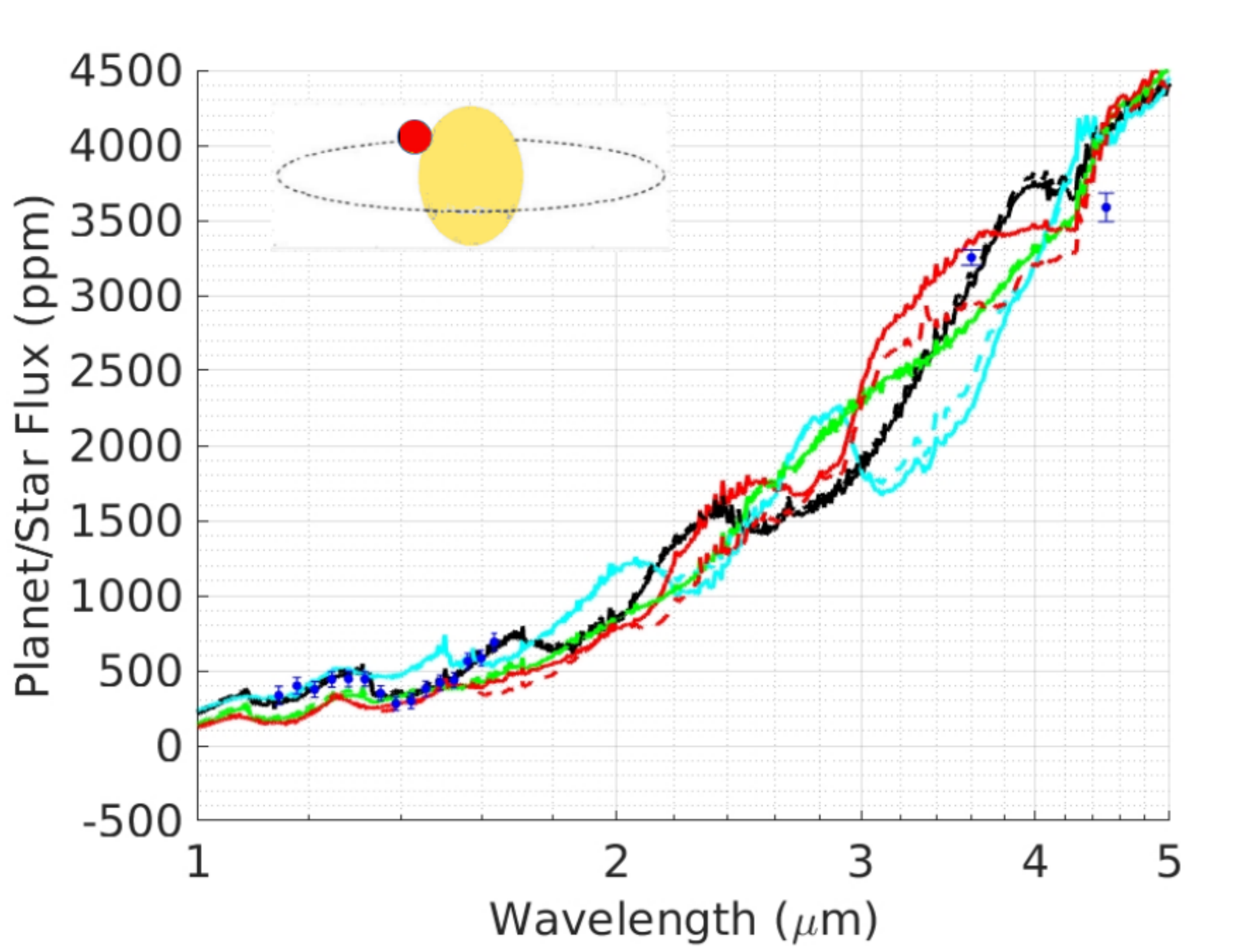}}
\subfigure[0.6875]{
\includegraphics[width=0.8\columnwidth]{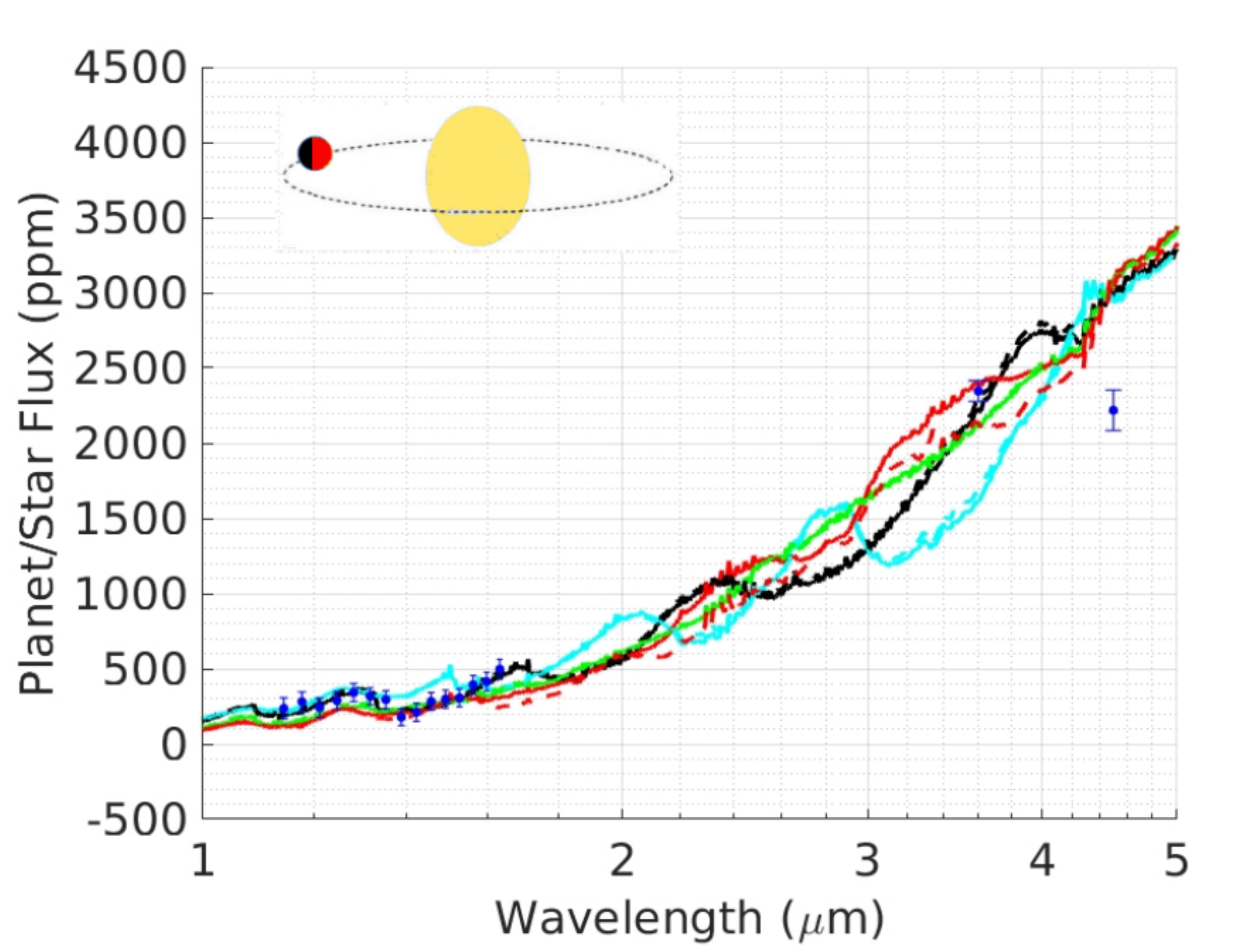}}
\subfigure[0.8750 ($\sim$ nightside)]{
\includegraphics[width=0.8\columnwidth]{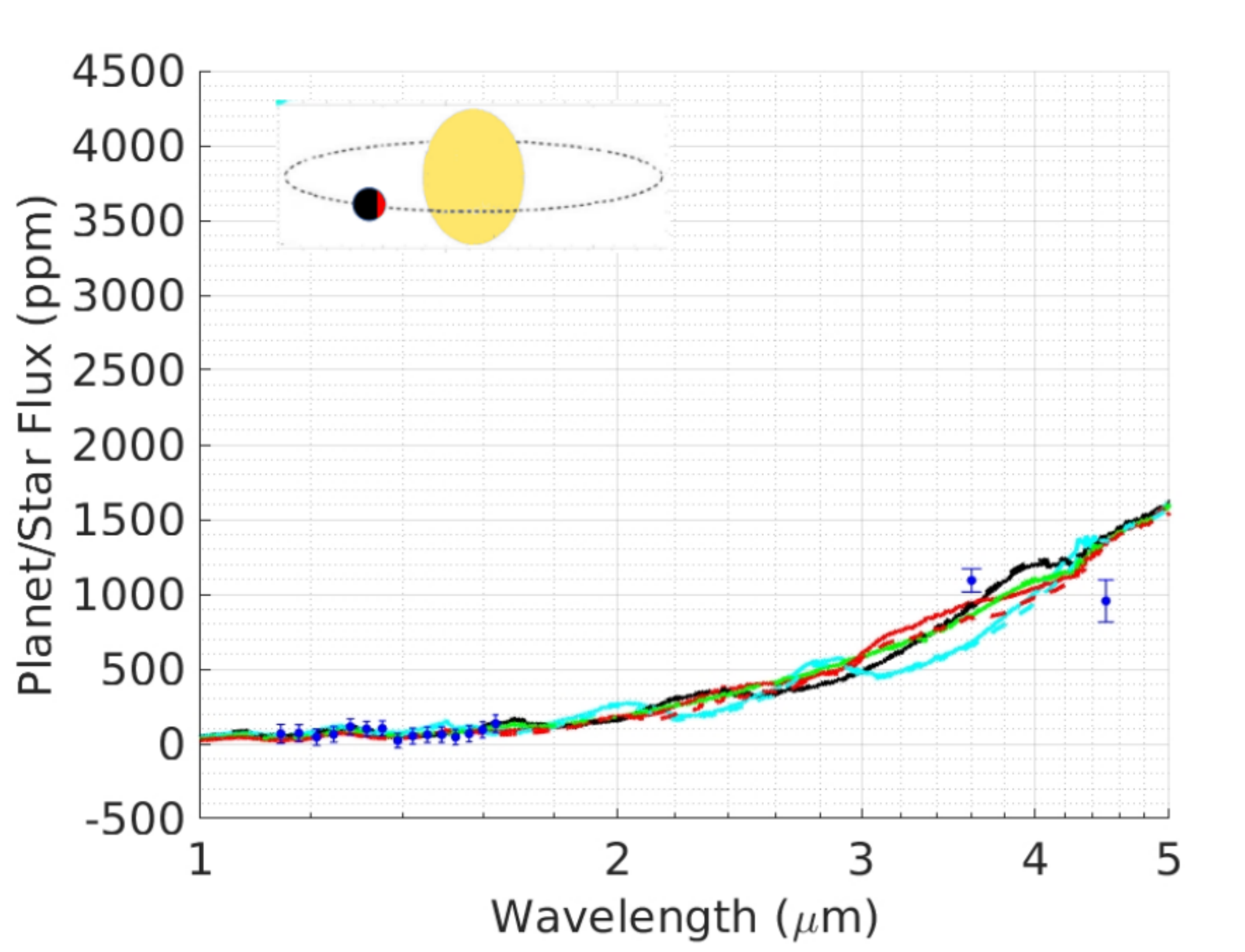}}
\caption{Emission spectra at different orbital phases (panels $a$ to $f$). The primary transit occurs at orbital phase 0.0 and the secondary eclipse at 0.5 (panel $e$). The blue points are \textit{WFC3} data from \cite{2014Stevenson} and \textit{Spitzer} data from \cite{2017Mendoncaa}. The different line colors represent the different experiments: $G = 0.5$ and $C/O = 0.5$ (black); $G = 0.5$ and $C/O = 2.0$ (cyan); $G = 2.0$ and $C/O = 0.5$ (green); $G = 2.0$ and $C/O = 2.0$ (red). The different line styles refer to different assumptions for the atmospheric chemical composition: solid lines assume disequilibrium abundances for the four molecules (CO$_2$, CO, H$_2$O and CH$_4$) studied in previous sections and the dashed lines assume equilibrium chemistry.}
\label{fig:planet_spec}
\end{centering}
\end{figure*}

The departures from chemical equilibrium driven by the circulation have an impact in the emission spectra of the planet. The four chemical species explored are the main gas absorbers in the infrared (IR). Fig. \ref{fig:planet_spec} shows the emission spectra at different orbital phases from the different experiments. The main goal of producing these plots was not to find the best fit scenario but to learn about the impact of disequilibrium chemistry, driven by the atmospheric circulation, in the emission spectra. A good match of the theoretical spectra with the observations requires a better representation of the cloud physics as mentioned in \cite{2017Mendoncaa}. Our GCM simulations also uses a gray radiative transfer, which does not take into account the changes in composition. Radiative transfer codes such as the ones used in \cite{2009Showman} and \cite{2014Amundsen}, would help to unveil the impact of the disequilibrium feedback in the radiative heating/cooling rates, which would refine the results found here. 

In Fig. \ref{fig:planet_spec}, the solid lines represent the atmospheres that are in chemical equilibrium except for CO, CH$_4$, H$_2$O and CO$_2$, and for the dashed lines all the gases are assumed in chemical equilibrium. When C/O = 0.5 there are no significant differences between the equilibrium and disequilibrium solutions. The main reason for this result is associated with the H$_2$O being the dominant IR absorber in the the upper atmosphere and its abundance is close to chemical equilibrium. When C/O is increased to 2, the abundance of water is reduced and the departures from chemical equilibrium become relatively larger. The largest differences in the emission spectrum for C/O = 2 is for the case with thermal inversion due to the enhanced levels of CH$_4$ in the dayside upper atmosphere in relation to its equilibrium (see the plots from the fourth column in Fig. \ref{fig:che_ref_v}). The largest differences between the dashed and solid lines are between 1.6 to 1.8 $\mu$m, 2.2 to 2.6 $\mu$m and 3 to 4 $\mu$m, which correspond to the strong absorption features of CH$_4$ (see for example \citealt{2017Malik}). The higher fluxes in the red solid line when compared with the dashed red line is due to the larger emission from the upper hot atmosphere (temperature increases with altitude). The opposite trend (absorption feature) is seen for the cyan cases, which does not have a thermal inversion. The fluxes closer to 1 $\mu$m come from deep regions in the atmosphere that exist close to equilibrium. The detection of departures from equilibrium, according to our results would be possible for hot Jupiter atmospheres with C/O higher than 0.5 and good spectral coverage between 1.6 to 1.8 $\mu$m, 2.2 to 2.6 $\mu$m and 3 to 4 $\mu$m. It is worth to highlight that the prescribed thick cloud deck largely reduces the impact of the disequilibrium chemistry in the nightside. In a cloud-free atmosphere the large departure from equilibrium in the night side would be easily found.

\cite{2017Mendoncaa}, reported the difficulty in explaining the absorption at 4.5 $\mu$m. In this spectral region CO and CO$_2$ have important absorption features and enhanced levels of these two molecules in the atmosphere could explain the discrepancy between theory and observations. H$_2$O is also efficient absorbing 4.5 $\mu$m, but higher levels of H$_2$O would make the spectrum inconsistent in other spectral regions. Our current work suggests that chemical disequilibrium produced by the atmospheric winds alone cannot explain the discrepancies between models and observations. The causes for the larger absorption at this particular wavelength are still not fully explained.

\section{Conclusions and future prospects}
\label{sec:conclu}

Consistent modelling of chemistry in exoplanets is still poorly explored. In the current work, we study the disequilibrium chemistry driven by atmospheric circulation using a 3D atmospheric model, which includes a complex representation of the atmospheric flow (solves the deep non-hydrostatic equations) and simplified representations of radiation and chemistry. We have chosen to study the planet WASP-43b and explore its main properties associated with the chemical distribution across the atmosphere. Our GCM solves the deep non-hydrostatic Euler equations and coupled with the radiative transfer code we explored two different scenarios: with and without thermal inversion in the dayside of the upper atmosphere. The main differences between these two simulations are the weaker jet and larger day-night contrast in temperature for the case with thermal inversion.

Our new formulation for the chemical timescale developed in \cite{2018Tsai}, allowed us to extend our disequilibrium chemistry study using a GCM to include more molecules than previous works due to the low cost of the parameterization. In the present work we study the chemical distribution of CO$_2$, CO, CH$_4$ and H$_2$O across the atmosphere of WASP-43b. The distribution of the chemical species depends on the strength of the atmospheric mixing and the efficiency of the chemical reactions. Different molecules have different chemical timescales, which result in different patterns of chemical distribution across the atmosphere. In general, the gas molecules are quenched longitudinally in regions near the equator, except for CO$_2$ when $C/O$ is 0.5. CO$_2$ has a shorter chemical timescale than the other molecules for the experiments studied, which resulted in a larger day-night contrast for the abundances of CO$_2$. In contrast with the regions at low latitude the polar vortexes have low levels of dynamical mixing and lower temperatures, which results in a distinct chemical distribution near and inside the polar vortexes. However, if the chemical timescales become very long the chemical species become well mixed over all latitudes and longitudes. The dynamical mixing in the latitudinal and vertical directions is much weaker than in the longitudinal direction, but in the nightside it is more efficient than the chemical production/destruction rates. 

The main mechanism driving the chemical space distribution in this work for WASP-43b is different than previous works, e.g., \cite{2006Cooper} and \cite{2018Drummond} for HD 209458b. In \cite{2006Cooper} and \cite{2018Drummond} the vertical mixing is the dominant mechanism, however, \cite{2018Drummond} found that the horizontal transport also has an important impact in the final chemical distribution of methane. In our work, we find the transport in the zonal direction (horizontal quenching) to be the dominant process. The disparity is mainly associated with the differences in the dynamical timescales, which can be associated with the different model parameters used: \cite{2006Cooper} and \cite{2018Drummond} explored HD 209458b with a clear atmosphere, and in our work, we explored WASP-43b with a cloudy nightside, which has stronger surface gravity ($\sim$5$\times$ stronger) and faster rotation rate ($\sim$4.4$\times$ faster) than HD 209458b. These results highlight the diversity of solutions and the need to consider 3D dynamical effects when studying disequilibrium chemistry in hot Jupiter planets.

The differences in the chemical distribution due to the atmospheric transport have an impact in the emission spectra of the planet. The largest differences exist for the atmosphere at C/O = 2, mainly due to the disequilibrium of CH$_4$. The CH$_4$ molecule has an important impact in the spectral regions between 1.6 to 1.8 $\mu$m, 2.2 to 2.6 $\mu$m and 3 to 4 $\mu$m. The differences due to CH$_4$ become clearly visible in the dayside of the planet. For C/O = 0.5 the atmosphere is dominated by the IR opacity of H$_2$O that is close to equilibrium.
 
Our model is currently being updated with a more sophisticated radiative transfer code capable of computing the feedback due to chemical disequilibrium, and cloud formation. The new \texttt{THOR} will be essential to analyse and interpret the data from \textit{James Webb Space Telescope}, which will have a better spectral resolution and wavelength coverage than current instruments to deal with the problem of disequilibrium chemistry in exoplanet atmospheres.

\section*{Acknowledgments}
J.M.M., S.T., M.M., S.L.G. and K.H. thank the Center for Space and Habitability (CSH), Swiss National Science Foundation, Swiss-based MERAC Foundation and the Space Research and Planetary Sciences Division (WP) of the University of Bern for financial, secretarial and logistical support.  J.M.M. also acknowledges financial support from the Villum
Foundation.
\appendix
\section{Sponge layer}
\label{apxd:sponge}
To formally conserve mass in the model's domain, the upper and lower boundary conditions are set to be rigid-lids. The impermeability of the upper boundary condition causes spurious wave reflections which can lead to unrealistic results in the upper part of the model's domain and compromise severely the numerical stability of the model. This problem is well known in the 3D climate model community, and the traditional approach to deal with this technical difficulty is to apply a ``sponge'' layer scheme that damps the eddy components in the upper layers of the model. There is no universal solution for the ``sponge'' layer scheme, which can have different forms (e.g., \citealt{2011Jablonowski}), and are usually poorly documented. Note that the cases studied in \cite{2016Mendoncab} did not require a sponge layer scheme. The sponge layer parameterizations are not physically based, however, they try to represent wave breaking phenomena in the upper atmosphere. Our simple and effective solution is based on a Rayleigh friction formulation: 
\begin{equation}
\frac{d\Psi}{dt}=-\frac{\Psi - \bar\Psi}{\tau},
\end{equation}
where $\Psi$ represents the wind velocity components. In this parameterization the horizontal wind is projected as a zonal and meridional component instead of the 3D Cartesian projection used in the dynamical core. The bar over the variable indicates longitudinal mean. To compute the longitudinal average of a quantity in a icosahedral grid we first decompose the grid into longitudinal rings with predefined latitudinal widths, and inside each ring we calculate the variable average. The longitudinal average is then calculated efficiently by interpolating linearly in latitude the average values of the rings. We assume a simplified formulation for the strength of the damping based on the formulation suggested in \cite{2008Skamarock}:  
\[ \frac{1}{\tau} =
  \begin{cases}
    0       & \quad \text{if } \eta < \eta_s\\
    k_s \sin^2(\frac{\pi}{2}\frac{\eta-\eta_s}{1.0-\eta_s})  & \quad \text{otherwise.}\\
  \end{cases}
\]
The value $k_s$ represents the largest damping at the top of the model's domain. The damping function increases slowly with altitude, otherwise, the sponge layer could be itself a source of spurious numerical waves. The parameter $k_s$ was set to $10^{-4}$ s$^{-1}$ and $\eta_s$ to 0.75. 

\begin{thebibliography}{}
\expandafter\ifx\csname natexlab\endcsname\relax\def\natexlab#1{#1}\fi

\bibitem[{{Ag{\'u}ndez} {et~al.}(2014){Ag{\'u}ndez}, {Parmentier}, {Venot},
  {Hersant}, \& {Selsis}}]{2014Agundez}
{Ag{\'u}ndez}, M., {Parmentier}, V., {Venot}, O., {Hersant}, F., \& {Selsis},
  F. 2014, \aap, 564, A73

\bibitem[{{Ag{\'u}ndez} {et~al.}(2012){Ag{\'u}ndez}, {Venot}, {Iro}, {Selsis},
  {Hersant}, {H{\'e}brard}, \& {Dobrijevic}}]{2012Agundez}
{Ag{\'u}ndez}, M., {Venot}, O., {Iro}, N., {et~al.} 2012, \aap, 548, A73

\bibitem[{{Allard} \& {Hauschildt}(1995)}]{1995Allard}
{Allard}, F., \& {Hauschildt}, P.~H. 1995, \apj, 445, 433

\bibitem[{{Amundsen} {et~al.}(2014){Amundsen}, {Baraffe}, {Tremblin},
  {Manners}, {Hayek}, {Mayne}, \& {Acreman}}]{2014Amundsen}
{Amundsen}, D.~S., {Baraffe}, I., {Tremblin}, P., {et~al.} 2014, \aap, 564, A59

\bibitem[{{Azzam} {et~al.}(2016){Azzam}, {Tennyson}, {Yurchenko}, \&
  {Naumenko}}]{2016Azzam}
{Azzam}, A.~A.~A., {Tennyson}, J., {Yurchenko}, S.~N., \& {Naumenko}, O.~V.
  2016, \mnras, 460, 4063

\bibitem[{{Barber} {et~al.}(2006){Barber}, {Tennyson}, {Harris}, \&
  {Tolchenov}}]{2006Barber}
{Barber}, R.~J., {Tennyson}, J., {Harris}, G.~J., \& {Tolchenov}, R.~N. 2006,
  VizieR Online Data Catalog, 6119

\bibitem[{{Blecic} {et~al.}(2014){Blecic}, {Harrington}, {Madhusudhan},
  {Stevenson}, {Hardy}, {Cubillos}, {Hardin}, {Bowman}, {Nymeyer}, {Anderson},
  {Hellier}, {Smith}, \& {Collier Cameron}}]{2014Blecic}
{Blecic}, J., {Harrington}, J., {Madhusudhan}, N., {et~al.} 2014, \apj, 781,
  116

\bibitem[{{Cooper} \& {Showman}(2005)}]{2005Cooper}
{Cooper}, C.~S., \& {Showman}, A.~P. 2005, \apjl, 629, L45

\bibitem[{{Cooper} \& {Showman}(2006)}]{2006Cooper}
---. 2006, \apj, 649, 1048

\bibitem[{{Draine}(2011)}]{2011Draine}
{Draine}, B.~T. 2011, {Physics of the Interstellar and Intergalactic Medium}

\bibitem[{{Drummond} {et~al.}(2018){Drummond}, {Mayne}, {Manners}, {Carter},
  {Boutle}, {Baraffe}, {H{\'e}brard}, {Tremblin}, {Sing}, {Amundsen}, \&
  {Acreman}}]{2018Drummond}
{Drummond}, B., {Mayne}, N.~J., {Manners}, J., {et~al.} 2018, \apjl, 855, L31

\bibitem[{{Gillon} {et~al.}(2012){Gillon}, {Triaud}, {Fortney}, {Demory},
  {Jehin}, {Lendl}, {Magain}, {Kabath}, {Queloz}, {Alonso}, {Anderson},
  {Collier Cameron}, {Fumel}, {Hebb}, {Hellier}, {Lanotte}, {Maxted},
  {Mowlavi}, \& {Smalley}}]{2012Gillon}
{Gillon}, M., {Triaud}, A.~H.~M.~J., {Fortney}, J.~J., {et~al.} 2012, \aap,
  542, A4

\bibitem[{{Guillot} {et~al.}(1996){Guillot}, {Burrows}, {Hubbard}, {Lunine}, \&
  {Saumon}}]{1996Guillot}
{Guillot}, T., {Burrows}, A., {Hubbard}, W.~B., {Lunine}, J.~I., \& {Saumon},
  D. 1996, \apjl, 459, L35

\bibitem[{{Harris} {et~al.}(2006){Harris}, {Tennyson}, {Kaminsky}, {Pavlenko},
  \& {Jones}}]{2006Harris}
{Harris}, G.~J., {Tennyson}, J., {Kaminsky}, B.~M., {Pavlenko}, Y.~V., \&
  {Jones}, H.~R.~A. 2006, \mnras, 367, 400

\bibitem[{{Haynes} \& {Shuckburgh}(2000{\natexlab{a}})}]{2000Haynes1}
{Haynes}, P., \& {Shuckburgh}, E. 2000{\natexlab{a}}, \jgr, 105, 22777

\bibitem[{{Haynes} \& {Shuckburgh}(2000{\natexlab{b}})}]{2000Haynes2}
---. 2000{\natexlab{b}}, \jgr, 105, 22795

\bibitem[{{Hellier} {et~al.}(2011){Hellier}, {Anderson}, {Collier Cameron},
  {Gillon}, {Jehin}, {Lendl}, {Maxted}, {Pepe}, {Pollacco}, {Queloz},
  {S{\'e}gransan}, {Smalley}, {Smith}, {Southworth}, {Triaud}, {Udry}, \&
  {West}}]{2011Hellier}
{Hellier}, C., {Anderson}, D.~R., {Collier Cameron}, A., {et~al.} 2011, \aap,
  535, L7

\bibitem[{{Heng} {et~al.}(2011){Heng}, {Menou}, \& {Phillipps}}]{2011aHeng}
{Heng}, K., {Menou}, K., \& {Phillipps}, P.~J. 2011, \mnras, 413, 2380

\bibitem[{{Heng} \& {Showman}(2015)}]{2015Heng}
{Heng}, K., \& {Showman}, A.~P. 2015, Annual Review of Earth and Planetary
  Sciences, 43, 509

\bibitem[{{Hoskins} {et~al.}(1985){Hoskins}, {McIntyre}, \&
  {Robertson}}]{1985Hoskins}
{Hoskins}, B.~J., {McIntyre}, M.~E., \& {Robertson}, A.~W. 1985, Quarterly
  Journal of the Royal Meteorological Society, 111, 877

\bibitem[{{Hu} \& {Seager}(2014)}]{2014Hu}
{Hu}, R., \& {Seager}, S. 2014, \apj, 784, 63

\bibitem[{{Husser} {et~al.}(2013){Husser}, {Wende-von Berg}, {Dreizler},
  {Homeier}, {Reiners}, {Barman}, \& {Hauschildt}}]{2013Husser}
{Husser}, T.-O., {Wende-von Berg}, S., {Dreizler}, S., {et~al.} 2013, \aap,
  553, A6

\bibitem[{{Iro} {et~al.}(2005){Iro}, {B{\'e}zard}, \& {Guillot}}]{2005Iro}
{Iro}, N., {B{\'e}zard}, B., \& {Guillot}, T. 2005, \aap, 436, 719

\bibitem[{{Jablonowski} \& {Williamson}(2011)}]{2011Jablonowski}
{Jablonowski}, C., \& {Williamson}, D. 2011, Numerical Techniques for Global
  Atmospheric Models, Springer Berlin Heidelberg, 381

\bibitem[{{Kataria} {et~al.}(2015){Kataria}, {Showman}, {Fortney}, {Stevenson},
  {Line}, {Kreidberg}, {Bean}, \& {D{\'e}sert}}]{2015Kataria}
{Kataria}, T., {Showman}, A.~P., {Fortney}, J.~J., {et~al.} 2015, \apj, 801, 86

\bibitem[{{Knutson} {et~al.}(2009){Knutson}, {Charbonneau}, {Cowan}, {Fortney},
  {Showman}, {Agol}, {Henry}, {Everett}, \& {Allen}}]{2009knutson}
{Knutson}, H.~A., {Charbonneau}, D., {Cowan}, N.~B., {et~al.} 2009, \apj, 690,
  822

\bibitem[{{Kopparapu} {et~al.}(2012){Kopparapu}, {Kasting}, \&
  {Zahnle}}]{2012Kopparapu}
{Kopparapu}, R.~k., {Kasting}, J.~F., \& {Zahnle}, K.~J. 2012, \apj, 745, 77

\bibitem[{{Kreidberg} {et~al.}(2014){Kreidberg}, {Bean}, {D{\'e}sert}, {Line},
  {Fortney}, {Madhusudhan}, {Stevenson}, {Showman}, {Charbonneau},
  {McCullough}, {Seager}, {Burrows}, {Henry}, {Williamson}, {Kataria}, \&
  {Homeier}}]{2014Kreidberg}
{Kreidberg}, L., {Bean}, J.~L., {D{\'e}sert}, J.-M., {et~al.} 2014, \apjl, 793,
  L27

\bibitem[{{Lebonnois} {et~al.}(2013){Lebonnois}, {Lee}, {Yamamoto}, {Dawson},
  {Lewis}, {Mendonca}, {Read}, {Parish}, {Schubert}, {Bengtsson}, {Grinspoon},
  {Limaye}, {Schmidt}, {Svedhem}, \& {Titov}}]{2013Lebonnois}
{Lebonnois}, S., {Lee}, C., {Yamamoto}, M., {et~al.} 2013, Towards
  Understanding the Climate of Venus, ISSI Scientific Report Series, Volume 11.
  ISBN 978-1-4614-5063-4, Springer Science+Business Media New York, p. 129

\bibitem[{{Madhusudhan}(2012)}]{2012Madhusudhan}
{Madhusudhan}, N. 2012, \apj, 758, 36

\bibitem[{{Malik} {et~al.}(2017){Malik}, {Grosheintz}, {Mendon{\c c}a},
  {Grimm}, {Lavie}, {Kitzmann}, {Tsai}, {Burrows}, {Kreidberg}, {Bedell},
  {Bean}, {Stevenson}, \& {Heng}}]{2017Malik}
{Malik}, M., {Grosheintz}, L., {Mendon{\c c}a}, J.~M., {et~al.} 2017, \aj, 153,
  56

\bibitem[{{Manabe} {et~al.}(1965){Manabe}, {Smagorinsky}, \&
  {Strickler}}]{1965Manabe}
{Manabe}, S., {Smagorinsky}, J., \& {Strickler}, R.~F. 1965, Monthly Weather
  Review, 93, 769

\bibitem[{{McIntyre}(1995)}]{1995Mcintyre}
{McIntyre}, M.~E. 1995, Philosophical Transactions of the Royal Society of
  London Series A, 352, 227

\bibitem[{{Mendon{\c c}a} {et~al.}(2016){Mendon{\c c}a}, {Grimm}, {Grosheintz},
  \& {Heng}}]{2016Mendoncab}
{Mendon{\c c}a}, J.~M., {Grimm}, S.~L., {Grosheintz}, L., \& {Heng}, K. 2016,
  \apj, 829, 115

\bibitem[{{Mendon{\c c}a} {et~al.}(2018){Mendon{\c c}a}, {Malik}, {Demory}, \&
  {Heng}}]{2017Mendoncaa}
{Mendon{\c c}a}, J.~M., {Malik}, M., {Demory}, B.-O., \& {Heng}, K. 2018, \aj,
  155, 150

\bibitem[{{Mendon{\c c}a} \& {Read}(2016)}]{2016Mendonca}
{Mendon{\c c}a}, J.~M., \& {Read}, P.~L. 2016, \planss, 134, 1

\bibitem[{{Mendon{\c c}a} {et~al.}(2015){Mendon{\c c}a}, {Read}, {Wilson}, \&
  {Lee}}]{2015Mendonca}
{Mendon{\c c}a}, J.~M., {Read}, P.~L., {Wilson}, C.~F., \& {Lee}, C. 2015,
  \planss, 105, 80

\bibitem[{{Menou} \& {Rauscher}(2009)}]{2009Menou}
{Menou}, K., \& {Rauscher}, E. 2009, \apj, 700, 887

\bibitem[{{Moses}(2014)}]{2014Moses}
{Moses}, J.~I. 2014, Philosophical Transactions of the Royal Society of London
  Series A, 372, 20130073

\bibitem[{{Moses} {et~al.}(2011){Moses}, {Visscher}, {Fortney}, {Showman},
  {Lewis}, {Griffith}, {Klippenstein}, {Shabram}, {Friedson}, {Marley}, \&
  {Freedman}}]{2011Moses}
{Moses}, J.~I., {Visscher}, C., {Fortney}, J.~J., {et~al.} 2011, \apj, 737, 15

\bibitem[{{Perez-Becker} \& {Showman}(2013)}]{2013Perez-Becker}
{Perez-Becker}, D., \& {Showman}, A.~P. 2013, \apj, 776, 134

\bibitem[{{Prinn} \& {Barshay}(1977)}]{1977Prinn}
{Prinn}, R.~G., \& {Barshay}, S.~S. 1977, Science, 198, 1031

\bibitem[{{Read}(1986)}]{1986Read}
{Read}, P.~L. 1986, Quarterly Journal of the Royal Meteorological Society, 112,
  253

\bibitem[{{Richard} {et~al.}(2012){Richard}, {Gordon}, {Rothman}, {Abel},
  {Frommhold}, {Gustafsson}, {Hartmann}, {Hermans}, {Lafferty}, {Orton},
  {Smith}, \& {Tran}}]{2012Richard}
{Richard}, C., {Gordon}, I.~E., {Rothman}, L.~S., {et~al.} 2012, \jqsrt, 113,
  1276

\bibitem[{{Rimmer} \& {Helling}(2016)}]{2016Rimmer}
{Rimmer}, P.~B., \& {Helling}, C. 2016, \apjs, 224, 9

\bibitem[{{Rothman} {et~al.}(2010){Rothman}, {Gordon}, {Barber}, {Dothe},
  {Gamache}, {Goldman}, {Perevalov}, {Tashkun}, \& {Tennyson}}]{2010Rothman}
{Rothman}, L.~S., {Gordon}, I.~E., {Barber}, R.~J., {et~al.} 2010, \jqsrt, 111,
  2139

\bibitem[{{Rothman} {et~al.}(2013){Rothman}, {Gordon}, {Babikov}, {Barbe},
  {Chris Benner}, {Bernath}, {Birk}, {Bizzocchi}, {Boudon}, {Brown},
  {Campargue}, {Chance}, {Cohen}, {Coudert}, {Devi}, {Drouin}, {Fayt}, {Flaud},
  {Gamache}, {Harrison}, {Hartmann}, {Hill}, {Hodges}, {Jacquemart}, {Jolly},
  {Lamouroux}, {Le Roy}, {Li}, {Long}, {Lyulin}, {Mackie}, {Massie},
  {Mikhailenko}, {M{\"u}ller}, {Naumenko}, {Nikitin}, {Orphal}, {Perevalov},
  {Perrin}, {Polovtseva}, {Richard}, {Smith}, {Starikova}, {Sung}, {Tashkun},
  {Tennyson}, {Toon}, {Tyuterev}, \& {Wagner}}]{2013Rothman}
{Rothman}, L.~S., {Gordon}, I.~E., {Babikov}, Y., {et~al.} 2013, \jqsrt, 130, 4

\bibitem[{{Showman} {et~al.}(2009){Showman}, {Fortney}, {Lian}, {Marley},
  {Freedman}, {Knutson}, \& {Charbonneau}}]{2009Showman}
{Showman}, A.~P., {Fortney}, J.~J., {Lian}, Y., {et~al.} 2009, \apj, 699, 564

\bibitem[{{Showman} {et~al.}(2014){Showman}, {Lewis}, \&
  {Fortney}}]{2014Showman}
{Showman}, A.~P., {Lewis}, N.~K., \& {Fortney}, J.~J. 2014, ArXiv e-prints,
  arXiv:1411.4731

\bibitem[{{Skamarock} \& {Klemp}(2008)}]{2008Skamarock}
{Skamarock}, W.~C., \& {Klemp}, J.~B. 2008, Journal of Computational Physics,
  227, 3465

\bibitem[{{Smith}(1998)}]{1998Smith}
{Smith}, M.~D. 1998, \icarus, 132, 176

\bibitem[{{Stevenson} {et~al.}(2014){Stevenson}, {D{\'e}sert}, {Line}, {Bean},
  {Fortney}, {Showman}, {Kataria}, {Kreidberg}, {McCullough}, {Henry},
  {Charbonneau}, {Burrows}, {Seager}, {Madhusudhan}, {Williamson}, \&
  {Homeier}}]{2014Stevenson}
{Stevenson}, K.~B., {D{\'e}sert}, J.-M., {Line}, M.~R., {et~al.} 2014, Science,
  346, 838

\bibitem[{{Stock} {et~al.}(2018){Stock}, {Kitzmann}, \& {Patzer}}]{2017Stock}
{Stock}, J., {Kitzmann}, D., \& {Patzer}, A. B.~C. 2018, in preparation to
  A$\&$A

\bibitem[{{Tomita} {et~al.}(2001){Tomita}, {Tsugawa}, {Satoh}, \&
  {Goto}}]{2001Tomita}
{Tomita}, H., {Tsugawa}, M., {Satoh}, M., \& {Goto}, K. 2001, Journal of
  Computational Physics, 174, 579

\bibitem[{Tsai {et~al.}(2018)Tsai, Kitzmann, Lyons, Mendonça, Grimm, \&
  Heng}]{2018Tsai}
Tsai, S.-M., Kitzmann, D., Lyons, J.~R., {et~al.} 2018, The Astrophysical
  Journal, 862, 31

\bibitem[{{Tsai} {et~al.}(2017){Tsai}, {Lyons}, {Grosheintz}, {Rimmer},
  {Kitzmann}, \& {Heng}}]{2017Tsai}
{Tsai}, S.-M., {Lyons}, J.~R., {Grosheintz}, L., {et~al.} 2017, \apjs, 228, 20

\bibitem[{{Venot} \& {Ag{\'u}ndez}(2015)}]{2015Venot}
{Venot}, O., \& {Ag{\'u}ndez}, M. 2015, Experimental Astronomy, 40, 469

\bibitem[{{Wang} {et~al.}(2013){Wang}, {van Boekel}, {Madhusudhan}, {Chen},
  {Zhao}, \& {Henning}}]{2013Wang}
{Wang}, W., {van Boekel}, R., {Madhusudhan}, N., {et~al.} 2013, \apj, 770, 70

\bibitem[{{Yurchenko} {et~al.}(2011){Yurchenko}, {Barber}, \&
  {Tennyson}}]{2011Yurchenko}
{Yurchenko}, S.~N., {Barber}, R.~J., \& {Tennyson}, J. 2011, \mnras, 413, 1828

\bibitem[{{Yurchenko} \& {Tennyson}(2014)}]{2014Yurchenko}
{Yurchenko}, S.~N., \& {Tennyson}, J. 2014, \mnras, 440, 1649

\end{thebibliography}

\end{document}